\smallskip

\documentclass[longauth]{aa}

\usepackage{siunitx}

\usepackage{CJK}

\usepackage{txfonts}
\usepackage{verbatim}
\usepackage[center]{caption}
\usepackage{multirow}

\usepackage{natbib}
\usepackage{float}
\usepackage{caption}
\usepackage{subfigure}
\usepackage{graphics}
\usepackage{geometry}
\usepackage{xcolor}
\usepackage{placeins} 
\usepackage{url}
\usepackage{appendix}
\usepackage{tabularx}
\usepackage{booktabs}
\usepackage{orcidlink}

\usepackage[version=4]{mhchem}

\begin{document} 
\begin{CJK*}{UTF8}{gbsn}

   \title{ALOHA IRDCs Molecular Line Follow-up: I. Gas properties and kinematics}

    \author{Jinjin Xie (谢津津)  \inst{1,2}, Yaoting Yan (闫耀庭)\inst{3,4}, Zhiyuan Ren \inst{2}, Jarken Esimbek \inst{1,5,6},  Di Li \inst{7}, Yan Duan (段言)\inst{8,2}, 
  Gary A. Fuller \inst{9, 10}, Nicolas Peretto \inst{11},  Jingwen Wu \inst{6,2}, Wenjin Yang \inst{12}
% ...
  }
   \institute{State Key Laboratory of Radio Astronomy and Technology, Xinjiang Astronomical Observatory, Chinese Academy of Sciences, 150 Science 1-Street, Urumqi, Xinjiang 830011, China\\          \email{\\dili@tsinghua.edu.cn,jarken@xao.ac.cn,renzy@nao.cas.cn}
     \and
            National Astronomical Observatories, Chinese Academy of Sciences, Beijing 100101, China
            \and
 Max-Planck-Institut f\"ur Radioastronomie, Auf dem H\"ugel 69, 53121 Bonn, Germany
\and
Division of Science, National Astronomical Observatory of Japan, 2-21-1 Osawa, Mitaka, Tokyo 181-8588, Japan
\and
Xinjiang Key Laboratory of Radio Astrophysics, Urumqi 830011, China
\and
University of the Chinese Academy of Sciences, Beijing 100080, China 
\and
New Cornerstone Science Laboratory, Department of Astronomy, Tsinghua University, Beijing 100084, China
\and
 Space Engineering University, Beijing 101416, China
\and 
Jodrell Bank Centre for Astrophysics, Department of Physics \& Astronomy, The University of Manchester, Manchester M13 9PL, UK
\and 
I. Physikalisches Institut, University of Cologne, Z\"ulpicher Str. 77, 50937 K\"oln, Germany
\and
School of Physics \& Astronomy, Cardiff University, Queen\textquotesingle s Building, The Parade, Cardiff, CF24 3AA, UK
\and
School of Astronomy and Space Science, Nanjing University, 163 Xianlin Avenue, Nanjing 210023, China
% ...
}

  \abstract
   {Infrared Dark Clouds (IRDCs) are ideal sites for investigating the initial conditions of massive star and cluster formation. The {\it A Lei Of the Habitat and Assembly of Infrared Dark Clouds} (ALOHA IRDCs), a James Clerk Maxwell Telescope (JCMT) Large Program, has mapped nearby IRDCs utilising the SCUBA-2 continuum camera at the highest achievable sensitivity. Complementary molecular line observations are needed to characterise the physical, kinematic, and chemical properties of the dense gas.}
   {We aim to determine the thermal, kinematic, and chemical properties of clumps identified in the ALOHA IRDCs, and to assess their evolutionary status and level of star-forming activity. } 
   {We performed single-pointing K-band and W-band observations towards 56 ALOHA IRDCs clumps using the Effelsberg 100-m and Yebes 40-m telescopes, respectively. We derived NH$_{3}$ kinetic temperatures using the hyperfine group ratio (HFGR) method and identified infall and shock signatures from HCO$^+$, H$^{13}$CO$^+$, SiO, and HNCO profiles. Water masers and NH$_{2}$D emission were used as complementary tracers of chemical evolution and star formation. 
   }
   {The clumps exhibit kinetic temperatures of 15--29\,K. We detect NH$_2$D emission towards 18 sources, with NH$_{2}$D centroid velocities consistent with NH$_{3}$, indicating both species trace the same dense gas component. More than half of the clumps display blue-asymmetric HCO$^+$ profiles, identifying them as infall candidates. Water masers are detected in 22 sources, showing prominent velocity ranges and variability. Broad SiO emission ($\gtrsim20\,\mathrm{km\,s^{-1}}$) indicates strong shocks, while narrower extents ($\lesssim6\,\mathrm{km\,s^{-1}}$) likely trace large-scale interactions or low-velocity shocks.
    }
   {The widespread infall signatures, shock tracers, masers, and NH$_{2}$D emission suggest that relatively quiescent, chemically young material can coexist with dynamically active gas affected by early protostellar feedback, providing insight into the coupled physical and chemical evolution of massive IRDC clumps.
  }

   \titlerunning{ALOHA IRDCs Molecular Line Follow-up}
   \authorrunning{Xie et al. }
   \maketitle

\section{Introduction}

Infrared Dark Clouds (IRDCs), characterised by their low temperatures and high densities, provide ideal environments for studying the earliest stages of massive star formation and the formation of associated stellar clusters \citep{2000carey,2006rathborne,2018motte}. Large IRDC catalogues have been compiled from the {\it Midcourse Space Experiment} (MSX) by \citet{2006simon} and from the {\it Spitzer} Space Telescope by \citet{2009perettofuller}. Subsequent continuum surveys with APEX \citep{2009schuller,2020peretto}, the James Clerk Maxwell Telescope \citep[JCMT; ][]{2009parsons}, {\it Herschel} \citep[e.g.,][]{2010peretto}, and the Atacama Large Millimeter/submillimeter Array \citep[ALMA; e.g.,][]{2019sanhueza,2023morii} have revealed that IRDCs contain networks of filaments, hubs, and compact clumps that host the initial conditions of massive star formation.

The ALOHA IRDCs (A Lei Of the Habitat and Assembly of Infrared Dark Clouds)\footnote{\url{https://www.eaobservatory.org/jcmt/science/large-programs/aloha-irdcs/}} project is a JCMT Large Program targeting a sample of IRDCs. Using the SCUBA-2 continuum receiver \citep{2013dempsey,2013chapin,2021mairs}, it aims to map these clouds to a sensitivity of approximately $2\,\mathrm{mJy\,beam^{-1}}$. The parent target list was constructed from the catalogue of \citet{2009perettofuller} using physically motivated criteria designed to select nearby, high-column-density clouds likely to host massive star formation, including a nominal distance limit of $\lesssim$1.4\,kpc based on the kinematic distance estimates available at the time of sample selection, together with column densities $N_{\rm H_2} > 1 \times 10^{22}$\,cm$^{-2}$. In the present work, however, we adopt revised kinematic distances derived from the NH$_3$ systemic velocities (Table~\ref{tab:sources}). For sources without detectable NH$_3$ emission, distances derived from CO observations of the associated cloud complexes were adopted (Ren et al., in prep.). These updated distances span a substantially broader range than the original selection values, and therefore some sources no longer satisfy the initial nominal distance criterion. We nevertheless include them here because they belong to the observed ALOHA sample. Although CO provides the cloud-scale velocities used here, it may not trace the entire molecular gas reservoir, particularly in low-density or strongly irradiated regions where C\,{\sc i} can reveal a substantial CO-dark component \citep{2026xia}. A first analysis of one ALOHA field has already demonstrated that this sensitivity reveals a rich network of filaments and compact clumps within IRDCs \citep{2024shen}. The dust-continuum properties of the full ALOHA IRDCs programme are being investigated through complementary studies of the overall survey characteristics, the multiscale structure of the clouds, and the physical properties of their dense cores (Ren et al., in prep.; Lyu et al., in prep.; \citealt{2026liu}). By combining ground- and space-based observations following \citet{2022jiao}, Ren et al. and Lyu et al. characterise the dust emission across a broad range of spatial scales.

The conversion of atomic gas into molecular gas establishes the reservoir available for star formation, and its efficiency depends on environmental conditions such as gas density and metallicity \citep{2024yu}. Ammonia (NH$_{3}$) and its deuterated isotopologue NH$_2$D are key probes of cold and dense gas in IRDCs. Ammonia is considered a reliable thermometer for interstellar clouds since its metastable inversion lines are sensitive to collisional excitation and insensitive to radiative excitation \citep{1983ho,1983walmsley,2002li}. Studies of IRDCs consistently find kinetic temperatures below 30\,K \citep{2006pillai,2011ragan,2013chira,2021xie,2022lishh}. The inversion transitions of NH$_{3}$ (1,1) and (2,2) allow resolving rotational and excitation temperatures, providing fundamental constraints on dense gas physical conditions and chemical evolution \citep{2020wang,2021xie,2022lishh,2024wang}. Deuterated ammonia (NH$_{2}$D) forms efficiently in cold CO-depleted gas through deuterium fractionation \citep{2002tafalla,2008caselli,2011pillai} and thus traces chemically young material in the earliest evolutionary phases within IRDCs \citep{2013sanhueza,2016lackington}.

Kinematic diagnostics further reveal how these cold clumps evolve dynamically. Optically thick tracers such as HCO$^+$, together with optically thin isotopologues like H$^{13}$CO$^+$, are widely used to identify infall motions, velocity gradients, and non-thermal support, and many IRDCs exhibit blue-skewed HCO$^+$ profiles and enhanced linewidths that indicate ongoing collapse \citep[e.g.,][]{2013sanhueza,2013peretto,2021xieraa}. As material accretes onto forming protostars, the infalling gas interacts with outflows and dense ambient clumps, naturally generating shocks and turbulent dissipation \citep{2010sanhueza}. Shock tracers such as SiO, HNCO, and SO selectively probe this dynamically processed gas and are widely interpreted as evidence for grain sputtering and early protostellar feedback \citep[e.g.,][]{1994chernin,2014leurini,2023xie,2024rigby}.

Masers provide additional signposts of early star formation and feedback \citep[e.g.,][]{2020chen,2022baan}. Water masers are observed in both low-mass and high-mass star-forming regions, whereas CH$_{3}$OH masers are much more closely associated with massive star formation \citep{1995codella,1997codella,1998walsh,2005goddi,2018kim,2019kim,2020moscadelli}. In IRDCs, water maser emission is detected in only a small fraction ($<10\%$) of dense cores, indicating that it traces the onset of embedded star formation \citep{2006wang}. Both H$_{2}$O masers and Class\,{\sc i} methanol masers are collisionally pumped in shocked gas \citep[e.g.,][]{1992cragg,2013hollenbach}. Water masers are typically located near the base of protostellar jets, whereas Class\,{\sc i} methanol masers arise at outflow--ambient interfaces \citep{2013kalenskii,2023yang}. The 25\,GHz Class\,{\sc i} methanol masers were first discovered in Orion by \citet{1971barrett} and \citet{1975hills}, and were later shown to be highly sensitive to local physical conditions \citep{1988menten}. Subsequent Effelsberg 100-m observations confirmed multiple 25\,GHz transitions and refined their rest frequencies \citep{2015gong}. These masers require higher temperatures and densities than the more common 36, 44, and 95\,GHz Class\,{\sc i} masers, making them selective tracers of strong shocks in outflows and jet--ambient interfaces \citep{1992cragg,2005sobolev,2019ladeyschikov}. Their occurrence in IRDCs, however, remains poorly constrained.

In this paper, we present follow-up molecular line observations of ALOHA IRDC clumps that characterise the thermal, kinematic, and chemical properties and their relation to early star formation. Sect. \ref{sect:obs} describes the observations, Sect. \ref{sect:results} presents the results, Sect. \ref{sect:discussion} discusses the thermal, kinematic, and chemical implications of the results, and Sect. \ref{sect:concl} summarises the conclusions.

\section{Observations and data reduction}\label{sect:obs}

The positions and physical properties of the 56 IRDC clumps analysed in this work are given in Table~\ref{tab:sources}, and the rest frequencies of the molecular lines are listed in Table~\ref{tab:linesfrequencies}. Both Effelsberg\footnote{This publication is based on observations with the 100-m telescope of the MPIfR (Max-Planck-Institut f\"ur Radioastronomie) at Effelsberg.} and Yebes\footnote{Based on observations with the 40-m radio telescope of the National Geographic Institute of Spain (IGN) at Yebes Observatory, under proposal 22A020.} observations were conducted towards the dust continuum peak positions of the clumps using the standard position-switching mode.

The Effelsberg K-band observations were performed in July 2023 using the S14mm double-beam secondary focus receiver to simultaneously cover the entire K-band frequency range (18.0--26.0\,GHz). The Effelsberg off position was 10 arcmin in azimuth away from the source. The integration time for each source was 9.8 min. For the whole frequency range, the full width at half maximum (FWHM) varied from 35" to 50", corresponding to frequencies from 26\,GHz to 18\,GHz. For the 8\,GHz frequency width spectra, the receiver band was divided into four 2.5\,GHz-wide subbands, each of which has 65536 channels with a channel width of 38.1\,kHz, yielding a velocity resolution of $0.62\,\mathrm{km\,s^{-1}}$ at 18.5\,GHz and $0.44\,\mathrm{km\,s^{-1}}$ at 26.0\,GHz. The main beam efficiency was 79\% at 22.85\,GHz \footnote{\url{https://eff100mwiki.mpifr-bonn.mpg.de}}. The focus was checked every two hours, additionally after sunrise and sunset. Pointing was obtained every hour towards nearby quasars and was found to be accurate to within 5". The system temperature was around 160\,K. We split the data into frequency ranges with small bandwidths of 300\,MHz and calibrated it based on continuum cross scans of NGC 7027 \citep{2012winkel,2024yan,2025alkhuja}. The adopted flux density of NGC 7027 follows \citet{1994ott}. The conversion factors from flux density to main-beam brightness temperature, $T_{\rm MB}/S$, are $1.95\,\mathrm{K\,Jy^{-1}}$, $1.73\,\mathrm{K\,Jy^{-1}}$, and $1.68\,\mathrm{K\,Jy^{-1}}$ at 18.5\,GHz, 22.2\,GHz, and 24.0\,GHz, respectively. Calibration uncertainties were estimated to be $\pm10$\% \citep{2024yan}.

W-band (72.8--91.3\,GHz) line survey observations were conducted towards the same sources with the Yebes 40-m radio telescope during February, March, and May 2022. The off-source positions were selected $+30$ arcmin offset in Right Ascension from the source centres. The system temperature was around 130\,K. The full bandwidth in the W band was 18.5\,GHz, with a spectral resolution of 38\,kHz. The spectra were smoothed to 153\,kHz, corresponding to velocity resolutions of $0.63\,\mathrm{km\,s^{-1}}$ at 72.8\,GHz and $0.51\,\mathrm{km\,s^{-1}}$ at 90\,GHz, to improve the signal-to-noise ratio of the measured line emission in a given channel. The half-power beam width (HPBW) varied from 19.9" at 88.5\,GHz to 24.3" at 72.8\,GHz. The telescope main beam efficiencies were 0.30, 0.29, 0.28, 0.27, 0.25, 0.24, 0.23, and 0.22 for spectral windows each covering $\sim$2.5\,GHz from 72.8\,GHz to 91.3\,GHz, with 200\,MHz of overlap at the edges of the different bands.

For data reduction of both the Effelsberg and Yebes observations, the GILDAS\footnote{\url{http://www.iram.fr/IRAMFR/GILDAS}} software package that includes CLASS was used \citep{2013gildas}. First-order baselines were subtracted from all spectra.

\section {Results}\label{sect:results}

In the Effelsberg observations, NH$_{3}$ (1,1) emission is detected towards 54 of 56 sources, while 48, 28, and two sources are detected in NH$_{3}$ (2,2), (3,3), and (4,4) transitions, respectively. Simultaneous K-band observations with the Effelsberg telescope also detected Class\,{\sc i} methanol masers towards one source and water maser emission towards 22 sources. Among the detected water masers, one (SDC33O) is newly identified in this work. A spectral line was considered detected when its signal-to-noise ratio satisfied $S/N\geq3$.

In the W-band Yebes observations, the dense gas tracers HCO$^+$ and H$^{13}$CO$^+$ are detected towards 53 and 49 of the 56 sources, respectively. H$^{13}$CO$^+$ features with $S/N\gtrsim3$
are included in this detection count but are identified as
tentative in Table~\ref{tab:hcopresidual}. SiO and HNCO emission is detected towards 19 and 20 sources, respectively. SO is detected in only six sources. Deuterated ammonia (NH$_{2}$D) is detected towards 18 sources.

\subsection{Kinetic temperature derived from NH$_{3}$}

We applied the hyperfine group ratio (HFGR) method \citep{2013li,2020wang} to derive rotational temperature $T_{\rm rot}$ and kinetic temperature $T_{\rm kin}$ from the observed NH$_{3}$ lines. The Gaussian-fitted parameters of NH$_{3}$ are listed in Table~\ref{tab:NH3class}. The HFGR technique has been successfully used in a variety of interstellar environments \citep[e.g.,][]{2020feng,2021xie,2021chen}, owing to its efficiency and robustness in temperature estimation. This method utilises the integrated intensities of hyperfine groups of NH$_{3}$ that relies solely on the observed line-intensity-ratios. It thus circumvents the hyperfine-fitting and is impervious to the coupling between linewidth and opacity. The statistical uncertainties are estimated directly from the signal-to-noise ratio of the observed spectra. The derived kinetic temperatures of the measured velocity components range between 14.8 and 29.4\,K, with a median of 21.4\,K. The Gaussian FWHM values of the central hyperfine group of NH$_{3}$ (1,1) have a range of $1.2-7.5\,\mathrm{km\,s^{-1}}$ with $2.3\,\mathrm{km\,s^{-1}}$ as the median value. 

The total column density of NH$_{3}$ is determined using {\it N}(1,1) $\times$Z/Z(1,1) where {\it Z} is the partition function:
\begin{equation}
\begin{aligned}
Z = \sum_{i}\left(2J+1\right){\it S}\left(J\right)\exp\frac{-h\left[BJ\left(J+1\right)+\left(C-B\right)J^{2}\right]}{kT_{\rm rot}},
\end{aligned}
\end{equation}
and 
\begin{equation}
\begin{aligned}
Z(1,1) = 3S\left(1\right)\exp\frac{-h\left[2B+\left(C-B\right)\right]}{kT_{\rm rot}},
\end{aligned}
\end{equation}
where {\it S(J)} is the extra statistical weight of ortho- over para-NH$_{3}$ states and equals 2 for {\it J} = 3, 6, 9, ... and equals 1 for other {\it J} values. The rotational constants are ${\it B}=298117\,\mathrm{MHz}$ and ${\it C}=186726\,\mathrm{MHz}$ \citep{1998pickett}, while $h$ and $k$ are Planck and Boltzmann constants, respectively. The column density of the NH$_{3}$ (1,1) transition is then calculated using the equation from \citet{2009friesen}:
\begin{equation}\label{eq:n11}
\begin{aligned}
{\it N}(1,1) = \frac{8\pi\nu_{0}^{2}}{c^{2}}\frac{g_{1}}{g_{2}}\frac{1}{A\left(1,1\right)}
\times \frac{1+\exp\left(-h\nu_{0}/kT_{\rm ex}\right)}{1-\exp\left(-h\nu_{0}/kT_{\rm ex}\right)} \int\tau\left(\nu\right)d\nu,
\end{aligned}
\end{equation}
where {\it g} is the degeneracy of the corresponding energy level. $\tau\left(\nu\right)$ is the optical depth as a function of the frequency $\nu$ determined from the hyperfine fit. The Einstein {\it A} coefficient {\it A}(1,1) is $1.68\times10^{-7}\,\mathrm{s^{-1}}$ \citep{1998pickett}. The excitation temperature, $T_{\rm ex}$ is related to $T_{\rm mb}$ through,
\begin{equation}\label{eq:tmb}
\begin{aligned}
{\it T_{\rm mb}} = \eta_{f}\left[J_{\nu}\left(T_{\rm ex}\right)-J_{\nu}\left(T_{\rm bg}\right)\right]\left[1-e^{-\tau(\nu)}\right],
\end{aligned}
\end{equation}
where $\eta_{\rm f}$ is the main beam filling factor. Assuming that the emission from the molecular-line is extended and uniformly distributed within the beam, $\eta_{\rm f}$ is then $\sim$1. $J_{\nu}(T)$ is the Rayleigh-Jeans equivalent temperature of a black body at a temperature {\it T},
\begin{equation}\label{eq:J}
\begin{aligned}
{\it J_{\nu}\left(T\right)} \equiv \frac{\frac{h\nu}{k}}{\exp\left(\frac{h\nu}{kT}\right)-1}.
\end{aligned}
\end{equation}

The kinetic temperature can be derived from the rotational temperature measured from the NH$_3$ (1,1) and (2,2) inversion transitions \citep{2004tafalla}, i.e., 
\begin{equation}
T_{\rm kin} =
\frac{T_{R}^{21}}
{1-\dfrac{T_{R}^{21}}{42}
\ln\!\left[1+1.1\exp\!\left(-16/T_{R}^{21}\right)\right]} ,
\end{equation}
where $T_{R}^{21}$ is the rotational temperature between the (1,1) and (2,2) metastable levels.

For velocity components without detectable NH$_3$(2,2) emission, the $3\sigma$ upper limits on the NH$_3$(2,2) peak brightness temperature listed in Table~\ref{tab:NH3class} were propagated through the same analysis to derive upper limits on $T_{\rm rot}$ and $T_{\rm kin}$. The derived values of these parameters are listed in Table~\ref{tab:NH3calculate}. For the 51 velocity components with measured column densities, $N_{\rm tot}(\mathrm{NH}_3)$ ranges from $(0.5$--$14.5)\times10^{14}\,\mathrm{cm^{-2}}$, with a mean of $3.8\times10^{14}\,\mathrm{cm^{-2}}$. The remaining nine components provide $3\sigma$ upper limits of $<(0.2$--$3.2)\times10^{14}\,\mathrm{cm^{-2}}$.

\subsection{Water and methanol masers}

The fitted parameters of the detected H$_{2}$O and Class\,{\sc i} methanol masers are listed in Table~\ref{tab:watermaser} and Table~\ref{tab:methanolmaser}, respectively. The RMS noise levels for non-detections in H$_{2}$O are given in Table~\ref{tab:nowatermaser}. The source SDC19A, which exhibits eight sequential transitions of Class\,{\sc i} methanol maser emission, also shows a water maser. Previous VLA (Very Large Array) observations of this source identified three methanol maser transitions, with flux densities that differ from those obtained by Effelsberg, as shown in Table~\ref{tab:methanolmaser}. The corresponding centroid velocity differences between the VLA and Effelsberg observations are $\lesssim1\,\mathrm{km\,s^{-1}}$. This source shows the highest water maser peak flux density within our sample and is the only source with detected Class\,{\sc i} methanol maser emission. The mean water peak flux density over the 22 detected sources is 5.7\,Jy. In SDC19A, the mean peak flux density over the eight detected Class\,{\sc i} methanol maser transitions is 2.5\,Jy.

In comparison with previous observations, the water masers detected in our data show clear variability in both spectral profiles and velocities. For example, \citet{2022ladeyschikov} reported a single water-maser component towards G19.364-0.030, located 10" from SDC19A, spanning a velocity range of approximately 0 to $\sim32\,\mathrm{km\,s^{-1}}$ observed with Effelsberg 100-m telescope, with a peak flux density of 1.51\,Jy at a velocity of $28.8\,\mathrm{km\,s^{-1}}$. In contrast, our observations reveal a substantially higher peak flux density of 30.3\,Jy at $32.53\,\mathrm{km\,s^{-1}}$, indicating pronounced temporal variability in the maser emission. 

The number of spectral components detected in the water masers ranges from one to ten, with an average of three. The water masers also show a wide velocity range and notable velocity shifts off the systemic velocity. Following \citet{2018kim}, we classify a source as containing high-velocity H$_2$O maser emission when at least one detected maser feature is offset from the NH$_3$ systemic velocity by more than $30\,\mathrm{km\,s^{-1}}$, which is
\begin{equation}
\max\!\left[
\left|V_{\min}({\rm H_2O})-V_{\rm sys}\right|,
\left|V_{\max}({\rm H_2O})-V_{\rm sys}\right|
\right]>30\,\mathrm{km\,s^{-1}}.
\end{equation} 
Nine of the 22 H$_2$O maser sources satisfy this criterion. Seven sources contain high-velocity blue-shifted features: SDC19C, SDC19E,
SDC33E, SDC33O, SDC35G, SDC35H, and SDC35S. Six contain
high-velocity red-shifted features: SDC19C, SDC35G, SDC35H,
SDC35I, SDC35L, and SDC35S. Four sources, SDC19C, SDC35G,
SDC35H, and SDC35S, contain both blue- and red-shifted
high-velocity features. The details of the comparisons among velocity ranges and velocity shifts, together with the implications for kinematics, are discussed in Sect.~\ref{sect:discussionmaser}.

\subsection{Kinematic tracers}

We use SiO, HNCO, SO, HCO$^+$, and H$^{13}$CO$^+$ to investigate the kinematic properties of the clumps. The full spectra are shown in Appendix~\ref{sec:kinematics}, and the Gaussian-fitted parameters are listed in Table~\ref{tab:shockfitted}. HCO$^+$ parameters are included for completeness, though this optically thick transition often shows non-Gaussian profiles. The velocity ranges are summarised in Table~\ref{tab:linewing}. For each velocity component, the systemic velocity was derived from the centroid velocity of the corresponding optically thin H$^{13}$CO$^{+}$ component. For sources without detectable H$^{13}$CO$^{+}$ emission, no H$^{13}$CO$^{+}$-based systemic velocity marker or central velocity interval was assigned. For each detected molecular species, the velocity interval was determined from the blue- and red-most contiguous channels with intensities exceeding $3\sigma$, where $\sigma$ is the rms noise level. For SiO, HNCO, and SO, the interval $[V_{\rm blue},V_{\rm red}]$ represents the full detected velocity range of the emission. We note that opacity and self-absorption can influence the observed HCO$^+$ profile. The velocity intervals listed in Table~\ref{tab:linewing} are used as descriptive measures of the detected profile extents rather than as direct measurements of intrinsic HCO$^+$ linewidths. Among sources detected in both H$^{13}$CO$^+$ and at least one shock tracer, SiO shows broader velocity extents than H$^{13}$CO$^{+}$ while HNCO shows slightly broader velocity extents than H$^{13}$CO$^{+}$, indicating the presence of additional dynamically perturbed gas components. Owing to the limited number of SO detections, the subsequent analysis focuses primarily on SiO and HNCO.

In the Yebes observations, broad SiO, HNCO, and SO emission is distinguished by comparing their detected velocity ranges and profile widths with those of the optically thin tracer H$^{13}$CO$^+$ where detected. Infall and outflow classifications were considered only for sources with detections of both HCO$^+$ and H$^{13}$CO$^+$. The number of velocity components was first determined from the optically thin H$^{13}$CO$^+$ profile. Sources with a single H$^{13}$CO$^+$ component were modelled with one Gaussian in HCO$^+$, whereas sources showing two resolved H$^{13}$CO$^+$ velocity components were fitted with two HCO$^+$ Gaussians, each associated with the corresponding H$^{13}$CO$^+$ component. Some H$^{13}$CO$^+$ spectra exhibit broadening or asymmetry. Such features are typically much weaker than those seen in HCO$^+$ and do not significantly affect the systemic velocity determination. Following standard practice \citep[e.g.,][]{2005fuller,2016wyrowski,2018yoo,2021xieraa}, we adopt H$^{13}$CO$^+$ as a reliable optically thin tracer of dense gas kinematics. The combination of HCO$^+$ and H$^{13}$CO$^+$ therefore provides an effective diagnostic of infall motions where blue-skewed asymmetries in the optically thick line relative to the thin reference are interpreted as evidence of inward motions in the dense envelopes.

To quantify whether HCO$^+$ profiles contain significant non-Gaussian structure, we performed the residual analysis summarised in Table~\ref{tab:hcopresidual}. The one- or two-component Gaussian model was subtracted from each observed HCO$^+$ profile. Profiles for which ($RMS_{\rm residual} > RMS_{\rm baseline}$) are considered to exhibit significant residual structure above the baseline noise. The inferred kinematic classifications are listed in Table~\ref{tab:linewing}. `S' denotes the presence of shock activity, as traced by SiO emission. Outflow (OF, e.g., SDC19C) and infall (IN, e.g., SDC19E) features are inferred from red-shifted or blue-shifted asymmetries in the HCO$^+$ profiles identified from the combination of the profile morphology relative to H$^{13}$CO$^+$ and the residual criterion described above.

The prevalence of shock signatures across the sample indicates ongoing star formation activity within these sources. Three sources, SDC19A, SDC19B, and SDC19E, exhibit broad SiO velocity ranges of $\gtrsim20\,\mathrm{km\,s^{-1}}$. Among them, SDC19E also shows an HCO$^+$ profile consistent with possible infall motions, while SDC19A and SDC19B are classified as shock-dominated sources \citep[e.g.,][]{2011lopezsepulcre,2014dc}. In contrast, shocks characterised by narrow SiO velocity extents ($\lesssim6\,\mathrm{km\,s^{-1}}$) are generally attributed to large-scale, low-velocity interactions such as converging flows rather than highly collimated jets \citep[e.g.,][]{2010jimenezserra,2016Csengeri}. Such narrow SiO emission is detected in seven velocity components. Several of these components also show infall signatures, suggesting that the narrow SiO emission may in some cases be associated with infall motions.

\subsection{Singly deuterated ammonia NH\texorpdfstring{$_{2}$}{2}D}

The fitted spectral parameters of 18 NH$_2$D detections are summarised in Table~\ref{tab:NH2D}. All NH$_2$D spectra in this work were fitted using a hyperfine-structure (HFS) method \citep{2016daniel}, which explicitly accounts for the full hyperfine splitting of the transition and yields more reliable estimates of the intrinsic line width and centroid velocity than single-component Gaussian fitting. The resulting NH$_2$D line widths range from 0.7 to $3.5\,\mathrm{km\,s^{-1}}$, with a median value of $2.2\,\mathrm{km\,s^{-1}}$. For the 17 NH$_2$D-detected sources with measured kinetic temperatures, the mean kinetic temperature is $21.5 \pm 4.6$\,K, consistent with the mean of $21.6\pm3.9$\,K for the full measured sample. SDC33O is detected in NH$_3$(1,1) but not in NH$_3$(2,2), and therefore has only an upper limit of $T_{\rm kin}<17.6$\,K, which is used to calculate a nominal NH$_2$D column density. The centroid velocity offsets between NH$_2$D and NH$_3$ (1,1) span from $-0.6$ to $+1.3\,\mathrm{km\,s^{-1}}$, with a median of $0.3\,\mathrm{km\,s^{-1}}$, indicating that both species predominantly trace the same bulk gas component. Previous studies of NH$_2$D in IRDCs have reported a wide range of line widths, from relatively narrow values of $\sim1.3\,\mathrm{km\,s^{-1}}$ \citep{2013sanhueza} to substantially broader profiles of up to $\sim8.9\,\mathrm{km\,s^{-1}}$ \citep{2016feng}, consistent with the diversity of kinematic environments sampled by deuterated gas.

Assuming optically thin emission, we estimated the NH$_2$D column density from the integrated intensity of the ortho-NH$_2$D $1_{11}$--$1_{01}$ transition following \citet{2021wienen}. Because only one NH$_2$D transition was observed, its excitation temperature cannot be determined directly. We therefore adopted the NH$_3$ kinetic temperature as a proxy for $T_{\rm ex}({\rm NH_2D})$. For sources with two measured NH$_3$ velocity components, we adopted their mean kinetic temperature. For SDC33O, the nominal NH$_2$D column density is included in the full column density range and is excluded from further deuteration fraction analysis. The beam-averaged NH$_2$D column density was calculated as
\begin{equation}
N({\rm NH_2D})=
1.94\times10^{3}\,
\nu^{2}A_{ul}^{-1}\,
\frac{T_{\rm ex}}{T_{\rm ex}-2.7}\,
\exp\!\left(\frac{E_u}{kT_{\rm ex}}\right)\,
\frac{Q(T_{\rm ex})}{g}\,
W,
\end{equation}
where $W=\int T_{\rm mb}\,dv$ is the integrated intensity in K\,km\,s$^{-1}$, $\nu=85.926$\,GHz, $A_{ul}=7.82\times10^{-6}$\,s$^{-1}$, $E_u/k=20.7$\,K, $g=27$ from the CDMS (Cologne Database for Molecular Spectroscopy) database \citep{2001muller,2005muller,2016endres}. The total ortho- and para-NH$_{2}$D partition function was approximated as $Q(T_{\rm ex})=1.04\,T_{\rm ex}^{1.41}$ \citep{2021wienen}. The resulting NH$_2$D column densities range from $4.6\times10^{12}$ to $8.7\times10^{13}$\,cm$^{-2}$, with a median value of $1.2\times10^{13}$\,cm$^{-2}$. The uncertainties were propagated from the errors in both the integrated intensity and the adopted NH$_3$ kinetic temperature. We note that, under the optically thin assumption, these NH$_2$D column densities should be regarded as lower limits if the true optical depth is non-negligible. 

We define the ammonia deuteration fraction as
\begin{equation}
D_{\rm frac}({\rm NH_3}) = \frac{N({\rm NH_2D})}{N({\rm NH_3})}.
\end{equation}
Using the NH$_2$D column densities derived from the Yebes data and the NH$_3$ column densities derived from the Effelsberg data, we first compute an apparent deuteration fraction, $D_{\rm frac}^{\rm app}$, directly from the observed column-density ratio. Because the two species were observed with different beam sizes, this ratio is formally beam-mismatched. If both transitions arise from the same unresolved emitting region, a simple beam correction gives
\begin{equation}
D_{\rm frac}^{\rm corr} = D_{\rm frac}^{\rm app}\left(\frac{\theta_{\rm Yebes}}{\theta_{\rm Eff}}\right)^2,
\end{equation}
where we adopt $\theta_{\rm Yebes}=21.9''$ at 85.926\,GHz and a representative $\theta_{\rm Eff}\approx40''$, yielding a correction factor of 0.30. In the opposite limit of extended emission filling both beams, no correction is required. The intrinsic deuteration fraction is therefore expected to lie between $D_{\rm frac}^{\rm corr}$ and $D_{\rm frac}^{\rm app}$. For the 15 unambiguous NH$_2$D detections with measured NH$_{3}$ column densities and unambiguous NH$_{3}$ component matching, the apparent ammonia deuteration fractions span $0.011$--$0.089$, with a median value of 0.038. Under the compact-source beam-correction assumption, the corresponding range becomes $0.003$--$0.027$, with a median value of 0.011.

\section{Discussion}\label{sect:discussion}

    Our observations probe the physical and chemical properties of the ALOHA IRDCs clumps through NH$_{3}$ kinetic temperatures that are broadly consistent with the values reported for IRDC samples in previous studies \citep[e.g.,][]{2011pillai,2012wienen,2016svoboda,2021xie}. The molecular-line detection rates are generally higher than those in \citet{2012sanhueza}, especially for SiO and NH$_{2}$D, of which the detection rates are increased by factors of four and ten, respectively. In the following subsections, we explore how these physical, kinematic, and chemical properties relate to one another and discuss their implications for the early evolution of massive star-forming clumps.

\subsection{Thermal decoupling of gas and dust in the presence of shocks and turbulence}

\begin{figure}[ht]
  \centering
  \captionsetup{justification=raggedright, singlelinecheck=false}
  \includegraphics[width=0.90\columnwidth]{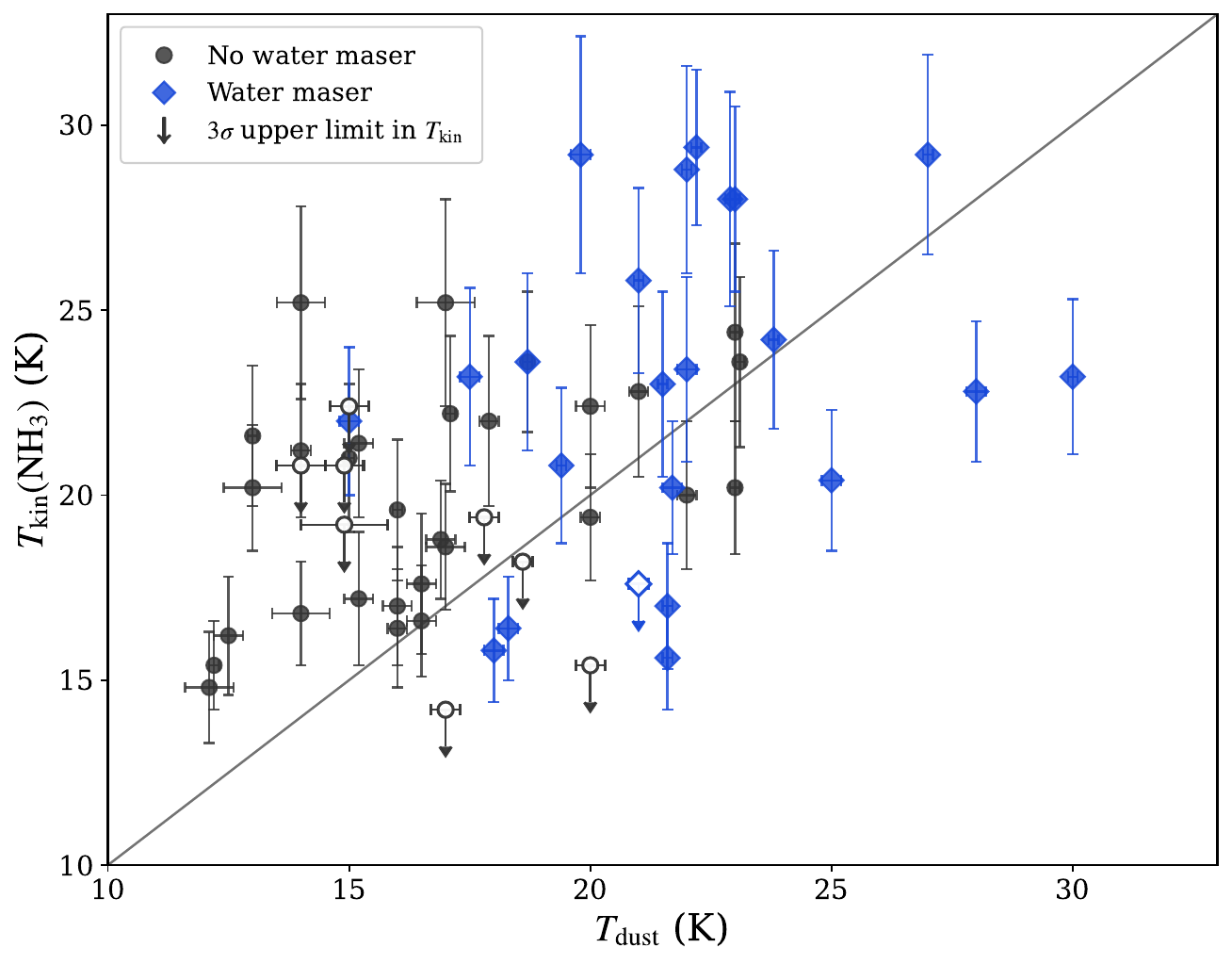} 
  \caption{Comparison between NH$_3$ kinetic temperatures and dust temperatures for the 56 ALOHA IRDCs clumps. For sources with multiple NH$_3$ velocity components, each component is plotted separately. Blue diamonds indicate sources associated with H$_2$O maser emission, while grey circles denote sources without maser detections. Open symbols with downward arrows represent the $3\sigma$ upper limits on $T_{\rm kin}(\mathrm{NH}_3)$. Error bars indicate the uncertainties in $T_{\rm dust}$ and $T_{\rm kin}(\mathrm{NH}_3)$. The dotted line marks equality between the gas and dust temperatures.}
  \label{fig:temperature}
\end{figure}

A comparison of the gas and dust temperatures is shown in Fig.~\ref{fig:temperature}. For the 48 clumps with measured NH$_3$ kinetic temperatures, the mean kinetic temperature is $21.6\pm3.9$\,K. For sources with multiple velocity components, the measured component temperatures were averaged and components with only upper limits were excluded. The average kinetic temperature is approximately 3\,K higher than the mean dust temperature of 19~K (Ren et al. in prep.). This offset is systematic across the sample, with 38 of the 48 sources showing lower dust temperatures than their gas kinetic temperatures. The observed trend is consistent with at least partial thermal decoupling between gas and dust on the scales traced by the NH$_3$ emission, although systematic uncertainties in either or both temperature estimates, particularly in the derivation of the dust temperatures, may also contribute to the offset. Similar gas--dust temperature offsets have been reported in previous studies of IRDCs and dense molecular clumps and have often been interpreted as evidence for inefficient thermal coupling between the two components \citep[e.g.,][]{2011pillai,2012wienen,2015urquhart}.

Independent evidence for dynamically perturbed gas is provided by the kinematic properties of shock tracers. The SiO profiles are typically about twice as broad as the corresponding H$^{13}$CO$^{+}$ profiles, while the HNCO line widths are enhanced by a factor of $\sim$1.2. Broad SiO and HNCO emission is widely associated with shocked gas and turbulent dissipation in star-forming regions \citep[e.g.,][]{1997schilke,2000zinchenko,2002codella,2008gusdorf}. This kinematic broadening indicates that the gas traced by SiO and HNCO is more dynamically perturbed than the dense gas traced by H$^{13}$CO$^{+}$. Together with the observed gas--dust temperature offset, this is consistent with shocks and turbulence contributing to partial thermal decoupling. Complex or disordered motions are also observed in filamentary systems, in which massive cores may be kinematically decoupled from their parent filaments, while local velocity gradients in supercritical filaments can show no preferred alignment with either the filament axes or the local gravitational field\citep[e.g.,][]{2023ren,2026zhangchao}.

At $n_{\rm H_2} \lesssim 10^{5}\,\mathrm{cm^{-3}}$, the timescale for collisional energy exchange between gas and dust is sufficiently long that the two components do not rapidly equilibrate \citep{2001goldsmith,2015steinacker}. Under these conditions, dust grains are primarily heated by the attenuated interstellar radiation field and cool through continuum emission, whereas the gas is subject to additional heating mechanisms, including cosmic rays, turbulent dissipation, and shocks, and cools predominantly via molecular-line radiation \citep[e.g.,][]{1983hildebrand,2012pon}. Outflow-driven shocks and turbulent energy injection can therefore raise the gas temperature locally without producing a corresponding increase in the dust temperature, leading to thermal decoupling between the two phases \citep[e.g.,][]{2001goldsmith,2010codella,2023duan}. The mean gas--dust temperature difference of approximately 3\,K in our sample is smaller than the differences reported in strongly heated regions such as L1157-B1 and S140 \citep{2010codella,2015koumpia}, although differences in spatial scale, molecular tracers, and source properties limit direct comparison.

Approximately half of the sample exhibits infall signatures in their HCO$^+$ spectra, including about one third of the full sample without SiO detections. Several sources exhibit infall and shock signatures together with NH$_2$D and/or H$_2$O maser emission, reflecting the coexistence of cold dense gas and dynamically active material within the same clump. Similar coexistence of cold chemically young gas and dynamically active shocked material is expected in filamentary systems undergoing rapid mass accretion and feedback \citep[e.g.,][]{2023ren,2023duan}. Dense clumps associated with maser emission are generally found to have higher luminosities, densities, and signatures of embedded protostellar activity in large surveys of massive star-forming regions \citep[e.g.,][]{2011urquhart,2015fontani,2018kim,2020billington}. These observational trends are broadly consistent with chemical evolutionary scenarios in which deuterated species are enhanced in cold and dense gas prior to significant heating by an embedded protostar, while maser activity and broadened line profiles indicate the onset of dynamical feedback \citep[e.g.,][]{2015gerner,2016barnes,2023cosentino}.

Consistently, sources associated with H$_2$O maser emission exhibit higher mean gas and dust temperatures than sources without maser detections (Fig.~\ref{fig:temperature}). Among the sources with measured kinetic temperatures, the 21 H$_2$O maser sources have mean temperatures of $T_{\rm kin}=23.5\pm4.3$\,K and $T_{\rm dust}=21.8\pm3.6$\,K. The 27 sources without H$_2$O masers have corresponding means of $20.2\pm3.0$\,K and $16.9\pm3.4$\,K, respectively. SDC33O, which hosts an H$_2$O maser but has only an upper limit on $T_{\rm kin}$, was excluded from these means. Similar trends have been reported in large surveys of dense clumps where maser activity is preferentially associated with more dynamically active and internally heated environments \citep[e.g.,][]{2006wang,2011urquhart,2018kim,2020yang}.

\subsection{Maser velocities, fluxes, and kinematics} \label{sect:discussionmaser}

The NH$_3$(1,1) and (2,2) inversion transitions provide a standard thermometer for dense molecular gas because their relative populations can be used to derive the NH$_3$ rotational temperature and, subsequently, the gas kinetic temperature \citep[e.g.,][]{1983ho,1983walmsley,2013li}. Since the NH$_3$(2,2) transition arises from a higher metastable level than NH$_3$(1,1), its detection indicates gas that is warm and/or dense enough to populate this level. Sources without detectable NH$_3$(2,2) emission may be among the coldest or least excited clumps in our sample. For these sources, only upper limits on the NH$_3$ rotational temperatures and kinetic temperatures can be derived. In this context, it is notable that none of the sources lacking NH$_3$(2,2) emission exhibits H$_2$O masers or Class~{\sc i} CH$_3$OH maser emission, except SDC33O, towards which a new H$_{2}$O maser is detected. Since these masers are generally associated with shocked, dense, and dynamically active gas, their absence in the NH$_3$(2,2)-non-detected sources is consistent with these clumps lacking the local heating and excitation conditions commonly associated with maser activity. Class\,{\sc i} methanol masers at 44, 84, and 95\,GHz are common tracers of shocked gas in massive star-forming regions \citep[e.g.,][]{2013gan,2014voronkov,2018kim,2019kim,2020yang}. In contrast, Class\,{\sc i} methanol masers at $24-25\,{\rm GHz}$ are relatively rare \citep[e.g.,][]{2007voronkov,2011walsh,2015gong} and require substantially denser environments \citep{2016leurini,2017towner}. SDC19A, the only source hosting 25\,GHz Class\,{\sc i} methanol maser emission in our sample, is also among the densest clumps, supporting the interpretation that these masers preferentially trace very compact and dense environments in the earliest phases of massive star formation, before strong radiative feedback develops.

\begin{figure}
  \centering
    \captionsetup{justification=raggedright, singlelinecheck=false}
  \includegraphics[width=0.90\columnwidth]
  {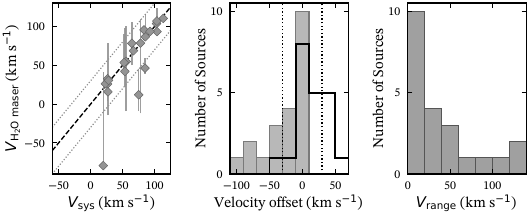}
    \caption{Left panel: Comparison between H$_2$O maser velocities and systemic velocities inferred from NH$_{3}$ (1,1). Grey diamonds show the peak velocity of the strongest maser component, and vertical grey segments show the full detected interval $[V_{\min},V_{\max}]$. The black dashed line marks $V(\mathrm{H_2O})=V_{\rm sys}$, and the grey dotted lines mark offsets of $\pm30\,\mathrm{km\,s^{-1}}$. Middle panel: Distribution of the most blue- and red-shifted detected H$_{2}$O maser velocities relative to the systemic velocity. The filled grey histogram represents $V_{\min}-V_{\rm sys}$, and the black outlined histogram represents $V_{\max}-V_{\rm sys}$. The vertical dotted lines mark the high-velocity thresholds. Right panel: Distribution of the full maser velocity ranges ($V_{\rm range}=\left|V_{\max}-V_{\min}\right|$).}
  \label{fig:maservelocity} 
\end{figure}

 The velocity range of the maser emission, $V_{\rm range}$, is a critical diagnostic of the velocity extent in shocked regions. A clear dichotomy is observed between Class\,{\sc i} methanol and water masers. In SDC19A, the peak velocities of all eight detected 25-GHz Class\,{\sc i} CH$_3$OH transitions are within approximately $0.6\,\mathrm{km\,s^{-1}}$ of the NH$_3$ systemic velocity. Methanol masers typically show narrow velocity ranges clustered around the systemic velocity \citep[e.g.,][]{2023yang}, suggesting that they trace gas motions closely linked to the clump-scale kinematics, potentially arising from large-scale shocks. For all detected transitions, Class\,{\sc i} methanol masers remain tightly aligned with systemic velocities. This agrees with previous large surveys showing that velocity offsets of Class\,{\sc i} masers relative to dense gas tracers are typically small ($\lesssim3\,\mathrm{km\,s^{-1}}$), with mean values near zero \citep[e.g.,][]{2017jordan,2017rodriguez,2018kim,2019kim,2023yang}. Even in low-mass regions, most Class\,{\sc i} maser velocities remain within a few $\mathrm{km\,s^{-1}}$ of systemic velocity \citep{2010kalenskii}. This reflects their origin in low-velocity shocks at outflow--cloud or cloud--cloud interfaces where the shocked gas remains dynamically coupled to the surrounding dense material rather than tracing the highest-velocity jet components. Projection effects and the requirement for velocity coherence in maser amplification further favour emission near systemic velocity. On the contrary, water masers exhibit much broader velocity spreads with half of the sources exceeding $20\,\mathrm{km\,s^{-1}}$ (Fig.~\ref{fig:maservelocity}), and significant velocity offsets from systemic velocity spanning from $-99$ to $+52\,\mathrm{km\,s^{-1}}$. This behaviour strongly supports their origin in high-velocity jets and compact outflows from deeply embedded massive protostars \citep{2010asanok,2013hollenbach}.

The numbers of blue- and red-shifted (H$_{2}$O maser peak velocities compared to systemic velocity traced by NH$_{3}$) are approximately balanced. Following previous H$_2$O maser studies, we classify high-velocity emission as the presence of at least one detected maser feature offset by more than $30\,\mathrm{km\,s^{-1}}$ from the systemic velocity \citep[e.g.,][]{2018kim}. Of the 22 H$_2$O maser sources, nine satisfy this criterion with seven showing blue-shifted features, six showing red-shifted features, and four showing both. Seven of these nine high-velocity maser sources are classified as HCO$^+$ infall candidates, while SDC19C and SDC35L are classified as outflow candidates. NH$_2$D is detected towards five of the nine sources: SDC19C, SDC19E, SDC33O, SDC35G, and SDC35S. The occurrence of high-velocity maser features does not correspond directly to the HCO$^+$ classifications because the two diagnostics trace different physical scales. Water masers arise in compact shocked gas associated with jets or outflows, whereas HCO$^+$ asymmetries probe larger-scale motions in the dense clump envelopes. This coexistence of high-velocity maser emission, infall kinematics and cold-gas tracers indicates that these sources are likely undergoing global collapse, representing the onset of massive star formation.

\begin{figure}
  \centering
    \captionsetup{justification=raggedright, singlelinecheck=false}
  \includegraphics[width=0.90\columnwidth]
  {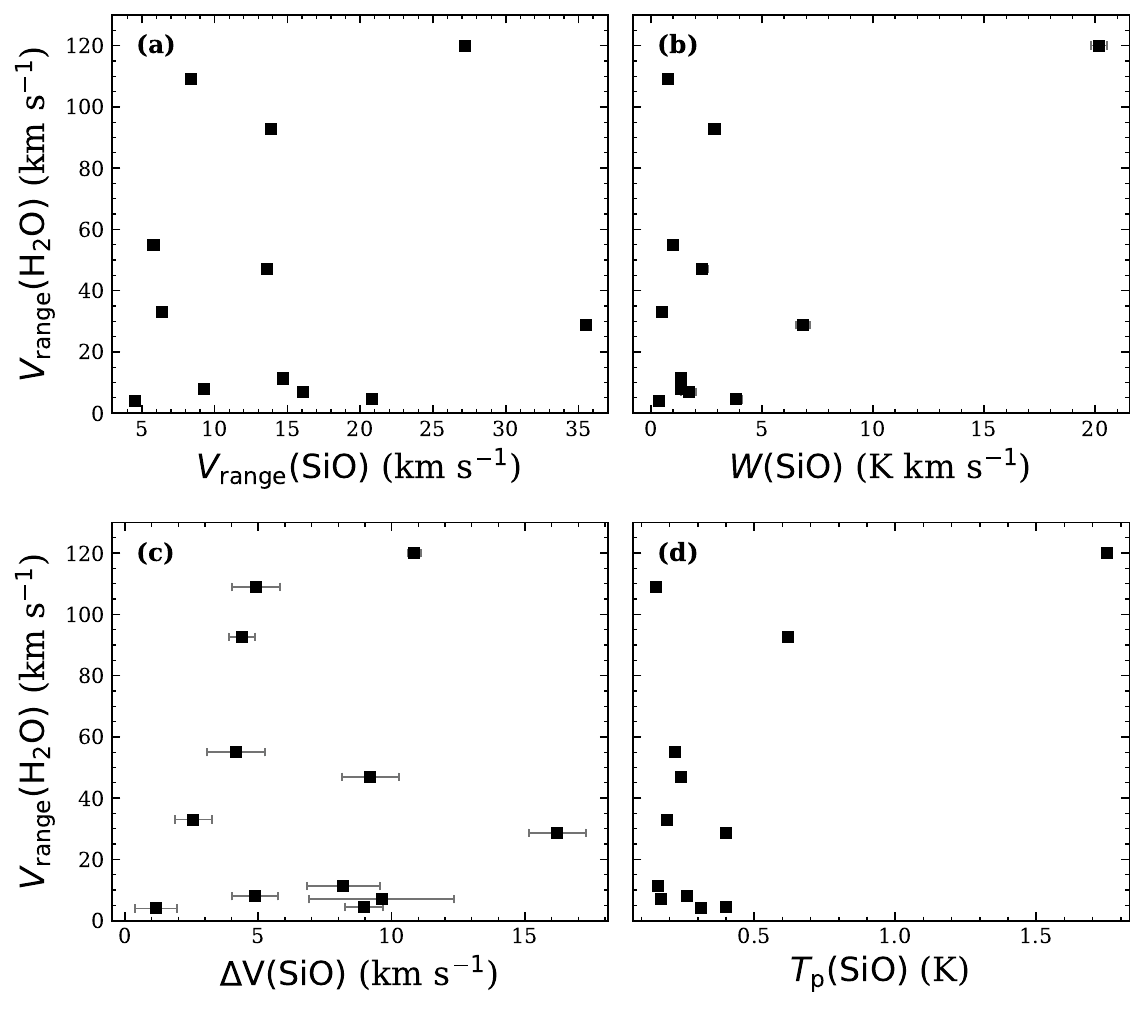}
    \caption{Relationship between the H$_2$O maser velocity range and SiO shock properties for sources detected in both tracers, with $V_{\rm range}=|V_{\max}-V_{\min}|$. The panels show $V_{\rm range}({\rm H_2O})$ compared with (a) $V_{\rm range}({\rm SiO})$, (b) $W({\rm SiO})$, (c) $\Delta {\rm V}({\rm SiO})$, and (d) $T_{\rm p}({\rm SiO})$. Error bars indicate uncertainties in $W({\rm SiO})$ and $\Delta {\rm V}({\rm SiO})$. No statistically significant correlations are found.
}
  \label{fig:H2OSiO} 
\end{figure}

Although both H$_2$O masers and SiO emission are widely used as tracers of shock activity, we find no statistically significant correlation between the kinematic properties of the SiO lines and the velocity extent of the H$_2$O masers (see Fig.~\ref{fig:H2OSiO}). The Spearman rank correlation analysis shows very weak correlations between $V_{\rm range}({\rm H_2O})$ and the SiO velocity range ($\rho_{\rm S}=0.10$, $p=0.770$), integrated intensity ($\rho_{\rm S}=0.25$, $p=0.467$), linewidth ($\rho_{\rm S}=0.17$, $p=0.612$), and peak temperature ($\rho_{\rm S}=0.06$, $p=0.863$). This is most plausibly explained by beam dilution in the single-dish data, in which multiple unresolved shock components are averaged within the beam, such that the two tracers probe different physical regions of the same shock system: compact, high-density post-shock gas in the case of H$_2$O masers, and more extended shock fronts traced by SiO. High angular resolution studies support this picture, showing that SiO and H$_2$O masers occupy distinct structures on scales of tens to hundreds of au \citep[e.g.,][]{2010matthews,2019moscadelli,2022beltran}. The absence of a correlation in our data therefore likely reflects unresolved internal substructure rather than an intrinsic physical disconnection between the tracers. In this picture, broad maser velocity ranges can coexist with enhanced NH$_2$D emission, indicating that localised protostellar shocks develop within a globally cold and dense environment.

\subsection{Singly deuterated ammonia in shocked environments}\label{sect:discussnh2d}

Deuterated molecules are widely regarded as tracers of cold and chemically young gas in star-forming regions. Their enhancement is driven by low-temperature ion--molecule chemistry. Cosmic-ray ionisation initiates the formation of H$_3^+$, which reacts with HD to form H$_2$D$^+$, while the backward reaction becomes inefficient at low temperatures. CO freeze-out onto dust grains further promotes deuterium fractionation, because gas-phase CO is one of the main destruction partners of H$_3^+$ and its deuterated isotopologues. As a result, deuterated ions and related neutral species, including NH$_2$D, can become enhanced in cold, dense, and CO-depleted gas \citep[e.g.,][]{1989millar,2003roberts,2012caselli}. Recent studies of massive clumps and IRDCs have also shown that deuterium fractionation is associated with cold dense gas, CO depletion, and evolutionary stage \citep[e.g.,][]{2020feng,2021wienen,2024sabatini}. The degree of deuteration, however, depends on several coupled parameters, including temperature, density, depletion, the H$_2$ ortho-to-para ratio, and the cosmic-ray ionisation rate. Therefore, NH$_2$D detections and NH$_2$D/NH$_3$ ratios should be interpreted as qualitative indicators of cold dense gas and chemical youth, rather than as direct measures of CO depletion or ionisation without dedicated chemical modelling.

In our sample, NH$_2$D is detected in 18 out of 56 clumps (32\%), comparable to the 39\% reported for ATLASGAL sources by \citet{2021wienen}. This agreement suggests that NH$_2$D emission is common, although not ubiquitous, in dense clumps at the sensitivities of these surveys. Our beam-averaged NH$_2$D column densities range from $4.6\times10^{12}$ to $8.7\times10^{13}\,\mathrm{cm^{-2}}$, with a median of $1.2\times10^{13}\,\mathrm{cm^{-2}}$, whereas \citet{2021wienen} reported a mean source-averaged value of approximately $1.6\times10^{15}\,\mathrm{cm^{-2}}$. For the 15 sources with measured NH$_3$ column densities and unambiguous NH$_3$ component matching, the apparent deuteration fraction ranges from 0.011 to 0.089, with a median of 0.038. Under the compact-source beam-correction assumption, the corresponding range is 0.003--0.027, with a median of 0.011. The median deuteration fraction in our sample, 0.038, is slightly below the median of $0.05\pm0.01$ reported by \citet{2021wienen} for sources without resolved NH$_{2}$D hyperfine structure and substantially below their median of $0.4\pm0.05$ for sources with resolved hyperfine structure, although the reported ranges overlap. The values in our sample are also smaller than the highest NH$_2$D/NH$_3$ ratios reported in massive clumps, including ratios up to $\sim$0.7 found by \citet{2007pillai}, yet are comparable to the lower NH$_2$D/NH$_3$ ratios reported by \citet{2022liyuqiang} for massive star-forming regions. This suggests that the ALOHA IRDCs clumps show moderate rather than extreme ammonia deuteration, although a direct comparison is limited by differences in beam size, excitation assumptions, and source averaging. The individual values are listed in Table~\ref{tab:NH2D}.

We note, however, that the derived values are subject to additional uncertainty. The NH$_2$D column densities were estimated under the optically thin assumption. If the NH$_2$D opacity is non-negligible, the column density and the corresponding deuteration fraction would be underestimated. Only one NH$_2$D transition was observed, so its excitation temperature could not be measured directly and was approximated using the NH$_3$ kinetic temperature. This may instead overestimate the NH$_2$D column density and deuteration fraction \citep{2021wienen}. The resulting deuteration fractions should therefore be regarded as approximate beam-averaged values rather than precise intrinsic abundances.

We find no statistically significant difference in either dust temperature or gas kinetic temperature between NH$_2$D detections and non-detections. Sources lacking detectable NH$_3$(2,2) emission also generally lack NH$_2$D emission. Although the absence of NH$_3$(2,2) may indicate low excitation, it may also reflect low molecular column densities, beam dilution, or limited sensitivity. Therefore, this association cannot be interpreted solely in terms of gas temperature. NH$_2$D frequently coexists with H$_2$O maser emission, with 12 of the 18 NH$_2$D detections also associated with H$_2$O masers. While strong shocks are often expected to suppress deuterated species through rapid heating and chemical destruction \citep{2011pillai}, recent studies have shown that deuterated molecules such as NH$_2$D, DCO$^{+}$, and DCN can remain detectable in IRDC cores with embedded protostellar activity and kinematic perturbations \citep{2022sakai,2024trofimova,2025sakai}. Our results therefore suggest that cold gas can persist in clumps that already show localised dynamical activity, even if the beam-averaged deuteration is only moderate.

Given the single-dish angular resolution, part of this coexistence may reflect beam averaging of cold dense gas and compact shocked regions within the same beam. Future observations of H\,{\sc i} Narrow Self-Absorption (HINSA), particularly with the high L-band sensitivity and spectral resolution of FAST \citep{2019zhangkai}, would complement the molecular-line tracers studied here by probing the cold atomic gas associated with the IRDCs. Combining HINSA with higher-resolution molecular-line observations may help constrain the roles of cosmic-ray ionisation, non-thermal heating, and chemical fractionation and disentangle the spatial coupling between chemistry, kinematics, and feedback \citep{2005goldsmith,2025luo,2025tang}.

\section{Conclusions}\label{sect:concl}

We conducted K-band and W-band molecular-line observations towards 56 clumps in the ALOHA IRDCs survey. Our main results can be summarised as follows:

\begin{enumerate}

\item
Ammonia emission in the (1,1) and (2,2) inversion transitions is detected in 54 and 48 of the 56 sources, respectively. Using the hyperfine group ratio (HFGR) method, we derive kinetic temperatures of 15--29\,K. These are, on average, $\sim$3\,K higher than the corresponding dust temperatures, suggesting partial thermal decoupling between gas and dust, although systematic uncertainties may also contribute to the observed offset.

\item
The dense gas tracers HCO$^+$ and H$^{13}$CO$^+$ are detected towards 53 and 49 of the 56 sources, respectively. Based on comparisons between the optically thick HCO$^+$ profiles and the H$^{13}$CO$^+$ systemic velocities, together with the residual analysis of the HCO$^+$ Gaussian fits, 34 kinematic components associated with 29 of the 56 sources are classified as infall candidates and 11 as outflow candidates. Shock tracers are also common, with SiO and HNCO detected towards 19 and 20 sources, respectively, indicating that a substantial fraction of the clumps are already dynamically active.

\item
Deuterated ammonia emission is detected in 18 sources (32\%). The derived NH$_2$D column densities and NH$_2$D$/$NH$_3$ ratios indicate moderate deuteration, although the exact values remain uncertain because of the optically thin assumption and the beam mismatch between the Yebes NH$_2$D and Effelsberg NH$_3$ observations. The close agreement between the NH$_2$D and NH$_3$ centroid velocities indicates that the deuterated gas traces the same bulk dense component.

\item
Water maser emission is detected in 22 sources, and one source also exhibits Class\,{\sc i} methanol maser emission. Their association with sources showing SiO and other shock tracers suggests that maser activity is linked to shocked gas and localised feedback from embedded protostars.

\item
The widespread occurrence of infall candidates, shock tracers, maser activity, and NH$_2$D emission indicates that a substantial fraction of the ALOHA IRDCs clumps are already undergoing active dynamical evolution while still retaining chemically young material.

\end{enumerate}

\begin{acknowledgements}
We thank the anonymous referee for many constructive comments and suggestions, which helped to improve the quality of this manuscript. This work is supported by the National Natural Science Foundation of China grant No. 12588202 and No. 12473023, National Key R\&D Program of China under grant Nos. (2022YFA1603103, 2023YFA1608002, and 2023YFA1608004), Natural Science Foundation of Xinjiang Uygur Autonomous Region No. 2025D01B173, and Tianshan Talent Training Program 2024TSYCTD0013. 
It was also partially funded by the NSFC under grants 12550003, 12373029, 12103082, 12563004 and 12403033, the XBZG-ZDSYS-202212, the Tianchi Talent Project of Xinjiang Uygur Autonomous Region. 

Y.T. Yan is supported by the Exploratory Research Fellow
in Division of Science, NAOJ.

G.A.F acknowledges support by the Deutsche Forschungsgemeinschaft (DFG, German Research Foundation) under Germany's Excellence Strategy via EXC 3037, the Cluster of Excellence “Our Dynamic Universe”, no. 533607693,  funding via the DFG Collaborative Research Center 1601 (SFB 1601, sub-project B1) and from the University of Cologne and its Global Faculty programme.  

J.W.W. thanks the support from the Tianchi Talent Program of Xinjiang Uygur Autonomous Region.

W.Y. acknowledges the support from the National Natural Science Foundation of China (12403027), China Postdoctoral Science Foundation (2024M751376), and Jiangsu Funding Program for Excellent Postdoctoral Talent (2024ZB347). 

Henkel has been funded by Chinese Academy of Sciences President's International Fellowship Initiative grant No. 2025PVA0048. 

X.-P.C. acknowledges support from the National Science Foundation of China (12041305), the Tianchi Talent Program of Xinjiang Uygur Autonomous Region, and the Tianshan Talent Training Program (2024TSYCTD0013). 

N.-Y. T. acknowledges support by the University Annual Scientific Research Plan of Anhui Province (No. 2023AH030052).

P.N.D. is acknowledged the support from Vietnam National Foundation for Science and Technology Development (NAFOSTED) under grant number 103.99-2024.36. 

MGR is supported by the National Radio Astronomy Observatory. The National Radio Astronomy Observatory and Green Bank Observatory are facilities of the U.S. National Science Foundation operated under cooperative agreement by Associated Universities, Inc. 

A.M. Sobolev has been funded by Chinese Academy of Sciences President's International Fellowship Initiative grant No. 2024VMA0002.

D.A.L. was supported by the Ministry of Science and Higher Education of the Russian Federation (theme № FEUZ-2025-0003).

W.K. is supported by the National Research Foundation of Korea (NRF) grant funded by the Korea government (MSIT) (RS-2024-00342488). 

E.C. acknowledges the support from Core Research Grant (CRG; sanction order number CRG/2023/008710) awarded by Anusandhan National Research Foundation (ANRF) under Science and Engineering Research Board (SERB), Govt. of India.

C.W.L is supported by the Basic Science Research Program through the NRF funded by the Ministry of Education, Science and Technology (grant No. NRF-2019R1A2C1010851) and by the Korea Astronomy and Space Science Institute grant funded by the Korea government (MSIT; project No. 2025-1-841-02).

G.W was supported by the Xinjiang Tianchi Talents Program, Central Guidance for Local Science and Technology Development Fund ZYYD2025ZY23, Youth Innovation Promotion Association CAS, the Tianchi Talent Project of Xinjiang Uygur Autonomous Region. 

This work is based in part on observations obtained with the James Clerk Maxwell Telescope under program ID M20AL021. These observations were obtained by the James Clerk Maxwell Telescope, operated by the East Asian Observatory on behalf of The National Astronomical Observatory of Japan; Academia Sinica Institute of Astronomy and Astrophysics; the Korea Astronomy and Space Science Institute; the National Astronomical Research Institute of Thailand. Additional funding support is provided by the Science and Technology Facilities Council of the United Kingdom and participating universities and organizations in the United Kingdom and Canada. Additional funds for the construction of SCUBA-2 were provided by the Canada Foundation for Innovation.

Based on observations with the 100-m telescope of the MPIfR (Max-Planck-Institut für Radioastronomie) at Effelsberg.

Yebes Observatory thanks the ERC for funding support under grant ERC-2013-Syg-610256-NANOCOSMOS. 

Based on observations carried out with the Yebes 40 m telescope 22A020. The 40 m radio telescope at Yebes Observatory is operated by the Spanish Geographic Institute (IGN; Ministerio de Transportes y Movilidad Sostenible). Maria Jesus Jimenez Donaire has carried out the observations and the first inspection of the data quality. 

\end{acknowledgements}

\bibliographystyle{aa}

\bibliography{reference}

\begin{appendix}

\section{Physical parameters and molecular-line detections of the observed IRDCs clumps}

\clearpage
\begin{table}
\footnotesize
\begin{center} \doublerulesep 0.1pt \tabcolsep 2.5pt
\begin{minipage}[]{150mm}
\caption{Physical parameters and molecular-line detections of the observed IRDCs clumps}\label{tab:sources}
\end{minipage}
\begin{tabular}{l cc c c c c c c c c c c c c c}
\hline
Source&  R.A.$^{(a)}$  &  Decl.$^{(a)}$  & Distance &   
<$N_{\rm H_{2}}$>$^{(b)}$ & Mass$^{(b)}$ &  $T_{\rm dust}$$^{(b)}$  & CH$_{3}$OH & H$_{2}$O & NH$_{3}$ & HCO$^+$ &  H$^{13}$CO$^+$ & SiO & HNCO & SO & NH$_{2}$D\\
 & (hh:mm:ss) & ($^\circ$ ' ") & (kpc)  & (10$^{22}$\,cm$^{-2}$) &  (M$_{\odot}$) & (K) &  &  & & & & & & &  \\
\hline
SDC19A & 18:26:25.9 & $-$12:03:54.1 & 1.5 (1) & 5.5 & 300 & 22.2 (1)  & Y  &Y  &  Y & Y  & Y & Y & Y & Y & Y \\
SDC19B & 18:25:58.8 & $-$12:03:57.2 & 1.5 (1) & 3.7 & 90 &  21.0 (1)  & N &  Y  &Y  & Y  & Y& Y & Y & N & Y \\
SDC19C & 18:25:52.4 & $-$12:04:49.1 & 1.5 (1) & 3.1 & 50 & 17.5 (2)    & N &  Y   & Y & Y & Y & Y &Y  & N & Y \\
SDC19D & 18:25:56.1 & $-$12:01:59.9 & 6.0 (1)& 1.1 & 440   & 17.0 (6) & N & N &  Y &  Y & Y& N & N & N & N\\
SDC19E & 18:25:54.5 & $-$11:52:34.6 & 1.5 (1) & 6.2 & 330 & 27.0 (1)  & N & Y  & Y & Y & Y & Y& Y& Y & Y \\
SDC19F & 18:26:24.2 & $-$12:01:32.2 & 1.5 (1) & 2.1 & 1360 & 16.0 (3)  & N &  N  & Y  & N & Y  & N & Y & N & N\\
SDC19G & 18:25:55.2 & $-$11:54:49.0 & 5.7 (4) & 1.8 & 4200 & 22.0 (2) & N & Y & Y &Y  & Y  & Y  & Y  & N & N \\
SDC19H & 18:26:02.5 & $-$11:52:39.7 & 1.5 (1)  & 4.3 & 15 & 28.0 (2)  & N & Y & Y & Y & Y & Y & N &  N & N\\ 
SDC33A & 18:52:26.2 & 00:32:09.3& 6.8 (5) & 1.5 & 1080 & 22.0 (2) & N & N &Y  & N & Y & N & N & N & N\\
SDC33B & 18:52:42.5 & 00:29:37.1& 5.2 (6) & 3.4 & 310 & 17.9 (2) & N & N & Y & Y & Y & N & N & N &N \\
SDC33C & 18:52:30.8 & 00:29:37.6& 6.8 (5) & 2.0 & 1560 & 18.7 (1)  & N & Y  & Y &Y  & N & N & N &N & N\\
SDC33D & 18:52:40.9 & 00:25:39.5& 6.9 (8) & 1.8 & 1150 & 16.9 (3)  & N & N & Y &Y  & N & N& N& N& N\\
SDC33E & 18:52:19.9 & 00:25:54.3& 5.3 (4) & 5.3 & 550 & 22.9 (1)  & N & Y & Y & Y & Y & N& N& N&N\\
SDC33F & 18:52:12.6 & 00:26:55.2& 6.8 (5) & 1.5 & 1580 & 19.8 (2)  & N & Y  &Y  &Y  & Y & N& N& N& N \\
SDC33G & 18:52:14.6 & 00:24:50.2& 6.8 (5) & 7.0 & 1550 & 21.5 (1)  & N & Y  & Y & Y  & Y & Y& N&N&N  \\
SDC33H & 18:52:15.2 & 00:23:01.0 & 5.2 (6) & 2.0 & 560 & 15.2 (3)  & N & N  & Y & Y & Y& Y& N&N&  N \\ 
SDC33I & 18:52:20.3 & 00:18:55.6& 6.6 (3) & 1.0 & 390 & 14.0 (5) & N & N &Y  &  Y & N& N& N&N&  N  \\
SDC33J & 18:52:09.6 & 00:18:27.3& 6.2 (6) & 1.5 & 1150 & 18.7 (1) & N & N  & Y & Y &Y & N& Y&N & N  \\
SDC33K & 18:52:04.3 & 00:15:39.2& 6.2 (6) & 6.0 & 690 & 17.1 (1)  & N & N & Y &Y  & Y & N& N& N &  N  \\
SDC33L & 18:51:58.3 & 00:14:21.2& 6.2 (6) & 3.0 & 1410 & 23.0 (1) & N  & N & Y &Y  & N& N& N&N& N   \\
SDC33M & 18:51:48.6 & 00:19:41.8& 6.8 (6) & 2.1 & 2780 & 21.0 (2)  & N & N  & Y &Y  & Y& N&Y &Y&  Y \\
SDC33N & 18:51:46.0 & 00:22:33.5& 6.0 (8) & 1.0 & 180 &
17.0 (4)  & N & N &  Y & Y & Y& Y& N&N& N \\
SDC33O & 18:51:35.5 & 00:26:13.0& 5.2 (6) & 2.4 & 270 &
21.0 (2)& N & Y & Y & Y & Y& N& N&N &Y  \\
SDC33P & 18:51:33.2 & 00:29:49.2& 5.2 (6) & 1.7 & 800 &
25.0 (2)  & N & Y  & Y & Y & Y& N& N&N &  Y  \\ 
SDC33Q & 18:51:40.7 & 00:29:01.1& 0.7 (1) & 1.9 & 10 &
23.0 (1) & N & N  & Y & Y &Y & N& N&N &  Y  \\
SDC33R & 18:51:52.8 & 00:29:29.3& 5.2 (6) & 1.1 & 930 &
18.0 (2)  & N  & Y & Y &Y  & Y& Y& Y&N&  Y  \\
SDC35A & 18:57:04.8 & 02:21:57.2& 2.1 (2) & 4.2 & 100 &
21.7 (2)  & N & Y & Y &Y  & Y& N& N& N&Y    \\
SDC35B & 18:57:09.5 & 02:16:11.1& 2.5 (2) & 1.6 & 90 &
17.0 (3)  & N & N  & Y & Y & Y& N& N&N &  N\\
SDC35C & 18:57:01.8 & 02:17:07.8& 2.6 (3) & 1.5 & 220 &
18.6 (2)  & N & N & Y  & Y &Y & N& N&  N  &  N  \\
SDC35E & 18:57:07.7 & 02:10:57.7& 2.5 (2) & 2.6 & 200 & 16.0 (2)   & N & N & Y  & Y &  Y& N& Y&  N &  Y  \\
SDC35F & 18:57:07.9 & 02:08:20.0 & 2.5 (2) & 2.7 & 40 & 16.5 (3) & N   & N & Y & Y  & Y & Y& Y&  N&  Y  \\
SDC35G & 18:57:05.1 & 02:06:30.5& 3.0 (1) & 1.7 & 200 &
15.0 (2)  & N  & Y & Y & Y  & Y  & Y  & N  &N  & Y  \\
SDC35H & 18:56:59.1 & 02:04:53.6& 2.8 (6) & 1.3 & 260 &
19.4 (1) & N & Y & Y & Y & Y& N& Y & N  &   N \\
SDC35I & 18:56:41.3 & 02:09:58.1& 4.7 (8) & 1.6 & 780 &
23.8 (1)  & N   &Y & Y  & Y & Y& Y & Y & N & N  \\
SDC35J & 18:56:17.6 & 02:16:56.2& 2.8 (6) & 2.4 & 70 &
20.0 (3) & N & N & Y & Y & Y & N  & N&  N  & N  \\
SDC35K & 18:56:09.2 & 02:16:48.1& 3.0 (1) & 2.0 & 500 &
20.0 (2)  & N & N & Y & Y & N & N& N &  Y & N  \\
SDC35L & 18:56:22.9 & 02:20:28.0& 2.9 (2) & 6.8 & 630 &
30.0 (1)  & N & Y  & Y & Y & Y & Y  &N  & Y  & N  \\
SDC35M & 18:56:14.8 & 02:21:36.0 & 5.0 (10) & 3.0 & 2150 &  21.6 (1)  & N   & Y & Y & Y & Y& N & N & N & Y  \\
SDC35N & 18:56:01.7 & 02:23:08.8& 2.5 (3) & 6.7 & 280 &
23.1 (1) & N & N & Y & Y  & Y& Y & Y & N & N \\
SDC35O & 18:56:03.0 & 02:13:50.2& 5.5 (6) & 1.3 & 1930 & 13.0 (1)  & N & N & Y & Y & Y& Y& Y&  N & Y  \\
SDC35P & 18:56:15.1 & 02:06:27.9& 5.2 (6) & 1.4 & 780 &
18.3 (2)  & N  & Y & Y & Y &  Y& N & Y &  N & Y  \\
SDC35Q & 18:55:50.1 & 02:12:04.0& 5.5 (6) &  2.4 & 450 & 16.0 (1)  & N  & N & Y &  Y & Y& N&N&  N &  N \\
SDC35R & 18:55:31.1 & 02:17:12.2& 5.0 (5) & 2.1 & 990 &
14.0 (2)  & N & N  &  Y &Y  & Y& N& N &   N  & N  \\
SDC35S & 18:55:34.3 & 02:19:23.1& 5.0 (5) & 4.3 & 3810 & 22.0 (1)  & N & Y & Y &Y  & Y& N& N&  N & Y  \\
SDC35T & 18:55:12.5 & 02:21:04.9& 2.5 (3) & 1.4 & 630 &
17.8 (3) & N & N  & Y  & Y & Y& N & N& N & N  \\
SDC37A & 19:10:46.5 & 08:02:24.0& 3.5 (3) & 1.1 & 80 &
12.1 (5) & N & N& Y & N & Y& N & N&  N &  N   \\
SDC37B & 19:10:33.2 & 08:01:58.1& 3.5 (3) & 1.2 & 60 &
12.2 (1) & N & N& Y & Y & Y & N& N &  N &N  \\
SDC37C & 19:10:36.6 & 08:00:06.6& 3.5 (3) & 1.1 & 50 &
13.0 (6) & N & N& Y &Y  &  Y & N  &  N   &  N & Y  \\
SDC37D & 19:10:53.1 & 07:58:22.3& 3.5 (3) & 1.2 & 230 &
12.5 (3) & N & N&  Y & Y & Y& Y & Y & N & N \\
SDC37E & 19:10:46.3 & 07:53:42.0& 3.5 (3) & 8.9 & 220 &
14.0 (6) & N & N& Y & Y & Y & N & Y & N & N  \\
SDC37F & 19:10:34.0 & 07:53:14.2& 3.5 (3) & 1.8 & 230 &
23.0 (1) & N & Y & Y & Y & Y &  N & N  & N & N  \\
SDC37G & 19:10:27.9 & 07:53:54.1& 3.5 (3) & 1.5 & 340 &
20.0 (3) & N & N & Y & Y & Y & N & N & N & N  \\
SDC37H & 19:10:22.7 & 07:55:12.0& 3.5 (3) & 6.9 & 90 &
15.0 (4) & N & N& Y & Y & Y & Y & N & N & N  \\
SDC37I & 19:10:40.5 & 07:50:44.5& 3.5 (3) &  7.6 & 100 & 14.9 (9) & N & N&N  & Y &  N & N & Y  & N & N  \\
SDC37J & 19:10:25.7 & 07:49:42.9& 3.5 (3) & 7.1 & 360 &
14.9 (4) & N & N& N  & Y & N & N & N &N & N  \\
SDC37K & 19:09:54.0 & 07:57:16.1& 3.4 (6) & 1.3 & 750 &
15.0 (1) & N & N & Y & Y &  Y& Y & Y & Y & N  \\
\hline
\multicolumn{16}{p{0.99\textwidth}}{Notes. $^{(a)}$ Equatorial coordinates are given in the J2000.0 reference frame. $^{(b)}$ The values of dust properties were derived from JCMT $850\,\mu\mathrm{m}$ and {\it Herschel} data (from Ren et al. (in prep.) \& Liu et al. (in prep.)), of which the methods are described in Appendix~\ref{sect:tdust}. Distances are estimated from NH$_{3}$ using the parallax based distance calculator \citep{2019reid}. For sources without detectable NH$_3$ emission, the distances of the associated cloud complex from CO data are adopted (Ren et al. (in prep.)). For the molecular lines and masers, Y and N indicate detections and non-detections in the present observations. Values in parentheses indicate the uncertainty in the last quoted digit.
}
\end{tabular}
\end{center}
\end{table}
\clearpage

\section{Summary of observed molecular lines}

    \begin{table}[H]
    \centering
    \begin{minipage}[]{60mm}
\caption{Summary of Observed Molecular Lines.}\label{tab:linesfrequencies}
    \end{minipage}
    \begin{tabular}{c*{4}{c}}
   \hline\noalign{\smallskip}
        Molecule \& Transition & Frequency (MHz) & Telescope & $\mathrm{RMS}$ \\
        \hline\noalign{\smallskip}
 NH$_{3}$ (1,1) &  23694.496 & Effelsberg & 0.039\,K\\
NH$_{3}$ (2,2) &  23722.634 & Effelsberg  & 0.039\,K\\
NH$_{3}$ (3,3) &  23870.129 & Effelsberg & 0.039\,K\\
NH$_{3}$ (4,4) &  24139.417 & Effelsberg & 0.046\,K\\
CH$_{3}$OH 3$_2$-3$_1$ E  &  24928.715  & Effelsberg  & 0.043\,Jy\\
CH$_{3}$OH 4$_2$-4$_1$ E  &  24933.468 & Effelsberg & 0.043\,Jy\\
CH$_{3}$OH 2$_2$-2$_1$ E  &  24934.382 & Effelsberg & 0.043\,Jy\\
CH$_{3}$OH 5$_2$-5$_1$ E  &  24959.079 & Effelsberg & 0.043\,Jy \\
CH$_{3}$OH 6$_2$-6$_1$ E  &  25018.123 & Effelsberg & 0.043\,Jy \\
CH$_{3}$OH 7$_2$-7$_1$ E  &  25124.872 & Effelsberg & 0.040\,Jy \\
CH$_{3}$OH 8$_2$-8$_1$ E  &  25294.417 & Effelsberg  & 0.040\,Jy \\
CH$_{3}$OH 9$_2$-9$_1$ E  &  25541.398  & Effelsberg & 0.042\,Jy \\
CH$_{3}$OH 10$_2$-10$_1$ E  &  25878.266 & Effelsberg & 0.043\,Jy \\
H$_{2}$O 6$_{(1,6)}$-5$_{(2,3)}$ &  22235.080 & Effelsberg & 0.047\,Jy \\
         SiO (2--1) & 86846.995 & Yebes & 0.050\,K \\
        HNCO (4--3) & 87925.238 & Yebes & 0.063\,K  \\
        SO (2--1) & 86093.983 & Yebes &0.062\,K \\
        HCO$^+$ (1--0) & 89188.525 & Yebes & 0.089\,K \\
        H$^{13}$CO$^+$ (1--0) & 86754.288& Yebes & 0.055\,K  \\
NH$_{2}$D (1$_{1,1,0s}$-1$_{0,1,0a}$) & 85926.278 & Yebes & 0.070\,K \\
\hline
\multicolumn{4}{p{.50\textwidth}}{Notes. The frequencies are taken from Splatalogue (https://www.cv.nrao.edu/php/splat/). The $\mathrm{RMS}$ are on $T_\mathrm{MB}$ scale and are converted to unit in Jy for maser lines. }
    \end{tabular}  
    \end{table}

\section{Fitted molecular-line parameters}\label{sec:fitted}

\newgeometry{left=1cm, right=1cm, top=1cm, bottom=1cm}
\begin{sidewaystable}[ht]
\scriptsize
\centering
\caption{Fitted \ce{NH_{3}} line parameters}\label{tab:NH3class}
\setlength{\tabcolsep}{2.0pt}
    \begin{tabular}{c*{17}{c}}
       \hline
         \noalign{\smallskip}
             & $I$(NH$_{3}$(1,1)) & $V$(NH$_{3}$(1,1))& $\Delta V$(NH$_{3}$(1,1)) & $T_{\rm p}$(NH$_{3}$(1,1)) & $I$(NH$_{3}$(2,2)) & $V$(NH$_{3}$(2,2)) & $\Delta V$(NH$_{3}$(2,2)) & $T_{\rm p}$(NH$_{3}$(2,2)) & $I$(NH$_{3}$(3,3)) & $V$(NH$_{3}$(3,3)) & $\Delta V$(NH$_{3}$(3,3)) & $T_{\rm p}$(NH$_{3}$(3,3))  & $I$(NH$_{3}$(4,4)) & $V$(NH$_{3}$(4,4)) &  $\Delta V$(NH$_{3}$(4,4)) & $T_{\rm p}$(NH$_{3}$(4,4)) \\
& (K\,km\,s$^{-1}$) & (km\,s$^{-1}$) & (km\,s$^{-1}$) & (K) & (K\,km\,s$^{-1}$) & (km\,s$^{-1}$) & (km\,s$^{-1}$) & (K) & (K\,km\,s$^{-1}$) & (km\,s$^{-1}$) & (km\,s$^{-1}$) & (K) & (K\,km\,s$^{-1}$) & (km\,s$^{-1}$) & (km\,s$^{-1}$) & (K) \\
              \hline\noalign{\smallskip}
SDC19A & 7.23 (55) & 26.47 (9)  & 2.86  (20)   & 2.39         & 4.64 (12)  & 26.49 (4)    & 2.73 (9)      & 1.60          & 4.47 (13)    & 26.59 (4)    & 2.23 (12)     & 1.30         &        -      &     -    &      -    &        <0.14    \\

SDC19B  & 7.23 (45) & 26.36 (8)   & 2.51 (20)   & 2.71           & 3.67 (13)  & 26.40 (4)    & 2.43 (11)     & 1.42           & 3.47 (20)  & 26.21 (10)   & 3.60 (25)     & 0.90           & 0.62 (17)   & 25.75 (40)    & 2.83 (117)     & 0.21    \\

SDC19C  & 5.47 (39) & 26.91 (8)  & 2.45 (20)   & 2.10   &               2.37 (10)  & 27.03 (5)  & 2.37 (12)  & 0.95   &               1.23 (21)  & 27.16 (19)  & 2.16 (53)  & 0.53   &                -      &    -   &     -     &  <0.14      \\

SDC19D & 0.52 (12) & 114.98 (16) & 1.37 (35)  & 0.36
&        0.41 (7) & 115.43 (26)  & 2.87 (49)  & 0.13  &   0.41 (12)    &   115.31 (24)     &   1.50 (43)   &     0.15    & 
                  -       &      -   &      -      &  <0.14    \\

SDC19E & 15.41 (21) & 19.76 (2) & 7.50 (13) & 1.93 & 
     8.67 (19) & 19.38 (7) & 6.35 (16)  & 1.28   & 
      11.28 (18)  & 19.78 (6) & 7.76 (15)  & 1.37 & 
             3.15 (16) & 20.34 (23) & 9.93 (62)  & 0.30 \\

SDC19F & 2.30 (14) & 26.96 (6) & 1.98 (14) & 1.09 & 
         0.63 (8) & 26.60 (13) & 1.98 (24) & 0.30  &          -           &     -        &       -     &  <0.12  &  
         -           &     -        &       -     &   <0.14 \\

SDC19G & 3.39 (15)& 113.30 (6) & 2.52 (14)  & 1.26   & 
         1.51 (9) & 113.27 (7) & 2.21 (16)  & 0.64  & 
         1.06 (9) & 113.49 (20) & 4.60 (47)  & 0.22  &         -      &      -       &        -      &    < 0.14   \\

SDC19H & 2.61 (16) & 23.04 (9) & 2.89 (21)  & 0.85   &
        0.95 (11) & 23.18 (17) & 2.75 (34) & 0.32    & 
        0.86 (11) & 23.12 (21) & 3.53 (51)   & 0.23   &         -      &      -      &        -      &          <0.14     \\

SDC33A & 2.26 (11) & 103.39 (6) & 2.12 (14)   & 1.00  &            0.91 (7) & 103.43 (9) & 2.33 (21)  & 0.37  & 
     0.35 (7) & 102.75 (13) & 1.47 (35)   & 0.22  &            -   &      -       &        -      &    <0.14       \\

SDC33B & 2.31 (14) & 84.55 (5) & 1.72 (12)  & 1.26    &              0.97 (10) & 84.71 (8) & 1.65 (20)  & 0.55     & 
       -            &      -      &      -       & <0.14    &         -      &        -       &         -     &  <0.13      \\

SDC33C & 2.05 (12) & 102.88 (7) & 2.63 (19)  & 0.73   & 
         1.09 (11) & 102.51 (15) & 3.53 (50) & 0.29    & 
        0.46 (7) & 101.95 (18) & 2.16 (40)  & 0.20    &           -    &     -      &        -      &    <0.14      \\

SDC33D & 2.80 (14) & 109.24 (6) & 2.37 (13) & 1.11    & 
         1.03 (12) & 109.42 (12) & 2.07 (29) & 0.47    & 
         0.99 (12) & 108.80 (26) & 4.21 (57)  & 0.22   &        -       &      -      &     -      &     <0.14       \\

SDC33E & 2.67 (14) & 75.05 (7) & 2.69 (16) & 0.93    & 
         1.52 (10) & 75.11 (7) & 2.16 (16) & 0.66    & 
        0.95 (10) & 75.03 (16) & 2.92 (36) & 0.30     &           -    &       -     &      -     &     <0.14    \\

SDC33F & 1.40 (9) & 103.37 (8) & 2.29 (19)  & 0.58   &          1.06 (8) & 103.11 (10) & 2.68 (26)  & 0.37   &         0.60 (6) & 103.10 (10) & 1.92 (24)  & 0.30   &              -    &      -       &     -     &      <0.14      \\

SDC33G & 5.00 (24) & 103.63 (6) & 2.48 (14)  & 1.89  &           2.75 (11) & 103.12 (8) & 3.02 (15)  & 0.90  &        3.07 (21) & 103.11 (19) & 7.26 (79) & 0.42   &
          -   & -   &   -     &    <0.14          \\

SDC33H[v1] & 1.80 (10)  & 72.96 (7)  &  2.72 (19)  &  0.62  &              0.72 (10) & 72.50 (19) & 2.98 (51) & 0.23 & 
            0.32 (6) & 72.79 (17) & 1.68 (38) & 0.18 & 
            -  & -  & -  & <0.10 \\

SDC33H[v2]  & 1.21 (8)  &  101.13 (7) & 2.01 (15) & 0.56  &             0.50 (7) & 101.13 (11) & 1.59 (26) & 0.29 & 
           -           &        -      &    -        & < 0.15 &
           -          &      -      &      -     &  <0.12 \\

SDC33I[v1]  & 0.82 (7) & 77.32 (8) & 1.87 (19) & 0.42  &               - &  - & - &   <0.18   &          
-   &-      &      -       &      <0.12     &  
       -      &    -     &      -      &      <0.14      \\

SDC33I[v2]  & 0.72 (10) & 99.26 (49) & 5.56 (80) & 0.13  &             0.35 (7)  &   100.30 (25)  & 2.19 (41) &   0.15    &           -  &        -       &        -     &      <0.12   &              -   &       -     &    -      &      <0.14         \\

SDC33J  & 1.62 (10) & 99.26 (7) & 2.47 (17)  & 0.62  & 
        0.95 (12) & 99.63 (15) & 1.99 (33) & 0.45  & 
        0.67 (10) & 98.33 (25) & 3.25 (66)  & 0.19  &            -   &       -      &        -      &    <0.14    \\
        
SDC33K  & 4.62 (24) & 100.20 (6) & 2.12 (13) & 2.04 & 
   2.75 (12) & 100.39 (5) & 2.31 (12)  & 1.11  & 
   1.00 (10) & 100.33 (11) & 2.35 (27)  & 0.40   &             -  &      -     &        -      &    <0.14      \\

SDC33L & 1.82 (11) & 99.55 (8) & 2.69 (21)  & 0.63  &          1.07 (8) & 99.07 (7) & 1.89 (13) & 0.53   & 
       0.75 (8) & 99.56 (12) & 2.17 (25)  & 0.32 &             -  &       -       &          -    &     <0.13         \\

SDC33M & 4.02 (21) & 98.68 (6) & 2.24 (13)  & 1.69 & 
        1.81 (7) & 98.82 (4) & 2.14 (9)   & 0.80 & 
        0.80 (9) & 99.25 (11) & 1.95 (30)  & 0.39  &            -   &        -       &       -       &     <0.14       \\
   
SDC33N & 1.06 (7) & 98.32 (7) & 1.90 (15)  & 0.53  & 
         0.81 (18) & 98.88 (19) & 1.69 (43)  & 0.46  & 
         -   &      -   &        -        &        <0.12   &         -   &      -     &      -     &     <0.14        \\

SDC33O  & 1.04 (12) & 84.84 (9) & 2.14 (22)  & 0.46  & 
        -     &    -    &    -     & <0.12  &  
        -     &   -     &     -     &     <0.12  &     
        -     &    -     &      -       &   <0.14           \\

SDC33P & 2.35 (12) & 85.26 (5) & 2.00 (12)  & 1.11  & 
         0.87 (9) & 85.11 (8) & 1.76 (20) & 0.47   & 
        -     &   -    &   -       &    <0.12   & 
        -     &   -      &   -     &      <0.15     \\

SDC33Q  & 2.15 (9) & 9.33 (3) & 1.46 (7)  & 1.38  & 
          0.63 (8) & 9.21 (10) & 1.58 (23)  & 0.38  &           -    &        -       &      -      &   <0.12    &        -    &       -      &      -    &      <0.14        \\

SDC33R  & 2.39 (13) & 83.20 (6) & 2.43 (16) & 0.92  & 
          0.43 (9) & 83.23 (18) & 1.63 (43)  & 0.26 &           -    &      -       &        -       &    <0.12  &          -   &       -        &       -       &      <0.15 \\

SDC35A & 3.70 (15) & 27.95 (3) & 1.60 (8)  & 2.17  & 
         1.53 (7) & 27.93 (3) & 1.79 (10)  & 0.81  & 
         0.93 (15) & 28.32 (24) & 3.32 (68)  & 0.26  &          -     &       -      &        -      &    <0.14    \\

SDC35B & 1.35 (10) & 45.02 (9) & 2.52 (26) & 0.50 &                        -      &      -      &     -     &    <0.12    &                 -    &      -    &      -      &     <0.13    &
       -     &      -     &  -     &   <0.10   \\
       
SDC35C & 1.19 (13) & 49.93 (11) & 2.10 (28) & 0.51 & 
            - & - & - & <0.12 & 
             -  &   -    &     -    &    <0.12        &  
             -    &  -    &    -   &    <0.14        \\

SDC35E  & 4.56 (23) & 45.85 (5) & 2.00 (12) & 2.14 & 
         0.91 (10) & 45.81 (9) & 1.66 (21) & 0.52 &             -  &     -   &     -    &     <0.12   &          
         -    &   -     &      -       &     <0.14         \\

SDC35F[v1]  & 2.19 (37) & 42.85 (15) & 1.93 (39) & 1.06 & 
             0.53 (10) & 42.75 (16) & 1.77 (42) & 0.29  &     
             -    &      -    &      -     &     <0.12      &         -    &     -    &      -      &     <0.14       \\
         
SDC35F[v2]  & 3.96 (35) & 45.46 (8) & 1.81 (18) & 2.04 &
              1.09 (11) & 45.46 (8) & 1.79 (22) & 0.57 &         -  &       -      &      -     &    <0.12     &       -      &     -     &     -      &  <0.14  \\

SDC35G  & 1.38 (11) & 55.47 (8) & 2.11 (18) & 0.62 & 
   0.47 (10) & 55.00 (15) & 1.49 (41) & 0.30 &
         -      &    -     &    -     & <0.12  & 
         -      &    -     &    -     &  <0.14         \\

SDC35H & 1.65 (10) & 53.45 (7) & 2.17 (15) & 0.71 & 
    0.95 (13) & 53.83 (19) & 2.77 (56) & 0.32 & 
    0.91 (20) & 52.50 (39) & 3.32 (79) &  0.26 & 
      -    &    -     &     -       &   <0.14       \\

SDC35I  & 1.40 (15) & 64.31 (16) & 2.94 (35)  & 0.45 &                  0.97 (12) & 64.12 (18) & 2.80 (39) & 0.32 &                  0.89 (13) & 64.31 (21) & 2.53 (43) & 0.33  &
            -   &  -   &  -   &   <0.14 \\

SDC35J & 0.65 (13) & 53.02 (20) & 2.00 (47) & 0.31  &                   0.85 (12) & 53.70 (21) & 2.53 (45) & 0.31  &
         -           &     -     &      -    &   <0.12    &
         -          &      -    &      -    &    <0.14       \\

SDC35K & 1.47 (16) & 58.98 (19) & 3.05 (39) & 0.43 & 
             0.60 (13) & 58.80 (52)  & 3.50 (93)  & 0.16 &     -   &     -    &   -       &     <0.12    & 
             -    &     -       &     -      &    <0.14        \\

SDC35L & 2.44 (14) & 52.29 (11) & 4.07 (29)  & 0.56 & 
       1.26 (13) & 52.81 (25) & 4.76 (52) & 0.25  & 
       1.26 (16) & 54.24 (32) & 4.81 (74) & 0.25  & 
            -   &   -   &    -    & <0.14         \\

SDC35M[v1] & 4.55 (13) & 53.08 (43) & 3.81 (47) & 1.12 & 
           1.50 (18) & 53.10 (25) & 4.36 (63) & 0.32 & 
           0.37 (18) & 52.25 (28) & 1.53 (51)  & 0.23 &           -    &      -       &      -     &    <0.14    \\

SDC35M[v2] & 1.99 (13) & 57.79 (57) & 3.32 (43) & 0.56 & 
           0.37 (10) & 57.91 (17) & 1.33 (42) & 0.26 & 
           1.21 (31) & 55.59 (32) & 3.74 (140)  & 0.30 &           -    &      -       &      -     &    <0.14    \\

SDC35N & 6.73 (30) & 50.06 (9) & 4.54 (30) & 1.39 & 
         3.30 (13) & 50.34 (8) & 4.17 (19) & 0.75 & 
        1.97 (13) & 49.87 (16) & 4.81 (40) & 0.38  &             -  &        -     &     -   &       <0.14    \\

SDC35O & 2.31 (13) & 95.00 (6) & 2.04 (13) & 1.06 & 
         1.09 (7) & 95.26 (8) & 2.52 (19)  & 0.41  & 
         0.69 (13) & 95.24 (29) & 2.97 (67)  & 0.22  &          -     &      -      &     -      &      <0.12     \\
         
SDC35P & 1.51 (10) & 92.77 (7) & 2.05 (15) & 0.69 & 
       0.44 (10) & 92.83 (31) & 2.71 (89)  & 0.15   &          -      &    -     &      -   &      <0.12     &   
       -     &    -       &     -      &     <0.12      \\
       
SDC35Q & 1.05 (8) & 94.46 (10) & 2.50 (23) & 0.39 & 
        0.30 (5) & 94.79 (12) & 1.23 (23)  & 0.23 &             -  &        -        &        -     &    <0.12    &          -   &     -     &    -      &     <0.14   \\

SDC35R & 2.38 (12) & 75.61 (5) & 2.07 (13) & 1.08 & 
        0.98 (9) & 75.45 (11) & 2.37 (23)  & 0.39 &             -  &      -     &       -     &        <0.12     &          -     &      -    &        -        &      <0.12        \\
        
SDC35S & 4.17 (18) & 77.77 (5) & 2.59 (13) & 1.52 & 
        2.62 (9) & 77.69 (5) & 2.88 (13) & 0.85 & 
        1.02 (11) & 77.47 (12) & 2.56 (35) & 0.37 &             -  &       -     &       -       &        <0.12     \\

SDC35T & 1.30 (10) & 49.22 (8) & 2.25 (23) & 0.54 & 
             -      &    -   &   -   & <0.12 & 
             -    &  -     &    -    &      <0.12       &            -   &      -      &     -      &      <0.09       \\

SDC37A & 1.12 (64) & 73.41 (4) & 1.50 (10) & 0.70 & 
         0.22 (5) & 74.09 (22) & 1.56 (43) & 0.13 &
           -        &    -    &      -    &   <0.12      & 
           -  &      -     &    -     &  <0.10         \\
    
SDC37B & 1.53 (72) & 72.35 (5) & 2.08 (11) & 0.69 & 
         0.31 (4) & 72.07 (24) & 3.00 (43) & 0.10  &            -   &     -     &   -    &    <0.09  & 
           -     &    -     &    -     &  <0.08    \\
           
SDC37C & 1.51 (8) & 72.51 (6) & 2.34 (12) & 0.61 & 
         0.55 (10) & 73.50 (26) & 2.25 (46)  & 0.16 &             -  &      -     &     -     &      <0.12    & 
           -    &      -      &       -       &      <0.10    \\

SDC37D & 2.58 (13) & 64.79 (6) & 2.61 (16) & 0.93 & 
    0.51 (6) & 64.62 (11) & 1.71 (20) & 0.26  & 
    -    &       -      &        -      &      <0.12    &             -   &     -       &      -   &     <0.11   \\

SDC37E & 1.52 (6) & 66.52 (4) & 1.85 (9) & 0.77 & 
        0.39 (5) & 66.58 (7) & 1.35 (15)  & 0.27 &            -   &     -      &      -     &      <0.10    &   
          -    &      -        &    -       &    <0.11      \\

SDC37F & 0.61 (5) & 66.86 (9)  & 2.33 (21) & 0.25 & 
         0.68 (11) & 67.05 (30) & 3.49 (68)  & 0.18 & 
         0.27 (3) & 66.96 (10) & 1.59 (23)  & 0.16 &            -   &      -        &       -     &      <0.10        \\

SDC37G & 0.86 (6) & 68.18 (6) & 1.84 (16) & 0.44 & 
         -    &     -     &      -      &   <0.12         &            -   &      -      &    -    &      <0.11    &         
         -     &      -      &     -     &    <0.10        \\

SDC37H & 0.22 (5) & 67.63 (12) & 1.17 (28) & 0.18   & 
          -     &       -    &     -     &      <0.11      &  
          -     &      -     &     -        &    <0.11      &          -     &     -     &     -     &     <0.11      \\

SDC37I  & -   &  -    & -    & <0.12   & 
        - &   -   &   -     & <0.12      &   
        -   &   -       &    -    &    <0.12       &   
        -    &      -    &    -     &   <0.12     \\

SDC37J & -     &    -    &    -    &   <0.12       & 
          -    &      -    &  -    &   <0.12      & 
           -    &   -     &    -      &   <0.12   & 
            -   &      -    &    -     &  <0.09      \\

SDC37K & 2.03 (10) & 55.09 (7) & 2.84 (16) & 0.67 & 
        1.30 (11) & 54.96 (13) & 3.39 (32) & 0.36 & 
        0.78 (11) & 55.33 (20) & 3.02 (53) & 0.24 & 
        -   &    -    &    -   &    <0.12 \\   
     \hline  
  \noalign{\smallskip}\hline
\end{tabular}
\parbox{0.90\textheight}{
\scriptsize
Notes. Values in parentheses indicate the uncertainty in the last quoted digit(s). The parameters $I$, $V$, $\Delta V$, and $T_{\rm p}$ denote the integrated intensity, centroid velocity, FWHM linewidth obtained from a single-Gaussian fit to the central hyperfine group of each NH$_{3}$ inversion transition, and peak main-beam brightness temperature, respectively. Dashes indicate non-detections or unavailable fit parameters. Upper limits are given as $3\sigma$ limits for $T_{\rm p}$. Labels [v1] and [v2] indicate different velocity components fitted toward the same source.
}
\end{sidewaystable}
\restoregeometry 
\clearpage

\clearpage
\begin{table*}[p]
\centering
\captionsetup{justification=raggedright}
\caption{Derived NH$_3$ physical parameters}
\label{tab:NH3calculate}

\footnotesize
\renewcommand{\arraystretch}{0.92}
\setlength{\tabcolsep}{7pt}

\begin{tabular}{lcccc}
\toprule
Source &
$T_{\rm rot}(\mathrm{NH}_3)$ &
$T_{\rm ex}(\mathrm{NH}_3)$ &
$T_{\rm kin}(\mathrm{NH}_3)$ &
$N_{\rm tot}(\mathrm{NH}_3)$ \\
&
(K) &
(K) &
(K) &
($\times10^{14}\,\mathrm{cm}^{-2}$) \\
\midrule

SDC19A & 24.2 (1) & 5.6 (4) & 29.4 (21) & 9.5 (16) \\
SDC19B & 22.0 (13) & 6.0 (4) & 25.8 (25) & 9.6 (15) \\
SDC19C & 20.2 (12) & 5.2 (3) & 23.2 (24) & 9.5 (17) \\
SDC19D & 21.5 (12)  & 3.6 (2) & 25.2 (28)  & 1.0 (2) \\
SDC19E & 24.1 (14) & 5.1 (4) & 29.2 (27) & 14.5 (30) \\
SDC19F & 15.7 (9) & 4.2 (2) & 17.0 (16) & 3.1 (6) \\
SDC19G & 20.4 (12) & 4.3 (3) & 23.4 (25) & 4.1 (6)   \\
SDC19H & 20.0 (12) & 3.8 (2) & 22.8 (19) & 2.5 (5) \\
SDC33A & 17.9 (11) & 4.2 (3) & 20.0 (20) & 2.3 (4)  \\
SDC33B & 19.4 (11) & 4.1 (2) & 22.0 (23) & 3.8 (6) \\
SDC33C & 20.5 (10) & 3.6 (3) & 23.6 (24) & 2.4 (4) \\
SDC33D & 17.0 (7) & 3.9 (2) & 18.8 (16) & 5.2 (10) \\
SDC33E & 23.4 (12) & 3.9 (2) & 28.0 (29) & 2.9 (6)  \\
SDC33F & 24.2 (10) & 3.7 (3) & 29.2 (32) & 1.6 (3)   \\
SDC33G & 20.1 (12) & 4.9 (3) & 23.0 (25) & 6.8 (10)\\
SDC33H[v1] & 18.9 (9) & 3.5 (3) & 21.4 (20) & 2.7 (3) \\
SDC33H[v2] & 15.8 (9) & 3.3 (2) & 17.2 (18) & 3.0 (1)  \\
SDC33I[v1] & $<18.5$ & 4.2\,(3) & $<20.8$ & $<0.7$ \\
SDC33I[v2] & 21.5 (10) & 3.0 (2) & 25.2 (26) & 1.0 (2) \\
SDC33J & 20.5 (11) & 3.7 (2) & 23.6 (19) & 2.0 (4)  \\
SDC33K & 19.5 (9) & 4.9 (4) & 22.2 (21) & 7.5 (13) \\
SDC33L & 21.0 (9) & 4.1 (3) & 24.4 (24) & 1.9 (3)  \\
SDC33M & 19.9 (9) & 4.7 (3) & 22.8 (23) & 5.4 (10)  \\
SDC33N & 16.8 (10) & 3.3 (2) & 18.6 (17) & 1.9 (5)\\
SDC33O & $<16.0$ & 3.6 (2) & $<17.6$ & $<1.0$   \\
SDC33P & 18.2 (10) & 4.3 (3) & 20.4 (19) & 2.5 (5) \\
SDC33Q & 18.1 (8) & 6.4 (5) & 20.2 (18) & 1.7 (3)  \\
SDC33R & 14.7 (9) & 3.7 (2) & 15.8 (14) & 3.7 (6) \\
SDC35A & 18.1 (11) & 5.8 (3) & 20.2 (18) & 3.9 (7)  \\
SDC35B & $<13.4$ & 3.5 (3) & $<14.2$ & $<1.9$ \\
SDC35C & $<16.5$ & 3.2 (2) & $<18.2$ & $<2.5$ \\
SDC35E & 15.1 (8) & 5.1 (3) & 16.4 (16) & 7.8 (2)   \\
SDC35F[v1] & 15.4 (8) & 4.3 (3) & 16.6 (15) & 2.8 (6)  \\
SDC35F[v2] & 16.1 (8) & 5.2 (3) & 17.6 (19) & 6.3 (10) \\
SDC35G & 19.4 (12) & 3.4 (3) & 22.0 (20) & 3.5 (6)   \\
SDC35H & 18.5 (12) & 4.1 (2) & 20.8 (21) & 2.0 (4) \\
SDC35I & 20.9 (12) & 3.7 (3) & 24.2 (24) & 1.8 (3)  \\
SDC35J & 19.6 (10) & 3.2 (2) & 22.4 (22) & 1.2 (2)  \\
SDC35K & 17.5 (8) & 3.7 (2) & 19.4 (17) & 0.8 (3) \\
SDC35L & 20.2 (12) & 4.5 (3) & 23.2 (21) & 2.2 (5) \\
SDC35M[v1] & 15.7 (9) & 4.6 (3) & 17.0 (17) & 4.1 (7) \\
SDC35M[v2] & 14.5 (8) & 4.7 (4) & 15.6 (14) & 1.3 (2)  \\
SDC35N & 20.5 (11) & 4.9 (3) & 23.6 (23) & 6.8 (12)  \\
SDC35O & 19.0 (9) & 3.9 (2) & 21.6 (19) & 6.5 (12)  \\
SDC35P & 15.1 (8) & 3.5 (2) & 16.4 (14) & 2.3 (4) \\
SDC35Q & 17.6 (10) & 3.5 (2) & 19.6 (19) & 1.2 (2) \\
SDC35R & 18.8 (10) & 4.0 (3) & 21.2 (18) & 4.5 (7)   \\
SDC35S & 23.9 (12) & 4.7 (4) & 28.8 (28) & 4.5 (9) \\
SDC35T & $<17.4$ & 3.4 (3) & $<19.4$ & $<3.2$ \\
SDC37A & 14.0 (8) & 3.6 (2) & 14.8 (15) & 2.6 (5)  \\
SDC37B & 14.3 (6) & 3.6 (2) & 15.4 (12) & 3.4 (7) \\
SDC37C & 18.0 (9) & 3.6 (2) & 20.2 (17) & 2.8 (6)  \\
SDC37D & 14.9 (6) & 3.9 (3) & 16.2 (16) & 4.3 (7)  \\
SDC37E & 15.5 (8) & 3.9 (3) & 16.8 (14) & 2.2 (4)  \\
SDC37F & 23.4 (11) & 3.8 (2) & 28.0 (25) & 0.5 (1)  \\
SDC37G & $<14.3$ & 3.4 (2) & $<15.4$ & $<0.9$ \\
SDC37H & $<19.6$ & 2.9 (2) & $<22.4$ & $<0.8$ \\
SDC37I & $<17.4$ & - &
$<19.2$ & $<0.2$ \\
SDC37J & $<18.5$ & - &
$<20.8$ & $<0.2$ \\
SDC37K & 18.7 (9) & 4.1 (3) & 21.0 (20) & 2.8 (5) \\

\bottomrule
\end{tabular}

\parbox{0.70\textwidth}{\scriptsize Notes. Values in parentheses indicate the uncertainty in the last quoted digit(s). The quantities $T_{\rm rot}$, $T_{\rm ex}$, $T_{\rm kin}$, and $N_{\rm tot}$ denote the rotational temperature, excitation temperature, kinetic temperature, and total NH$_3$ column density, respectively. For velocity components without detectable NH$_3$(2,2) emission, the $3\sigma$ upper limits on the NH$_3$(2,2) peak brightness temperature listed in Table~\ref{tab:NH3class} were propagated through the analysis. The resulting upper limits are marked by $<$. For SDC37I and SDC37J, which are undetected in all observed NH$_3$ transitions, the reported limits were derived from the corresponding $3\sigma$ limits in Table~\ref{tab:NH3class}. Dashes indicate unavailable fit parameters. }

\end{table*}
\clearpage

\clearpage
\begin{table*}
\scriptsize
\caption{Water maser detection features}\label{tab:watermaser}
\centering
    \begin{tabular}{c*{9}{c}}
    \hline\hline\noalign{\smallskip}
         Source    & Date  & RMS & $V_{\rm peak}$ & [$V_{\rm min}$, $V_{\rm max}$]  &  $F_{\rm peak}$ & $N_{\rm comp}$   \\
& (2023) & (Jy) & (km\,s$^{-1}$) & (km\,s$^{-1}$)  & (Jy) &  \\           
              \hline\noalign{\smallskip}
SDC19A  &     05-July    &   0.04    &    32.53 &   [3.8, 32.5]     & 30.3   & 3        \\
SDC19B  & 05-July    &    0.09    &    25.80  &   
[23.5, 28.0]    &  2.13    &     1     \\
SDC19C  &  05-July  &  0.10 &  15.71  & [$-$14.1, 78.6]  &   1.68 &    7   \\
SDC19E & 05-July   & 0.03 & $-$79.2 &    [$-$79.3, 40.7]  & 26.3 &   7      \\
SDC19G  &  05-July  &  0.03  & 110.13 & [106, 114]   & 0.16 &   1           \\
SDC19H &  07-July  & 0.03 & 26.1  & [24.0, 28.0] &  0.13  & 1  \\
SDC33C & 06-July  &  0.03 & 103.56 & [97.5, 110.0]  & 1.91  &  3 \\
SDC33E&   06-July & 0.06 & 11.89 & [$-$9.0, 45.0] & 0.43 &  6  \\
SDC33F &    06-July & 0.05  & 106.72 & [102.0, 113.0]  & 0.17 &  1   \\
SDC33G &  06-July  & 0.04 & 93.23 & [90.0, 123.0] & 0.23 &  3 \\
SDC33O &  07-July  &  0.04 & 46.17 & [39.0, 59.0]  &   0.20  &  2 \\
SDC33P &  07-July  &  0.03 & 86.53 & [83.5, 115.0]  &   2.02  &  2 \\
SDC33R &  07-July  & 0.05 & 95.43 & [93.0, 100.0] &0.28  & 1 \\
SDC35A & 06-July & 0.03  & 30.42 & [15.0, 33.0]   & 21.76 &  3 \\
SDC35G  & 06-July & 0.09 & 54.62 & [$-$9.0, 100.0] & 0.75 &  5 \\
SDC35H  & 06-July & 0.09 & 42.03 & [22.5, 98.0] & 0.66 &  6  \\
SDC35I & 06-July & 0.04 & 78.09 & [50.0, 105.0]  & 1.12 &  4 \\
SDC35L&  06-July & 0.03 & 53.31  & [43.0, 90.0] &  20.0 & 10 \\
SDC35M &  06-July & 0.05 & 54.74  & [49.3, 60.6]  & 0.12 & 1  \\
SDC35P & 06-July  &  0.03  & 93.47 & [90.0, 99.0] & 15.0 & 1  \\
SDC35S & 06-July & 0.03 & 78.35 & [$-$12.0, 110.0] & 0.16 & 2  \\
SDC37F & 07-July & 0.03 & 68.61 & [63.0, 80.0]  & 0.13 &  2 \\    
     \hline  
  \noalign{\smallskip}\hline    
         \multicolumn{7}{p{.50\textwidth}}{Notes. The RMS were measured on 38.1\,kHz channel, corresponding to approximately $0.51\,\mathrm{km\,s^{-1}}$ at H$_{2}$O maser frequency, and were converted from the $T_{MB}$ scale to flux density in units of Jy. The $V_\mathrm{peak}$ is the peak velocity of the strongest component of water maser.}
\end{tabular} 
\end{table*}

\begin{table*}
\scriptsize
\centering
    \begin{minipage}[]{150mm}
\caption{Parameters of the detected Class\,{\sc i} methanol maser transitions towards SDC19A}\label{tab:methanolmaser}
\end{minipage}
    \begin{tabular}{c*{6}{c}}
    \hline\hline
          Transition   & Date  & RMS & $I_{\rm peak}$ & $V_{\rm peak}$ &  $F_{\rm peak}$ & Flux in \citet{2017towner} \\
 & (2023) & (Jy) & (Jy\,km\,s$^{-1}$) & (km\,s$^{-1}$)   & (Jy)  & (Jy)  \\           
              \hline\noalign{\smallskip}
 CH$_{3}$OH 3$_2$-3$_1$ E  &   05-July    &   0.03 & 4.38 (4) & 26.78 (1) &1.93  & 2.87  \\
 CH$_{3}$OH 4$_2$-4$_1$ E   &   05-July   &  0.03     &  8.75 (6) &  26.78 (1)   &  3.83  & - \\
  CH$_{3}$OH 2$_2$-2$_1$ E  &   05-July    &    0.03   & 0.25 (4)  &  26.33 (14)  &  0.15   & -   \\
 CH$_{3}$OH 5$_2$-5$_1$ E  & 05-July  &  0.02   &  9.61 (4) & 26.83 (1)  &  4.50  & 6.39 \\
 CH$_{3}$OH 6$_2$-6$_1$ E & 05-July  & 0.03 & 
9.12 (3) & 26.85 (1)  &  4.31  & -  \\
 CH$_{3}$OH 7$_2$-7$_1$ E &   05-July  & 0.03  & 
 5.75 (5) &  26.93 (1)      &  3.05  & - \\
 CH$_{3}$OH 8$_2$-8$_1$ E  &   05-July    &   0.03    &  3.06 (4) &  26.98 (1)    &  1.79  & 1.37 \\
  CH$_{3}$OH 9$_2$-9$_1$ E  &   05-July    &  0.07 &    1.02 (3) &  26.89 (2)    &  0.68   & -  \\
\hline  
  \noalign{\smallskip}\hline
\end{tabular}
\vspace{1mm}
\parbox{0.7\textwidth}{
\scriptsize
Notes. Values in parentheses indicate the uncertainty in the last quoted digit(s). The RMS were measured on 38.1\,kHz channel, corresponding to approximately $0.46\,\mathrm{km\,s^{-1}}$ at CH$_{3}$OH maser frequency, and were converted from the $T_{MB}$ scale to flux density in units of Jy. The $I_{\rm peak}$ is the integrated flux density of the maser component, and the $V_{\rm peak}$ is the velocity of the peak channel, and the $F_{\rm peak}$ is the peak flux density.
}
\end{table*}

\begin{table*}
\scriptsize
\centering
\caption{Water maser non-detections}\label{tab:nowatermaser}
    \begin{tabular}{c*{6}{c}}
    \hline\hline
         Source    & Date  & RMS  & Source & Date & RMS  \\    
          & (2023) & (Jy)  & (2023) & (Jy) \\   
              \hline\noalign{\smallskip}
SDC19D & 05-July & 0.030  & SDC19F & 05-July & 0.027  \\  
SDC33A & 06-July & 0.022  & SDC33B & 06-July & 0.028  \\ 
SDC33D & 06-July & 0.029  & SDC33H & 07-July & 0.020   \\  
SDC33I & 06-July & 0.028  & SDC33J & 07-July & 0.030  \\ 
SDC33K & 07-July & 0.027   & SDC33L & 07-July & 0.028  \\  
SDC33M & 07-July & 0.023  &  SDC33N & 07-July & 0.026  \\  
SDC33Q & 07-July & 0.026   & SDC35B & 07-July & 0.025   \\  
SDC35C & 06-July & 0.055  &  SDC35E & 06-July & 0.040   \\ 
SDC35F & 06-July & 0.066  &SDC35J & 06-July & 0.062    \\  
SDC35K & 06-July & 0.028  & SDC35N & 06-July & 0.038   \\  
SDC35O & 06-July & 0.030  & SDC35Q & 06-July & 0.036  \\   
SDC35R & 06-July & 0.045  & SDC35T & 07-July & 0.026  \\ 
SDC37A & 05-July & 0.018  & SDC37B & 07-July & 0.015  \\ 
SDC37C & 05-July & 0.018  &  SDC37D & 06-July & 0.010 \\   
SDC37E & 07-July & 0.018  & SDC37G & 06-July & 0.021 \\  
SDC37H & 06-July & 0.019 &  SDC37I & 06-July & 0.028   \\ 
SDC37J & 06-July & 0.020 &    SDC37K & 06-July & 0.020  \\ 
     \hline  
  \noalign{\smallskip}\hline    
         \multicolumn{6}{p{.38\textwidth}}{Notes. The RMS were measured on 38.1\,kHz channel, corresponding to approximately $0.51\,\mathrm{km\,s^{-1}}$ at H$_{2}$O maser frequency, and were converted from the $T_{MB}$ scale to flux density in units of Jy.}
\end{tabular}
\end{table*}

\clearpage
\newgeometry{left=1cm, right=1cm, top=1cm, bottom=1cm}
\begin{sidewaystable}[p]
\scriptsize
\centering
\caption{CLASS Gaussian fitted parameters of HCO$^+$, H$^{13}$CO$^+$, HNCO, SiO, and SO}\label{tab:shockfitted}
\setlength{\tabcolsep}{0.2pt}
\renewcommand{\arraystretch}{0.58}
\vspace{-3mm}

    \begin{tabular}{c*{21}{c}}
       \hline
         \noalign{\smallskip}
             & $I$(HCO$^+$)& $V$(HCO$^+$) & $\Delta V$(HCO$^+$) & $T_{\rm p}$(HCO$^+$) & 
             $I$(H$^{13}$CO$^+$) & $V$(H$^{13}$CO$^+$) & $\Delta V$(H$^{13}$CO$^+$) & $T_{\rm p}$(H$^{13}$CO$^+$) & $I$(HNCO) & $V$(HNCO) & $\Delta V$(HNCO) & $T_{\rm p}$(HNCO) & $I$(SiO) & $V$(SiO)& $\Delta V$(SiO) & $T_{\rm p}$(SiO)  & $I$(SO) & $V$(SO) & $\Delta V$(SO) & $T_{\rm p}$(SO) \\
& (K\,km\,s$^{-1}$) & (km\,s$^{-1}$) & (km\,s$^{-1}$) & (K) &
(K\,km\,s$^{-1}$) & (km\,s$^{-1}$) & (km\,s$^{-1}$) & (K) & (K\,km\,s$^{-1}$) & (km\,s$^{-1}$) & (km\,s$^{-1}$) & (K) & (K\,km\,s$^{-1}$) & (km\,s$^{-1}$) & (km\,s$^{-1}$) & (K) & (K\,km\,s$^{-1}$) & (km\,s$^{-1}$) & (km\,s$^{-1}$) & (K) \\
              \hline\noalign{\smallskip}
SDC19A & 10.58 (25) & 30.82 (5) & 4.01 (12) & 2.49 &
6.18 (17) & 26.55 (33)  & 2.90 (7)  & 2.21 & 
2.21 (12)  & 26.39 (8)  & 2.85 (20)  & 0.76  &  
6.85 (33) & 25.68 (35) & 16.21 (107) & 0.40 &
1.15 (13) & 26.38 (17) & 3.09 (42) & 0.35 
\\

SDC19B  & 2.03 (15) & 29.47 (9) & 2.42 (22) & 0.79 & 
2.65 (11) &  26.22 (44) & 2.13 (10) & 1.17 &  
2.53 (17) & 26.22 (10) & 3.33 (30) & 0.71 &
3.85 (24) & 26.98 (26) & 8.96 (70) & 0.40 &
- & - & - & - 
 \\

SDC19C  &  5.11 (23) & 28.77 (13) & 4.23 (24) & 1.15 & 
2.52 (12) & 27.04 (6) & 2.36 (15) & 1.00 &  
2.53 (12) & 27.29 (6) & 2.61 (14) & 0.91 &
2.87 (22) & 27.09 (15) & 4.39 (50) & 0.62 &
- & - & - & - 
\\

SDC19D[v1] &  0.68 (21) & 121.43 (12) & 1.00 (41) & 0.64 & 
0.25 (9) & 121.67 (17) & 0.74 (33) & 0.31 & 
- & - & - & - & 
- & - & - & - & 
- & - & - & - 
\\

SDC19D[v2] &  0.73 (18) & 122.82 (13) & 1.04 (28) & 0.66 & 
0.17 (7) & 122.81 (10) & 0.44 (19) & 0.37 &
- & - & - & - & 
- & - & - & - & 
- & - & - & - 
\\

SDC19E & 20.21 (46) & 16.53 (3) & 3.92 (7) & 4.89 & 
11.39 (26) & 19.51 (7) & 6.24 (16) & 1.71 & 
8.80 (30) & 20.15 (11) & 6.87 (27) & 1.20 &
20.17 (36) & 19.37 (10) & 10.85 (24) & 1.75 &
5.97 (27) & 19.13 (14) & 6.53 (35) & 0.86 
\\

SDC19F &  - & - & - & <0.24 & 
1.36 (13) & 26.65 (6) & 1.40 (15) & 0.91 &
0.58 (13) & 26.99 (14) & 1.35 (39) & 0.40 &
- & - & - & - & 
- & - & - & - 
\\

SDC19G & 20.61 (18) & 112.80 (1) & 3.29 (3) & 5.89 &
3.45 (15) & 113.06 (6) & 2.88 (15) & 1.13 & 
0.92 (17) & 113.27 (21) & 2.58 (63) & 0.34 &
1.36 (20) & 113.12 (35) & 4.87 (86) & 0.26 &
- & - & - & -
\\

SDC19H[v1] &  6.58 (17) & 22.65 (2) & 2.44 (8) & 2.55 & 
1.01 (15) & 22.77 (11) & 1.62 (32) & 0.59 & 
- & - & - & - &
- & - & - & - & 
- & - & - & -
\\

SDC19H[v2] &  2.91 (20) & 25.50 (9) & 2.63 (22) & 1.04 & 
0.30 (11) & 24.46 (9) & 0.50 (23) & 0.57 & 
- & - & - & - &
0.38 (16) & 27.47 (28) & 1.16 (80) & 0.31 & 
- & - & - & -
\\

SDC33A & - & - & - & <0.26 & 
0.90 (13) & 103.34 (12) & 1.78 (37) & 0.47 &
- & - & - & - & 
- & - & - & - & 
- & - & - & -
\\

SDC33B & 4.22 (15) & 84.69 (5) & 3.14 (12) & 1.26 &
0.49 (11) & 84.43 (35) & 3.14 (71) & 0.15 & 
- & - & - & - & 
- & - & - & - &
- & - & - & -
\\

SDC33C & 2.04 (19) & 86.66 (19) & 4.12 (48) & 0.47 & 
- & - & - & <0.16   &
- & - & - & -& 
- & - & - & - &
- & - & - & -
\\

SDC33D & 1.96 (22) & 109.38 (18) & 3.58 (60) & 0.51 & 
- & - & - & <0.16 &  
- & - & - & - & 
- & - & - & - & 
- & - & - & -
\\

SDC33E & 8.05 (15) & 74.38 (3) & 3.78 (8) & 2.00 &
0.53 (12) & 73.91 (30) & 2.71 (64) & 0.18 &  
- & - & - & - &
- & - & - & - & 
- & - & - & -
\\

SDC33F & 2.40 (29) & 103.35 (27) & 4.60 (62) & 0.49 &
0.47 (16) & 104.26 (41) & 2.50 (91) & 0.18  & 
- & - & - & - &
- & - & - & - & 
- & - & - & -
\\

SDC33G[v1] &  6.95 (22) & 73.97 (7) & 4.54 (16) & 1.44 &
0.79 (18) & 74.29 (42) & 3.17 (75) & 0.23  & 
- & - & - &- & 
- & - & - &- & 
- & - & - & -
\\

SDC33G[v2] &  5.18 (25) & 102.88 (15) & 6.62 (38) & 0.74 & 
0.66 (15)  & 104.02 (53) & 3.33 (99) & 0.17 & 
- & - & - & - & 
0.51 (13) & 104.98 (35) & 2.56 (70) & 0.19 &
- & - & - & -
\\

SDC33H[v1] & 17.47 (43) & 74.17 (6) & 4.78 (13) & 3.50 & 
1.17 (26) & 74.53 (31) & 3.76 (133) & 0.29 &
- & - & - & - &
0.35 (9) & 67.32 (15) & 1.08 (30) & 0.30 &
- & - & - & -
\\

SDC33H[v2]  & 6.63 (21) & 101.90 (7) & 4.59 (18) & 1.35 & 
0.65 (13) & 101.53 (25) & 2.11 (49) & 0.29  & 
- & - & - & - & 
- & - & - & - &
- & - & - & -
\\

SDC33I  & 0.99 (16) & 102.40 (23) & 2.68 (40) & 0.35 & 
- & -  &- &<0.16 &
- & - & - & - &
- & - & - & - &
- & - & - & -
\\

SDC33J  &  1.97 (22) & 100.56 (29) & 4.67 (56) & 0.39 & 
0.27 (9) & 101.32 (17) & 1.11 (47) & 0.24 & 
0.28 (10) & 101.33 (18) & 1.07 (37) & 0.25 &
- & - & - & - &
- & - & - & -
\\

SDC33K  &   1.39 (17) & 100.23 (27) & 3.99 (44) & 0.33 &
0.23 (11) & 99.57 (34) & 1.44 (69)& 0.15 &  
- & - & - & - & 
- & - & - & - &
- & - & - & -\\

SDC33L & 1.85 (18) & 101.49 (13) & 2.87 (34) & 0.60 & 
- & - & - & <0.24 &  
- & - & - & - &
- & - & - & - &
- & - & - & -
\\

SDC33M &  2.88 (16) & 100.59 (6) & 2.41 (18) & 1.12 &
1.85 (10) & 98.75 (5) & 1.95 (12) & 0.89 &
0.76 (10) & 99.19 (13) & 2.00 (33) & 0.36 &
- & - & - & - &
0.25 (9) & 99.23 (15) & 0.93 (41) & 0.25
\\
   
SDC33N &  2.92 (13) & 99.13 (4) & 1.89 (11) & 1.45 &
0.79 (10) & 98.42 (10) & 1.52 (22) & 0.49 & 
- & - & - & - &
0.45 (11) & 99.93 (32) & 2.64 (64) & 0.16 & 
- & - & - & -
\\

SDC33O  & 9.42 (15) & 84.92 (2) & 2.95 (5) & 3.01 & 
1.13 (11) & 84.86 (9) & 1.98 (23) & 0.53  & 
- & - & - & - & 
- & - & - & - &
- & - & - & -
\\
        
SDC33P & 16.20 (15) & 85.46 (1) & 2.93 (3) & 5.19 & 
2.06 (10) & 85.27 (3) & 1.61 (9) & 1.19 & 
- & - & - & - & 
- & - & - & - &
- & - & - & -
\\

SDC33Q  &  7.10 (11) & 9.43 (1) & 1.32 (2) & 5.07 &
3.02 (8) & 9.31 (1) & 0.99 (3) & 2.87 & 
-& - & - & - & 
- & - & - & - &
- & - & - & -
\\

SDC33R  & 9.39 (19) & 83.12 (3) & 3.36 (7) & 2.63 & 
1.66 (13) & 83.24 (8) & 2.08 (18) & 0.75 & 
1.08 (22) & 84.59 (57) & 5.58 (131) & 0.18 & 
1.71 (34) & 83.82 (4) & 9.63 (272) & 0.17 &
- & - & - & -
\\

SDC35A & 10.45 (34) & 28.70 (5) & 3.63 (16) & 2.70 & 
1.83 (19) & 28.02 (7) & 1.32 (17) & 1.30& 
- & - & - & - & 
- & - & - & - &
- & - & - & -
\\

SDC35B & 6.77 (20) & 44.82 (3) & 2.41 (9) & 2.64 & 
0.51 (13) & 45.19 (10) & 0.90 (29) & 0.54  & 
- & - & - & - & 
- & - & - & - &
- & - & - & -
\\
       
SDC35C[v1] &   4.91 (16) & 49.77 (10) & 3.22 (19) & 1.43 &
0.95 (13) & 50.13 (14) & 1.79 (30) & 0.46 & 
- & - & - & - & 
- & - & - & - &
- & - & - & -\\

SDC35C[v2] &  3.90 (26) & 53.39 (5) & 1.85 (15) & 1.98 &
0.74 (14) & 53.63 (11) & 1.17 (23) & 0.60  & 
- & - & - & - & 
- & - & - & - &
- & - & - & -
\\

SDC35E  &  15.61 (17) & 45.50 (2) & 3.88 (5) & 3.78 & 
2.87 (11) & 45.87 (4) & 2.08 (9) & 1.30 & 
0.65 (9) & 45.87 (12) & 1.69 (26) & 0.36 &
- & - & - & - &
- & - & - & -
\\

SDC35F[v1]  & 6.71 (3) & 42.97 (4) & 2.34 (5) & 2.69 &
1.33 (10) & 42.85 (5) & 1.44 (13) & 0.80  &
0.36 (7) & 42.77 (13)  & 1.27 (28) & 0.27 &
0.17 (6) & 41.26 (16) & 0.79 (26) & 0.20 &
- & - & - & -
\\
         
SDC35F[v2]  & 10.75 (8) & 44.63 (4) & 4.27 (11) & 2.37 &
1.95 (2) & 45.37 (4) & 1.64 (10) & 1.11  &
0.55 (9) & 45.50 (12) & 1.63 (29) & 0.32 &
0.53 (11) & 45.07 (34) & 3.26 (85) & 0.15 &
- & - & - & -
\\

SDC35G  &  3.99 (17) & 54.20 (5) & 2.50 (12) & 1.50 &
1.92 (11) & 55.51 (8) & 2.57 (18) & 0.70  & 
- & - & - & - & 
0.78 (12) & 54.67 (38) & 4.91 (91) & 0.15 &
- & - & - & -
\\

SDC35H & 12.04 (20) & 52.64 (3) & 3.68 (7) & 3.08 &
2.03 (13) & 53.61 (7) & 2.33 (17) & 0.81 &
0.25 (7) & 53.72 (9) & 0.64 (22) & 0.36 &
- & - & - & - &
- & - & - & -
\\

SDC35I  & 11.52 (22) & 63.74 (6) & 6.09 (12) & 1.78 & 
2.33 (14) & 64.41 (10) & 3.22 (23) & 0.68 &
1.61 (19) & 64.28 (35) & 6.07 (81) & 0.25 &
0.99 (18) & 64.10 (32) & 4.17 (108) & 0.22 &
- & - & - & -
\\

SDC35J &  5.62 (23) & 54.08 (13) & 6.34 (28) & 0.83 &
0.92 (11) & 54.05 (13) & 2.10 (29) & 0.41 & 
- & - & - & - & 
- & - & - & - &
- & - & - & -
\\

SDC35K & 5.27 (23) & 60.19 (10) & 4.79 (28) & 1.03 & 
- & - & - & -  &
- & - & - & - &
- & - & - & - & 
0.15 (7) & 58.56 (23) & 0.64 (45) & 0.22
\\

SDC35L &  25.03 (30) & 50.97 (6) & 10.17 (13) & 2.31 & 
4.85 (19) & 52.54 (10) & 5.08 (22) & 0.90 & 
- & - & - & - & 
2.25 (21) & 52.29 (43) & 9.05 (95) & 0.23 &
2.10 (19) & 53.30 (33) & 7.28 (71) & 0.27
\\

SDC35M[v1] & 6.01 (35) & 53.12 (8) & 4.60 (18) & 1.23 & 
2.75 (21) & 53.31 (13) & 3.53 (30) & 0.74 &
- & - & - & - &
- & - & - &  - &
- & - & - & -
\\

SDC35M[v2] & 7.55 (51) & 56.97 (53) & 16.89 (91) & 0.42 & 
1.30 (20) & 57.56 (21) & 2.87 (46) & 0.43&
-  & - & - & - &
- & - & - & - &
- & - & - & -
\\

SDC35N &  15.79 (37) & 48.13 (6) & 5.64 (16) & 2.63 &
3.45 (16) & 50.04 (10) & 4.45 (24) & 0.71 &
1.74 (17) & 49.69 (29) & 5.77 (62) & 0.28 &
1.19 (22) & 49.37 (49) & 5.87 (155) & 0.19 &
- & - & - & -
\\

SDC35O & 7.31 (14) & 95.56 (3) & 3.06 (7) & 2.25 &
1.49 (10) & 95.14 (7) & 2.04 (15) & 0.69 &
0.60 (10) & 94.98 (15) & 1.81 (35) & 0.31 &
1.39 (15) & 95.13 (26) & 4.81 (62) & 0.27 &
- & - & - & -
\\
         
SDC35P &  8.57 (16) & 92.25 (4) & 3.97 (9) & 2.03 & 
1.45 (11) & 92.56 (8) & 2.21 (21) & 0.62 & 
0.31 (8) & 92.63 (17) & 1.31 (38) & 0.22 &
- & - & - & - &
- & - & - & -
\\
       
SDC35Q & 6.11 (21) & 94.66 (8) & 4.65 (18) & 1.24 & 
0.81 (11) & 94.41 (13) & 1.69 (33) & 0.45 &
- & - & - & -  & 
- & - & - & - &
- & - & - & -
\\

SDC35R[v1] &  0.88 (39) & 74.46 (7) & 1.22 (24) & 0.68 &
0.53 (23) & 75.50 (26) & 1.32 (40) & 0.38 & 
- & - & - & - & 
- & - & - & - &
- & - & - & -\\

SDC35R[v2] & 7.51 (53) & 76.19 (11) & 3.37 (20) & 2.10 &
0.45 (25) & 76.94 (32) & 1.39 (69) & 0.30 & 
- & - & - & - & 
- & - & - & - &
- & - & - & -\\
        
SDC35S &  17.24 (16) & 77.41 (1) & 2.34 (3) & 6.92 &
2.17 (12) & 77.92 (6) & 2.05 (13) & 1.00  & 
- & - & - & - &
- & - & - & - &
- & - & - & -\\

SDC35T[v1] & 4.74 (20) & 49.06 (9) & 3.93 (17) & 1.13 & 
1.34 (11) & 49.46 (9) & 1.98 (17) & 0.63 &
- & - & - & - &
- & - & - & - &
- & - & - & -
\\

SDC35T[v2] & 1.93 (15) & 59.55 (7) & 1.72 (15) & 1.06 & 
0.76 (1) & 59.68 (4) & 0.90 (13) & 0.80 &
- & - & - & - & 
- & - & - & - &
- & - & - & -
\\

SDC37A &  - & - & - & <0.24 & 
0.94 (10) & 73.32 (7) & 1.30 (16) & 0.68 &
- & -  & - & - &
- & -  & - & - &
- & - & - & -
\\

SDC37B & 3.00 (15) & 72.17 (5) & 1.99 (13) & 1.42 & 
1.52 (14) & 72.45 (11) & 2.32 (29) & 0.62  & 
- & - & - & - &
- & -  & - & - &
- & - & - & -
\\
           
SDC37C &  1.27 (93) & 71.79 (26) & 1.42 (31) & 0.83 &
1.13 (11) & 72.39 (9) & 1.70 (16) & 0.61 &
- & - & - & - &
- & - & - & - &
- & - & - & -
\\

SDC37D  & 21.72 (17) & 65.00 (1) & 3.12 (3) & 6.53 &
2.59 (13) & 64.85 (5) & 2.21 (12) & 1.09 & 
0.93 (16) & 64.69 (21) & 2.69 (68) & 0.32 & 
1.41 (13) & 64.37 (10) & 2.25 (26) & 0.59 &
- & - & - & -
\\

SDC37E & 5.73 (13) & 66.40 (3) & 2.25 (6) & 2.40 &
0.89 (10) & 66.63 (6) & 1.22 (17) & 0.69 & 
0.33 (7) & 66.56 (12) & 1.05 (23) & 0.29 &
- & -  & - & - &
- & - & - & -
\\

SDC37F &  9.21 (12) & 66.82 (1) & 1.57 (2) & 5.51 & 
0.81 (10) & 66.81 (11) & 1.88 (27) & 0.41 &
- & - & - & - & 
- & -  & - & - &
- & -  & - & - 
\\

SDC37G & 6.85 (18) & 67.13 (4) & 3.51 (13) & 1.83 & 
0.78 (9) &68.13 (10) & 1.68 (24) & 0.44  & 
- & - & - & - & 
- & -  & - & - &
- & - & - & -
\\

SDC37H &  2.71 (14) & 67.21 (13) & 5.13 (32) & 0.50 & 
0.22 (5) & 67.88 (6) & 0.50 (12) & 0.42 & 
- & - & - & - & 
0.28 (8) & 67.11 (17) & 1.19 (37) & 0.22 &
- & - & - & -
\\

SDC37I  & 1.72 (11) & 66.55 (9) & 2.74 (17) & 0.59 & 
- &- & -& <0.16 &
0.49 (10) & 66.61 (12) & 2.08 (47) & 0.22 & 
- & - & - & - &
- & - & - & -
\\

SDC37J & 1.35 (10) & 67.30 (6) & 1.66 (16) & 0.76 &
- & - & - &  <0.16&
- & - & - & - &
- & -  & - & - &
- & - & - & -
\\

SDC37K & 12.60 (17) & 54.92 (2) & 3.25 (5) & 3.64 & 
1.59 (12) & 54.98 (11) & 2.91 (26) & 0.51 &
0.17 (7) & 56.05 (15) & 0.72 (31) & 0.22 & 
0.28 (8) & 58.30 (19) & 1.31 (49) & 0.20 &
0.23 (6) & 54.49 (17) & 1.22 (31) & 0.18
\\   
     \hline  
  \noalign{\smallskip}\hline
\end{tabular}
\vspace{1mm}

\parbox{0.95\textheight}{
\tiny
Notes. Values in parentheses indicate the uncertainty in the last quoted digit(s). The parameters $I$, $V$, $\Delta V$, and $T_{\rm p}$ denote the integrated intensity, centroid velocity, FWHM linewidth, and peak main-beam brightness temperature, respectively.
Dashes indicate non-detections or unavailable fit parameters.
Upper limits are given as $3\sigma$ limits for $T_{\rm p}$. The number of Gaussian components fitted to HCO$^+$ was determined from the number of resolved H$^{13}$CO$^+$ components. One HCO$^+$ Gaussian was fitted for a single H$^{13}$CO$^+$ component, and two HCO$^+$ Gaussians were fitted for two resolved H$^{13}$CO$^+$ components. The labels [v1] and [v2] identify the corresponding velocity components. Because HCO$^+$ is optically thick and may exhibit self-absorption, asymmetry, or line wings, its Gaussian parameters are descriptive and are not used as systemic velocities or intrinsic linewidths.
}
\end{sidewaystable}
\restoregeometry 
\clearpage

\clearpage
\begin{table*}
\scriptsize
    \centering
    \begin{minipage}[]{150mm}
\caption{Velocity ranges and kinematic features of the kinematic tracers}\label{tab:linewing}
\end{minipage}
    \begin{tabular}{c*{7}{c}}
         \hline\hline
                 Source  & HCO$^+$ & H$^{13}$CO$^+$ & HNCO & SiO & 
                SO & Kinematic Features \\   
               & (km\,s$^{-1}$) &(km\,s$^{-1}$) & (km\,s$^{-1}$) & (km\,s$^{-1}$) & 
               (km\,s$^{-1}$)  & \\  \hline 
SDC19A &  [6.8, 42.9]  & [21.9, 32.5] & [23.5, 29.9] & [10.1, 45.6] & [21.6, 31.1] & S \\
SDC19B &  [16.7, 38.3] & [23.5, 29.2] & [21.0, 30.5] & [16.7, 37.5] & -  & S \\
SDC19C &  [15.2, 41.7] & [24.1, 30.9] & [23.9, 30.5] & [22.7, 36.6] & -  & OF; S \\
SDC19D &  [119.5, 125.0] & [121.1, 123.2] & - & - & -  & -  \\
SDC19E &  [5.6, 37.0] & [12.1, 28.3] & [11.2, 28.6] & [6.6, 33.8] & [11.3, 27.7] & IN; S \\
SDC19F &  -   & [24.9, 28.8] & [25.2, 28.8] & - & -  & - \\
SDC19G &  [107.3, 118.6]  & [109.3, 116.6] & [111.2, 116.5] & [108.2, 117.5] & - & OF; S\\
SDC19H[v1] & [19.0, 23.0] &  [20.8, 23.0] & - & - & - & IN \\
SDC19H[v2] & [23.0, 33.1] & [23.0, 26.2] & - & [25.1, 29.6] & - & IN; S \\
SDC33A & - & [100.1, 105.6] & - & - & - & - \\
SDC33B & [79.3, 88.9] &  [82.4, 87.4] & - & - & - & IN \\
SDC33C & [79.3, 91.6] & - & - & - & - & - \\
SDC33D & [107.1, 111.5] & - & - & - & - & - \\
SDC33E & [69.4, 80.9] & [70.8, 77.6] & - & - & - & IN \\
SDC33F & [97.8, 108.3] & [98.7, 107.5] & - & - & - & - \\
SDC33G[v1] & [66.9, 79.9] & [69.9, 77.7] & - & - & -  & IN \\
SDC33G[v2] & [95.1, 111.5] & [101.2, 110.5] & - & [102.1, 108.5] & - & IN; S \\
SDC33H[v1] & [68.8, 81.7] & [69.0, 81.0] & - & [65.6, 69.7] & - & IN; S \\
SDC33H[v2] & [96.5, 110.1] & [99.5, 104.6] & - & - & - & IN \\
SDC33I & [71.5, 107.0] & - & - & - & - & - \\
SDC33J & [96.7, 105.1] & [98.5, 103.3] & [98.5, 103.1] & - & - & - \\
SDC33K & [96.9, 103.5] & [96.0, 100.9] & - & - & - & - \\
SDC33L & [95.1, 105.5] & - & - & - & - & -  \\
SDC33M & [96.7, 104.7] & [95.1, 101.9] & [97.0, 102.1] & - & [97.2, 100.5] & IN \\
SDC33N & [96.3, 102.3] & [96.6, 100.5] & - & [97.6, 102.3] & - & OF; S \\
SDC33O & [80.9, 89.7] & [83.0, 86.5] & - & - & - & IN\\
SDC33P & [79.7, 90.3] & [82.0, 88.9] & - & - & - & IN\\
SDC33Q & [4.2, 12.9] & [7.6, 10.6] & - & - & -  & IN \\
SDC33R & [80.0, 88.7] & [80.5, 86.2] & [79.8, 89.6] & [74.9, 91.0] & - & IN; S  \\
SDC35A & [21.5, 33.2] & [26.8, 29.6] & - & - & - & OF \\
SDC35B & [41.3, 51.5] & [43.6, 46.0] & - & - & - & IN \\
SDC35C[v1] & [40.1, 52.2] & [48.0, 51.6] & - & - & - & IN \\
SDC35C[v2] & [52.2, 56.6] & [52.5, 55.0] & - & - & - & IN \\
SDC35E & [39.0, 51.6] & [43.0, 49.0] & [44.7, 48.1] & - & - & IN \\
SDC35F[v1] & [39.7, 43.7] & [40.5, 44.2] & [41.0, 43.8] & [38.9, 41.6] & - &  IN; S  \\
SDC35F[v2] & [43.6, 50.5] & [44.2, 46.8] & [43.9, 48.1] & [42.3, 48.3] & - & OF; S \\
SDC35G & [40.2, 61.5] & [52.2, 59.3] & - & [51.1, 59.5] & - & IN; S \\
SDC35H & [49.0, 56.8] & [50.2, 56.7] & [52.9, 54.9] & - & - & IN \\
SDC35I & [58.2, 70.2] & [60.6, 67.8] & [59.9, 69.6] & [61.9, 67.7] & - & IN; S\\
SDC35J & [46.5, 60.2] & [51.7, 56.4] & - & - & - & OF \\
SDC35K & [49.2, 66.2] & - & - & - & [57.5, 59.6] & - \\
SDC35L & [33.0, 57.1] & [48.1, 57.2] & - & [45.5, 59.1] & [46.1, 59.6] & OF; S \\
SDC35M[v1] & [42.1, 53.5] & [48.6, 55.1] & - & - & - & IN \\
SDC35M[v2] & [52.7, 77.2] & [56.0, 61.5] & - & - & - & OF \\
SDC35N & [39.7, 56.0] & [45.2, 54.3] & [44.2, 55.5] & [45.0, 56.7] & - & OF; S \\
SDC35O & [89.2, 99.8] & [93.3, 97.7] & [92.6, 97.5] & [92.9, 100.3] & - & OF; S \\
SDC35P & [87.7, 97.1] & [89.7, 95.0] & [91.6, 95.6] & - & - & IN \\
SDC35Q & [90.9, 99.0] & [92.6, 97.1] & - & - & - & IN \\
SDC35R & [69.2, 81.8] & [73.9, 78.8] & - & - & - & - \\
SDC35S & [74.6, 79.9] & [76.0, 81.0] & - & - & - & IN \\
SDC35T[v1] & [44.3, 52.2] & [47.5, 51.1] & - & - & - & IN\\
SDC35T[v2] & [57.7, 61.8] & [58.9, 60.4] & - & - & - & IN \\
SDC37A & - & [71.7, 74.4] & - & - & - & - \\
SDC37B &  [70.2, 76.7] & [70.5, 75.2] & - & - & -  & IN \\
SDC37C & [69.9, 77.1] & [70.1, 74.2] & - & - & - & - \\
SDC37D & [60.0, 69.9] & [62.2, 69.0] & [62.4, 67.4] & [61.7, 68.2] & - & IN; S \\
SDC37E & [62.2, 70.9] & [64.5, 69.0] & [65.4, 68.7] & - & - & IN \\
SDC37F & [62.7, 68.2] & [65.1, 69.1] & - & - & - & OF  \\
SDC37G & [62.5, 75.9] & [65.1, 70.1] & - & - & - & IN \\
SDC37H & [62.5, 73.5] & [67.5, 68.7] & - & [65.0, 70.1] & -  & IN; S \\
SDC37I & [63.7, 69.0] & - & [64.2, 69.5] & - & -  & - \\
SDC37J & [65.3, 69.7] & - & - & - & -  & - \\
SDC37K & [49.2, 63.0] & [52.2, 58.2] & [52.2, 57.5] & [51.2, 65.0] & [52.9, 55.8] & IN; S \\ 
         \hline\hline      
         \multicolumn{7}{p{.75\textwidth}}{Notes. The HCO$^+$ and H$^{13}$CO$^+$ velocity ranges give the full detected emission intervals above $3\sigma$. These quantities are descriptive measures of the observed profile extents. For SiO, HNCO, and SO, the interval $[V_{\rm blue},V_{\rm red}]$ denotes the full detected velocity range between the blue- and red-most channels with emission above $3\sigma$. The IN and OF classifications require both significant HCO$^+$ residual structure and profile morphology relative to H$^{13}$CO$^+$. S (Shock) is inferred from SiO detections. Sources labelled in [v1] and [v2] are those showing two velocity components on H$^{13}$CO$^+$ profiles. For SDC19D and SDC35R, the spectrum may contain two velocity components in HCO$^+$ and H$^{13}$CO$^+$ while both components are within the same continuous velocity interval listed here and are therefore not separated into [v1] and [v2] entries.    
         }
     \end{tabular} 
    \end{table*}
\clearpage

\clearpage
\begin{table*}
\scriptsize
\caption{Residuals of HCO$^+$ spectra after subtraction of one- or two-component Gaussian fits}\label{tab:hcopresidual}
\centering
    \begin{tabular}{c*{4}{c}}
    \hline\hline\noalign{\smallskip}
Source & $\mathrm{RMS}_{\rm residual}$ & $\mathrm{RMS}_{\rm baseline}$ & Notes \\
& (K) & (K) & & \\
  \hline\noalign{\smallskip}

SDC19A & 0.40 & 0.09 & Significant Residual \\
SDC19B & 0.11 & 0.08 & Significant Residual \\
SDC19C & 0.19 & 0.06 &  Significant Residual \\
SDC19D & - & - &  Two Velocity Components\\
SDC19E & 0.71 & 0.11 & Significant Residual  \\
SDC19F & - & - & no HCO$^+$ emission \\
SDC19G & 0.17 & 0.08 & Significant Residual\\
SDC19H[v1] & 0.13 & 0.10 & Significant Residual\\
SDC19H[v2] & 0.13 & 0.10 & Significant Residual\\
SDC33A & - & - & no HCO$^+$ emission\\
SDC33B & 0.18 & 0.08 &Significant Residual; Tentative H$^{13}$CO$^+$ Emission \\
SDC33C & - & - & No H$^{13}$CO$^+$ Emission \\
SDC33D & - & - & No H$^{13}$CO$^+$ Emission \\
SDC33E & 0.08 & 0.07 &  Significant Residual \\
SDC33F & 0.08 & 0.09 &   Tentative H$^{13}$CO$^+$ Emission\\
SDC33G[v1] & 0.11 & 0.09 & Significant Residual  \\
SDC33G[v2] & 0.11 & 0.09 & Significant Residual; Tentative H$^{13}$CO$^+$ Emission  \\
SDC33H[v1] & 0.13 & 0.09 & Significant Residual \\
SDC33H[v2] & 0.10 & 0.09 & Significant Residual \\
SDC33I & 0.09 & 0.09 & No H$^{13}$CO$^{+}$ Emission \\
SDC33J & 0.07 & 0.11 & Tentative H$^{13}$CO$^+$ Emission \\
SDC33K & 0.10 & 0.08 & Tentative H$^{13}$CO$^+$ emission \\
SDC33L & 0.10 & 0.10 & No H$^{13}$CO$^+$ emission \\
SDC33M & 0.10 & 0.09 & Significant Residual  \\
SDC33N & 0.09 & 0.07 & Significant Residual  \\
SDC33O & 0.12 & 0.07 & Significant Residual  \\
SDC33P & 0.11 & 0.07 & Significant Residual \\
SDC33Q & 0.15 & 0.07 &   Significant Residual\\
SDC33R & 0.18 & 0.10 &  Significant Residual \\
SDC35A & 0.30 & 0.16 & Significant Residual \\
SDC35B & 0.29 & 0.12 & Significant Residual  \\
SDC35C & 0.21 & 0.09 & Significant Residual; Two Velocity Components\\
SDC35E & 0.23 & 0.08 & Significant Residual \\
SDC35F & 0.12 & 0.07 &Significant Residual; Two Velocity Components \\
SDC35G & 0.22 & 0.07 &Significant Residual   \\
SDC35H & 0.17 & 0.08 &  Significant Residual\\
SDC35I & 0.47 & 0.10 &Significant Residual   \\
SDC35J & 0.13 & 0.10 & Significant Residual \\
SDC35K & 0.08 & 0.07 &  Significant Residual; No H$^{13}$CO$^+$ Emission\\
SDC35L & 0.75 & 0.11 & Significant Residual  \\
SDC35M & 0.15 & 0.06 & Significant Residual; Two Velocity Components  \\
SDC35N & 0.13 & 0.10 &  Significant Residual  \\
SDC35O & 0.17 & 0.07 & Significant Residual  \\
SDC35P & 0.12 & 0.07 &  Significant Residual \\
SDC35Q & 0.17 & 0.06 & Significant Residual \\
SDC35R & 0.08 & 0.08 & Two Velocity Components\\
SDC35S & 0.11 & 0.07 &   Significant Residual\\
SDC35T & 0.17 & 0.10 &   Significant Residual; Two Velocity Components \\
SDC37A & - & -  & No HCO$^+$ Emission\\
SDC37B & 0.15 & 0.09 &  Significant Residual\\
SDC37C & 0.07 & 0.10 & No Significant Residual \\
SDC37D & 0.29 & 0.08 & Significant Residual  \\
SDC37E & 0.17 & 0.09 &  Significant Residual  \\
SDC37F & 0.23 & 0.07 & Significant Residual  \\
SDC37G & 0.17 & 0.07 &  Significant Residual  \\
SDC37H & 0.08 & 0.06 &   Significant Residual  \\
SDC37I & 0.11 &  0.06 & Significant Residual; No H$^{13}$CO$^+$ Emission  \\
SDC37J & 0.06 & 0.08 & No Significant Residual; No H$^{13}$CO$^+$ Emission  \\
SDC37K & 0.18 & 0.08 & Significant Residual \\ 
     \hline  
  \noalign{\smallskip}\hline    
          \multicolumn{4}{p{.50\textwidth}}{Notes. $RMS_{\rm residual}$ is the rms measured over the HCO$^+$ emission interval after subtraction of the appropriate one- or two-component Gaussian model. $RMS_{\rm baseline}$ is the rms measured in line-free channels on the $T_{\rm MB}$ scale. Profiles satisfying $RMS_{\rm residual}>RMS_{\rm baseline}$ are considered to contain non-Gaussian residual structure above the baseline noise. H$^{13}$CO$^+$ features labelled as tentative have $S/N\gtrsim3$ and are included in the quoted H$^{13}$CO$^+$ detection count.}
\end{tabular} 
\end{table*}
\clearpage

     \clearpage
\begin{table*}
\scriptsize
\caption{Fitted parameters for NH$_2$D and the derived ammonia deuteration fractions}\label{tab:NH2D}
\centering
\begin{tabular}{c*{7}{c}}
\hline\hline\noalign{\smallskip}
Source & Velocity & Width & Integrated Intensity & $T_{\rm pk}$ & $N({\rm NH_2D})$ & $D_{\rm frac}^{\rm app}$ & $D_{\rm frac}^{\rm corr}$ \\
& (km\,s$^{-1}$) & (km\,s$^{-1}$) & (K\,km\,s$^{-1}$) & (K) & ($10^{13}$\,cm$^{-2}$) & & \\
\hline\noalign{\smallskip}
SDC19A & 26.2 (1) & 3.50 (4) & 4.60 (2) & 0.66 & 8.50 (52) & 0.089 (16) & 0.027 (5) \\
SDC19B & 27.0 (1) & 3.20 (6) & 0.60 (1) & 0.17 & 1.03 (18) & 0.011 (3) & 0.003 (1) \\
SDC19C & 27.4 (2) & 2.20 (4) & 0.80 (1) & 0.33 & 1.31 (17) & 0.014 (3) & 0.004 (1) \\
SDC19E & 20.7 (5) & 2.20 (2) & 4.70 (3) & 0.62 & 8.65 (73) & 0.060 (13) & 0.018 (4) \\
SDC33M & 100.0 (2) & 1.80 (3) & 1.8 (2) & 0.41 & 2.93 (34) & 0.054 (12) & 0.016 (4) \\
SDC33O & 85.1 (5) & 1.30 (6) & 0.30 (8) & 0.24 & 0.46 (12) & $>0.046$ & $>0.014$ \\
SDC33P & 85.2 (2) & 2.10 (8) & 0.60 (2) & 0.29 & 0.95 (32) & 0.038 (15) & 0.011 (4) \\
SDC33Q & 9.5 (7) & 0.70 (2) & 0.40 (7) & 0.47 & 0.63 (10) & 0.037 (65) & 0.011 (20) \\
SDC33R & 83.7 (2) & 1.30 (5) & 0.40 (1) & 0.31 & 0.62 (15) & 0.0168 (49) & 0.0051 (15) \\
SDC35A & 28.2 (3) & 3.50 (8) & 2.00 (3) & 0.41 & 3.14 (48) & 0.081 (19) & 0.024 (6) \\
SDC35E & 45.3 (3) & 2.90 (5) & 2.00 (2) & 0.27 & 3.08 (31) & 0.0395 (41) & 0.0119 (12) \\
SDC35F & 46.7 (2) & 1.20 (2) & 1.20 (2) & 0.20 & 1.85 (31) & 0.029 (7) / 0.066 (18) & 0.0088 (20) / 0.0199 (54) \\
SDC35G & 55.2 (2) & 1.40 (5) & 0.30 (1) & 0.19 & 0.48 (16) & 0.0137 (51) & 0.0041 (15) \\
SDC35M & 53.5 (3) & 3.00 (6) & 2.20 (3) & 0.21 & 3.39 (46)  & 0.083 (18) / 0.261 (53) & 0.0249 (54) / 0.0785 (161) \\
SDC35O & 96.1 (3) & 3.00 (9) & 0.60 (2) & 0.17 & 0.96 (32) & 0.0148 (56) & 0.0045 (17) \\
SDC35P & 92.7 (5) & 3.10 (8) & 0.50 (2) & 0.16 & 0.77 (31) & 0.033 (15) & 0.010 (4) \\
SDC35S & 78.1 (2) & 1.90 (6) & 1.10 (3) & 0.20 & 2.01 (56) & 0.0447 (153) & 0.0135 (46) \\
SDC37C & 72.3 (3) & 2.60 (8) & 0.70 (2) & 0.24 & 1.10 (32) & 0.039 (14) & 0.0118 (43) \\
\hline
\noalign{\smallskip}\hline
\multicolumn{8}{p{.88\textwidth}}{Notes. $D_{\rm frac}^{\rm app}=N({\rm NH_2D})_{\rm Yebes}/N({\rm NH_3})_{\rm Eff}$ is the apparent beam-mismatched ammonia deuteration fraction. $D_{\rm frac}^{\rm corr}=D_{\rm frac}^{\rm app}(\theta_{\rm Yebes}/\theta_{\rm Eff})^2$ adopts $\theta_{\rm Yebes}=21.9''$ at 85.926\,GHz and $\theta_{\rm Eff}=40''$, assuming that NH$_2$D and NH$_3$ arise from the same unresolved region. For SDC35F and SDC35M, two values are given because the NH$_3$ spectra show two velocity components and the component matching is ambiguous. Values in parentheses indicate the uncertainty in the last quoted digit(s). The listed width is the common intrinsic line width of the underlying NH$_2$D velocity component obtained from the simultaneous HFS fit. For SDC33O, NH$_3$(1,1) is detected but NH$_3$(2,2) is not. The listed $N({\rm NH_2D})$ is a nominal estimate obtained by adopting the upper limit $T_{\rm kin}=17.6$\,K as $T_{\rm ex}({\rm NH_2D})$.}
\end{tabular}
\end{table*}
\clearpage

\section{Dust properties derivation}\label{sect:tdust}

The core properties are computed from ASCUMET method on the two-dimensional dust continuum map by aggregating all pixels belonging to each structure (Liu et al. in prep.). The dust mass is calculated pixel by pixel as
\begin{equation}
M = \sum \frac{S_{\nu, \mathrm{pix}} \, D^2}{\kappa_\nu \, B_\nu(T_{d,\mathrm{pix}})} ,
\end{equation}
where $S_{\nu, \mathrm{pix}}$ is the flux of each pixel within the structure, $B_\nu$ is the Planck function, and $T_d$ is adopted from Ren et al. (in prep.). We adopt $\kappa_\nu = 0.015 \ \mathrm{cm^2\,g^{-1}}$ at $850\,\mu$m, and $D$ is the distance. The column density is derived as
\begin{equation}
N(\mathrm{H}_2) = \frac{I_\nu}{\kappa_\nu \, B_\nu(T_d) \, \mu_{\mathrm{H}_2} \, m_{\mathrm{H}}} ,
\end{equation}
where $I_\nu$ is the intensity. 

The dust temperatures were obtained using the SED-fitting procedure  \citep{1983hildebrand} and is described below, as part of the analysis presented in Ren et al. (in prep.). The spectral energy distribution (SED) is described by
\begin{equation}
S_{\nu} = \Omega\, B_{\nu}(T_{\rm d}) \left[1 - \exp(-\tau_{\nu})\right],
\end{equation}
where $S_{\nu}$ is the mean flux density at frequency $\nu$, $\Omega$ is the solid angle of the clump or selected region, $B_{\nu}(T_{\rm d})$ is the Planck function at dust temperature $T_{\rm d}$, and $\tau_{\nu}$ is the optical depth. The optical depth is given by
\begin{equation}
\tau_{\nu} = \kappa_{\nu}\,\mu\,m_{\rm H}\,N_{\rm tot}/g,
\end{equation}
where $N_{\rm tot}$ is the total gas column density (dominated by H$_2$), $\mu = 2.33$ is the mean molecular weight \citep{1983myers}, $m_{\rm H}$ is the hydrogen atom mass, and $g = 100$ is the gas-to-dust mass ratio. The dust opacity is parameterised as $\kappa_{\nu} = \kappa_{230\,\mathrm{GHz}}(\nu/230\,\mathrm{GHz})^{\beta}$, adopting $\kappa_{230\,\mathrm{GHz}} = 0.9\,\mathrm{cm^{2}\,g^{-1}}$ from the dust model of \citet{1993ossenkopf}, appropriate for coagulated grains with thin ice mantles at densities of $\sim10^{6}\,\mathrm{cm^{-3}}$.

\begin{figure}[h]
  \centering
    \captionsetup{justification=raggedright, singlelinecheck=false}
  \includegraphics[width=0.90\columnwidth]
{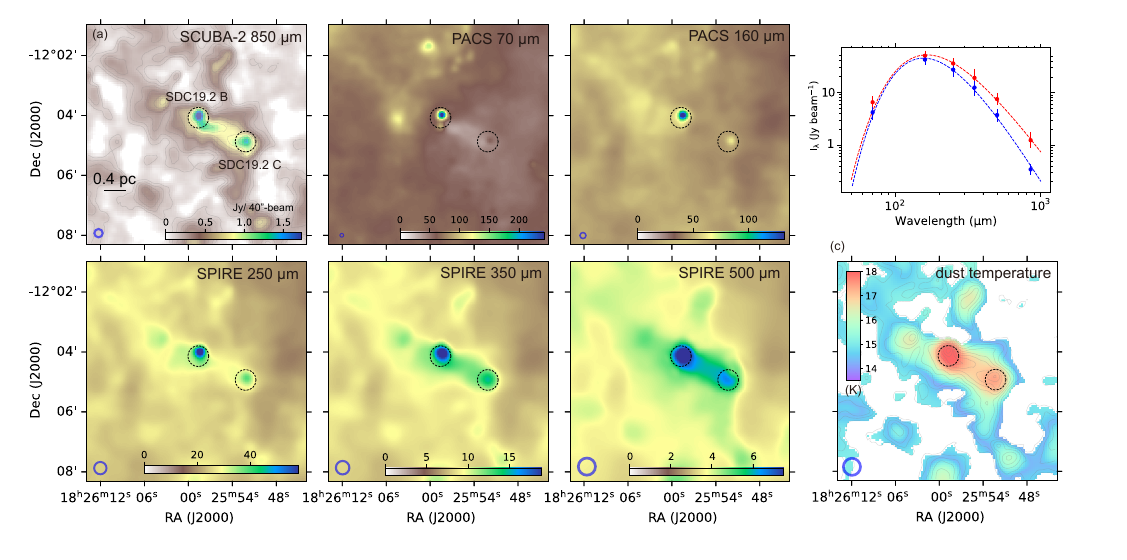}
  \caption{Example SED fitting for the clumps SDC19B and SDC19C. {\it Left panels:} JCMT SCUBA-2 $850\,\mu\mathrm{m}$ and {\it Herschel} PACS (70 and $160\,\mu\mathrm{m}$) and SPIRE (250, 350, and $500\,\mu\mathrm{m}$) images of the SDC19 region, obtained from the {\it Herschel} Science Archive\protect\footnote{\protect\url{http://archives.esac.esa.int/hsa/whsa/}}. For the SED analysis, all images were convolved to a common angular resolution of $36''$, corresponding to the SPIRE $500\,\mu\mathrm{m}$ beam. Dashed circles indicate the $40''$ apertures used to extract representative flux densities and dust temperatures for comparison with the molecular-line data. {\it Right panels:} The upper panel shows the best-fitting single-temperature greybody SEDs for SDC19B and SDC19C after subtraction of the estimated background and foreground emission. The lower panel shows the dust-temperature map derived from pixel-by-pixel SED fitting.}
  \label{fig:sdc19b_sed} 
\end{figure}

Flux densities were measured from JCMT SCUBA-2 $850\,\mu\mathrm{m}$ data and from {\it Herschel} PACS (70 and $160\,\mu\mathrm{m}$) and SPIRE (250, 350, and $500\,\mu\mathrm{m}$) observations. To derive the dust-temperature distribution, all images were first convolved to the angular resolution of the SPIRE $500\,\mu\mathrm{m}$ data and converted to units of mJy beam$^{-1}$ for a common $36''$ beam. The average sensitivity of the $850\,\mu\mathrm{m}$ map is $\sim10$\,mJy\,beam$^{-1}$, and pixels with emission above the $4\sigma$ level (40\,mJy\,beam$^{-1}$) were adopted to define the source region. Pixels below this threshold were treated as the surrounding external region. For each {\it Herschel} band, the average intensity of these external pixels was used as an approximate estimate of the background and foreground emission, which was subtracted prior to the SED fitting. A pixel-by-pixel SED fit was then performed to derive the $T_{\rm dust}$ map. For comparison with the molecular-line analysis, representative dust temperatures for individual clumps were extracted from this map using a common aperture of $\theta_{\rm FWHM}=40''$, comparable to the Effelsberg K-band beam. From SED fitting, the dust temperature, opacity index $\beta$, can be derived together. Example SED fits for SDC19B and SDC19C are shown in Fig.~\ref{fig:sdc19b_sed}, and the corresponding $T_{\rm dust}$ values were adopted as estimates of the core temperatures. The SEDs are well reproduced by a single dust temperature component. The derived values therefore represent beam-averaged dust temperatures and do not account for possible temperature gradients or multiple dust components within the clumps. They nevertheless provide a consistent basis for comparison with the gas kinetic temperatures derived from NH$_3$. Typical fitting uncertainties are $\pm0.2$\,K in $T_{\rm d}$ and $\pm0.2$ in the emissivity index $\beta$.

\clearpage

\section{Maser figures}
\begin{figure*}
  \centering
\includegraphics[width=0.80\textwidth,height=1.30\textwidth]{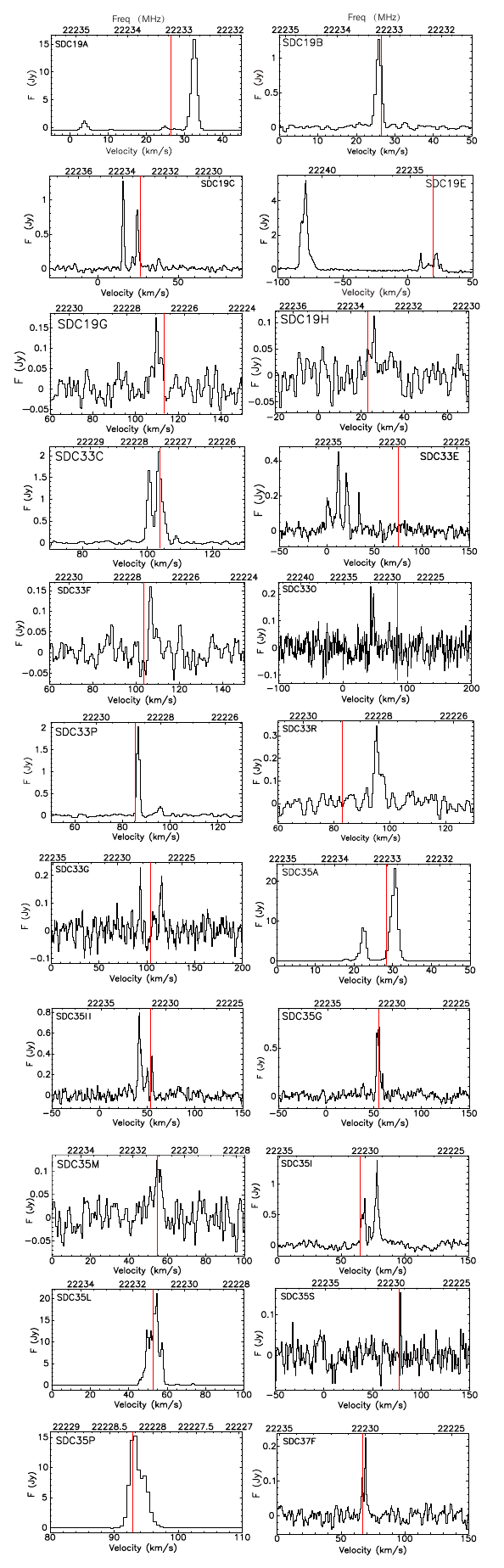}
  \caption{Water maser spectra observed by the Effelsberg 100-m telescope. The red vertical lines indicate the NH$_{3}$ systemic velocities. }
  \label{fig:watermasers} 
\end{figure*}

\begin{figure*}
  \centering
\includegraphics[width=0.35\textwidth,height=0.35\textwidth]{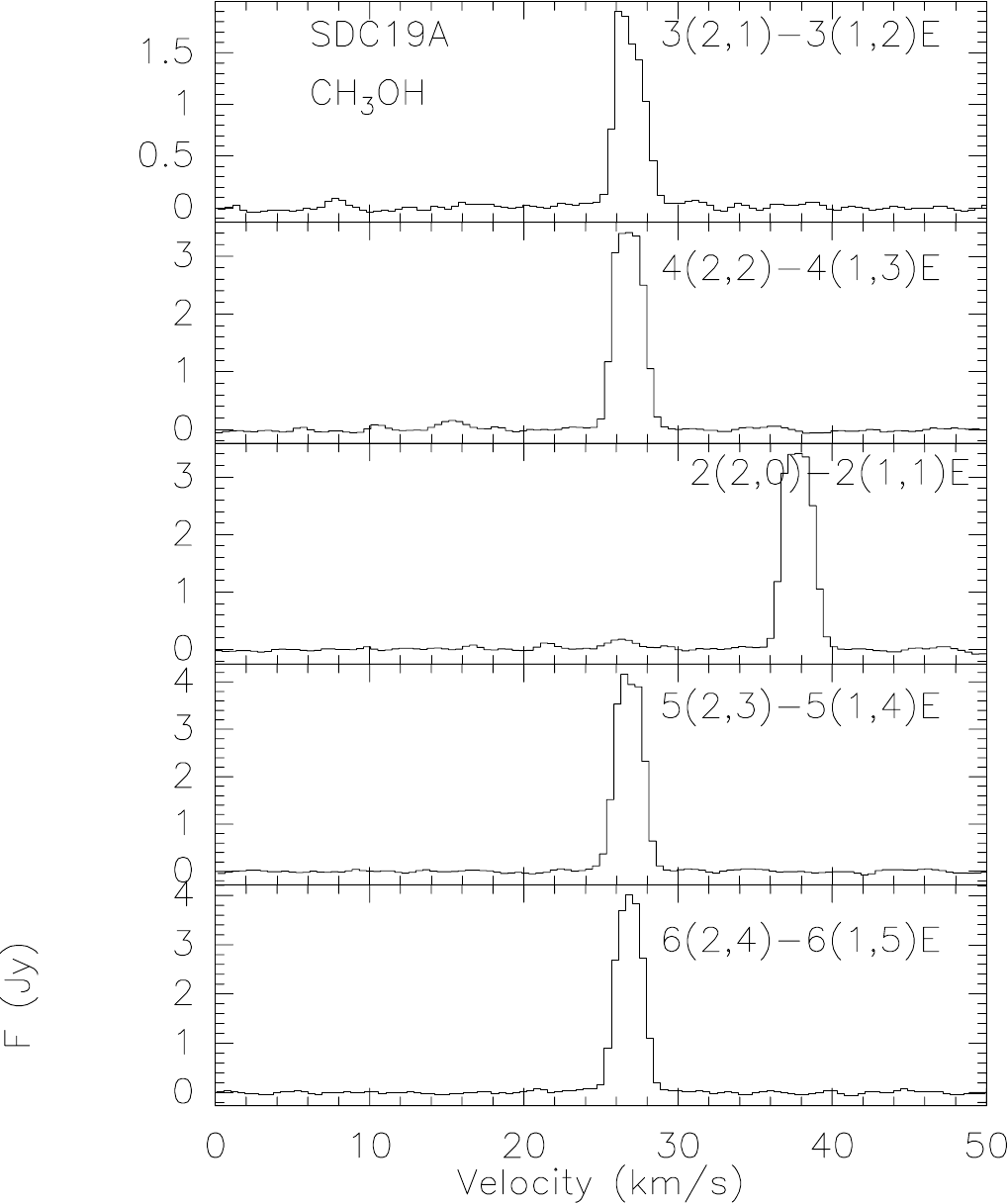}
\includegraphics[width=0.35\textwidth,height=0.35\textwidth]{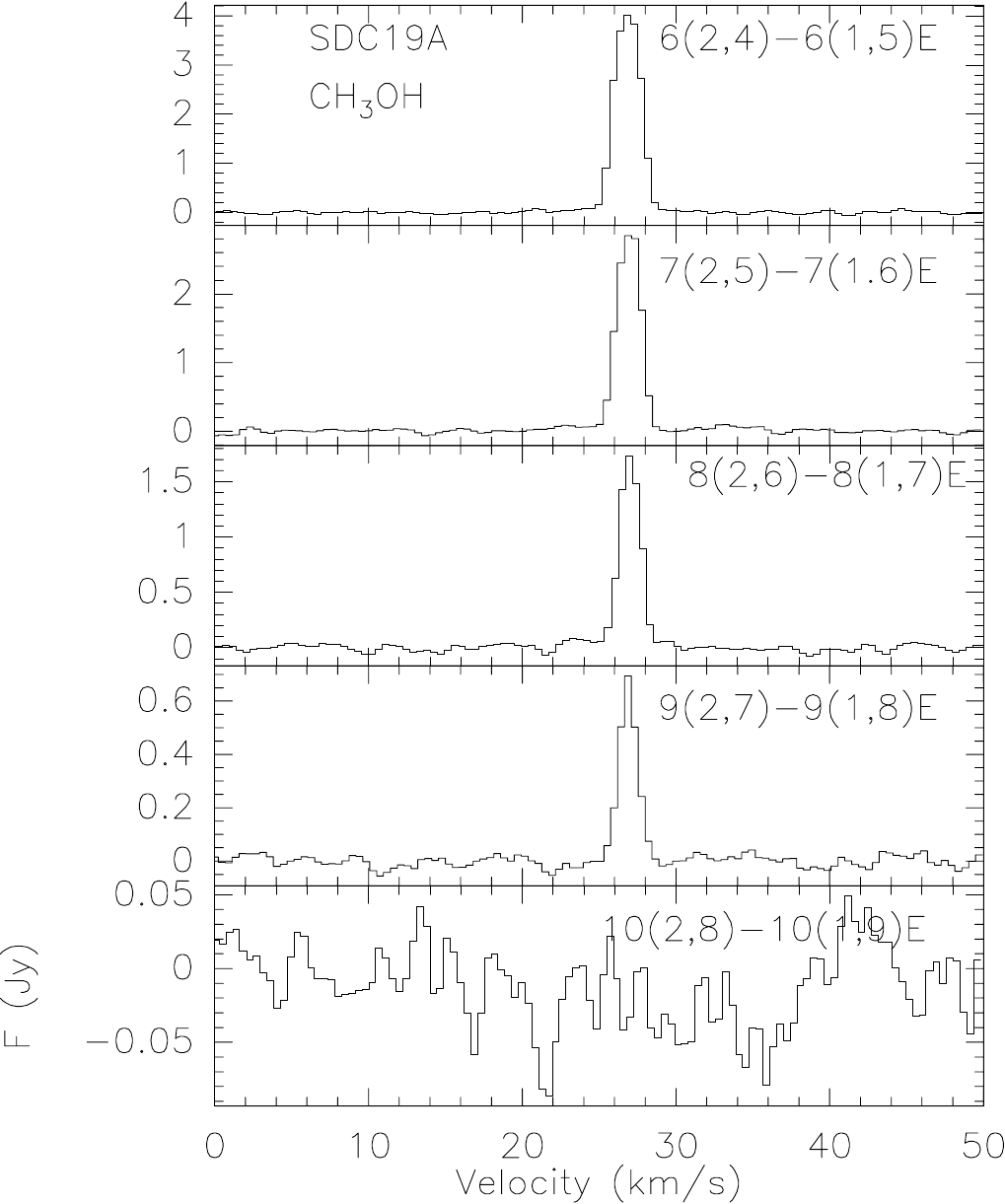}
  \caption{Class\,{\sc i} methanol maser spectra observed by the Effelsberg 100-m telescope. The CH$_{3}$OH 10$_2$-10$_1$ $E$ was not detected.}
  \label{fig:methanolmasers} 
\end{figure*}

Figures for water masers and Class\,{\sc i} methanol masers are shown in Fig.~\ref{fig:watermasers} and Fig.~\ref{fig:methanolmasers}, respectively.

\clearpage
\onecolumn

\section{Kinematics figures} \label{sec:kinematics}

 \begin{figure}[H]
  \centering
\includegraphics[width=0.33\textwidth,height=0.39\textwidth, trim={0 0  0 0}, clip]{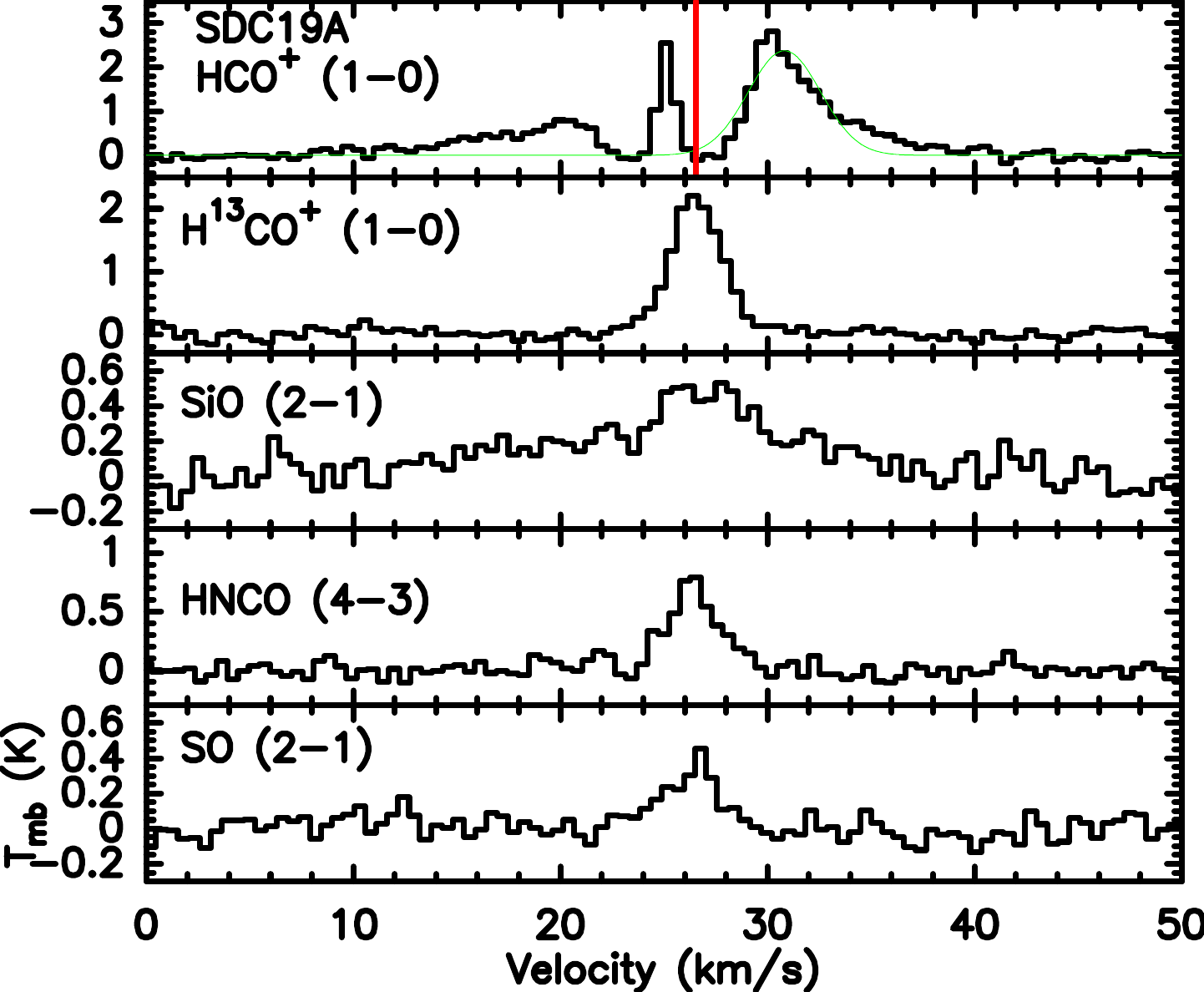}
\includegraphics[width=0.33\textwidth,height=0.39\textwidth,trim={0 0  0 0}, clip]{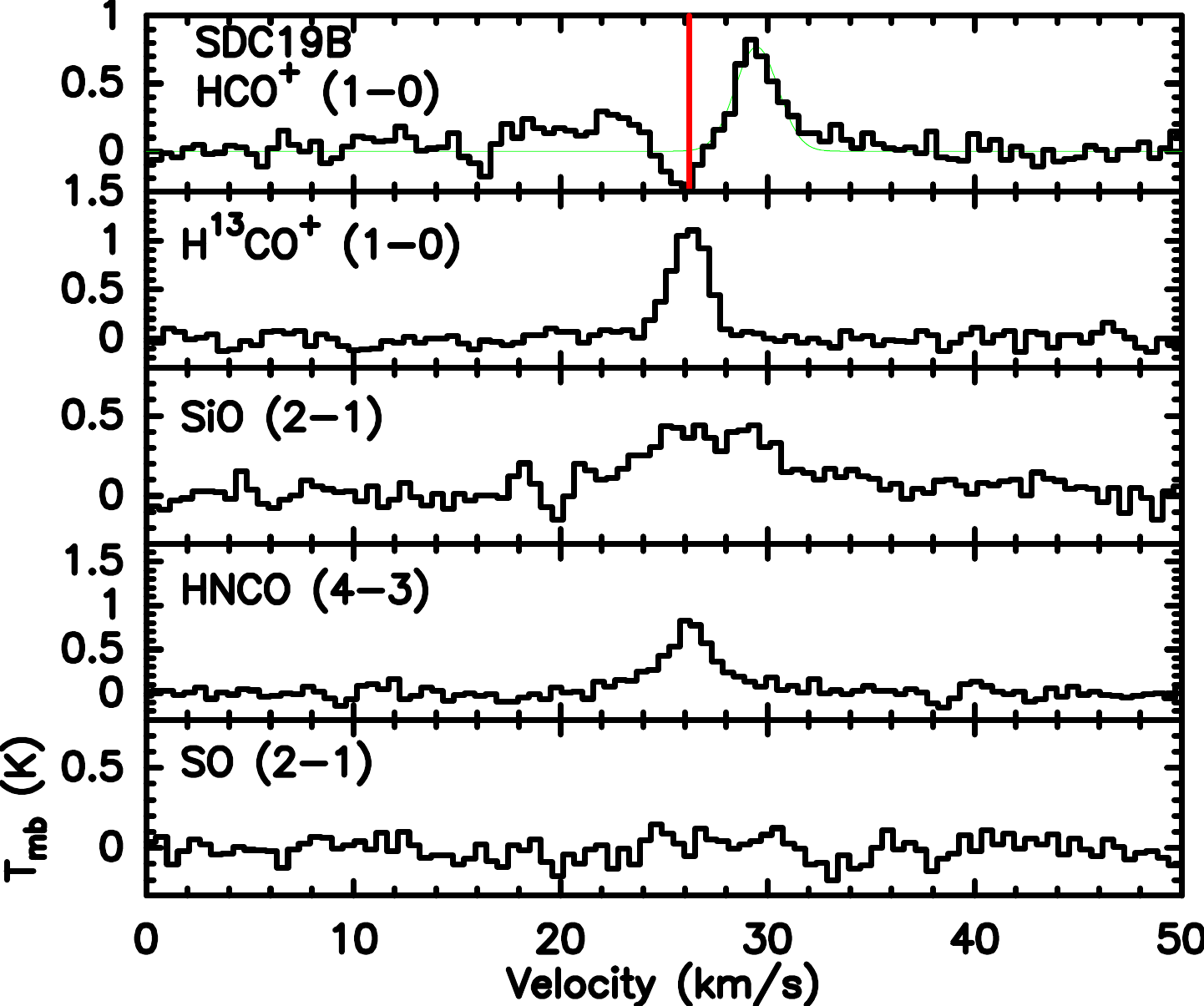}
\includegraphics[width=0.33\textwidth,height=0.39\textwidth,trim={0 0  0 0}, clip]{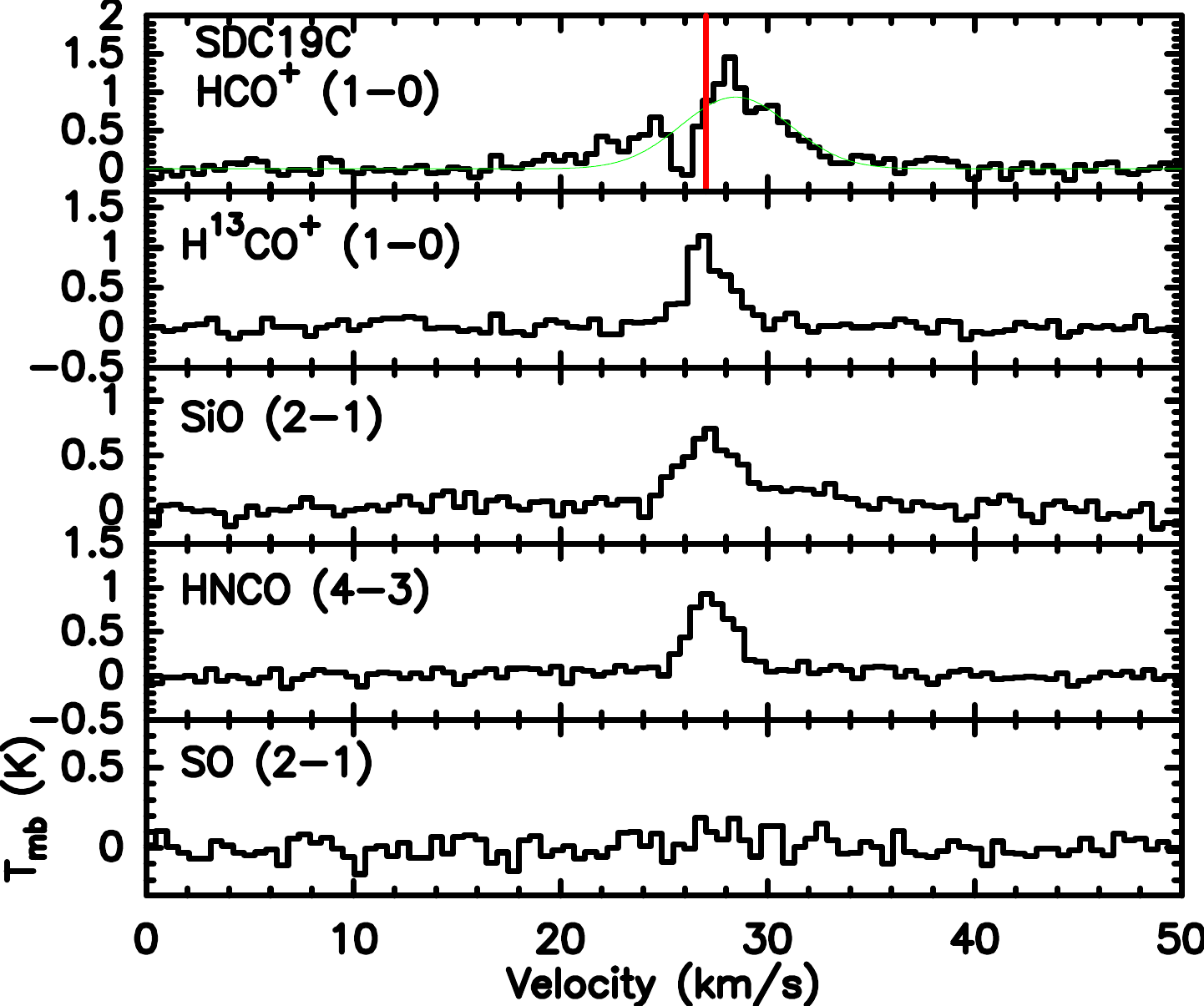}\\[1mm]

\includegraphics[width=0.33\textwidth,height=0.39\textwidth,trim={0 0  0 0}, clip]{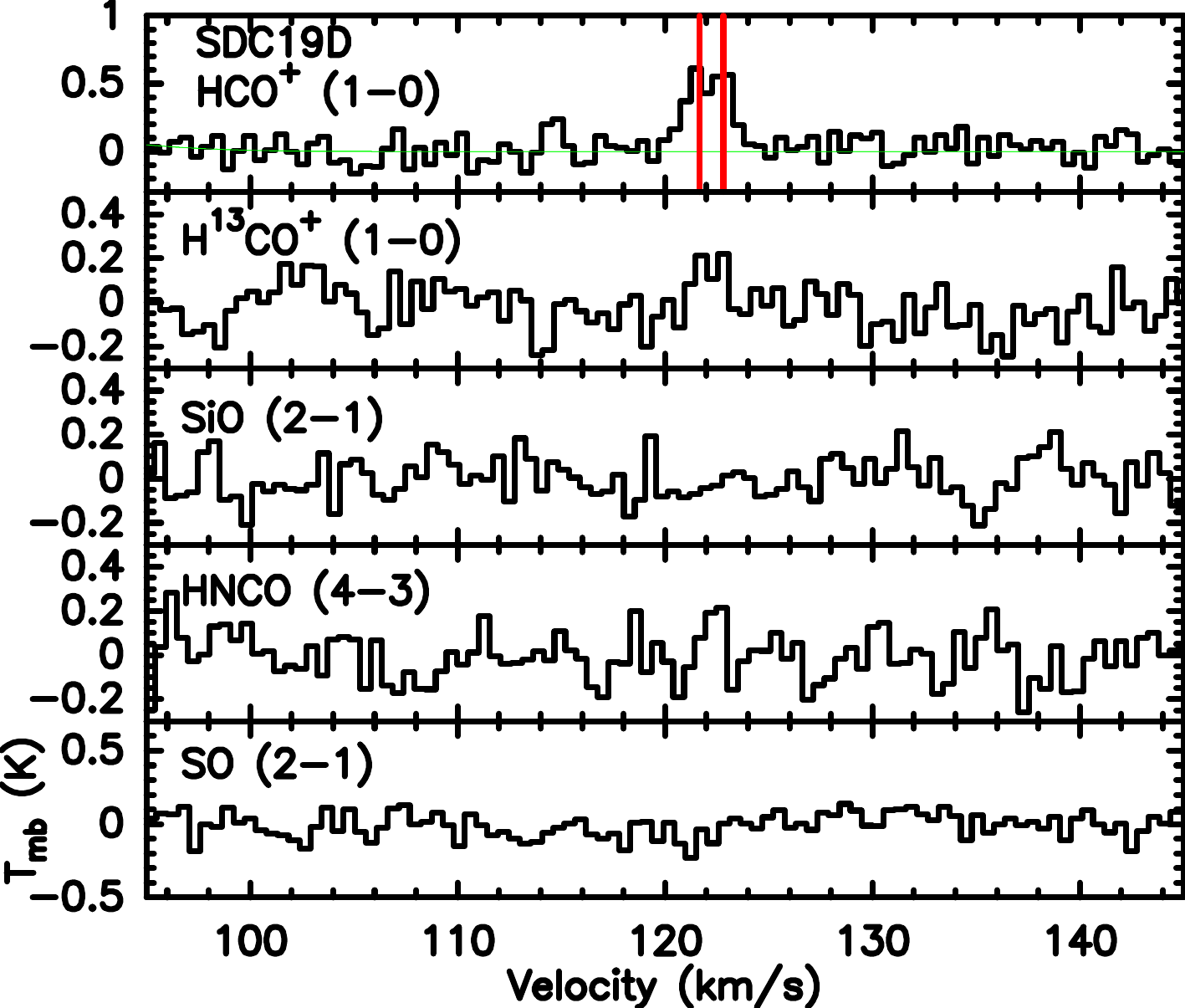}
\includegraphics[width=0.33\textwidth,height=0.39\textwidth,trim={0 0  0 0}, clip]{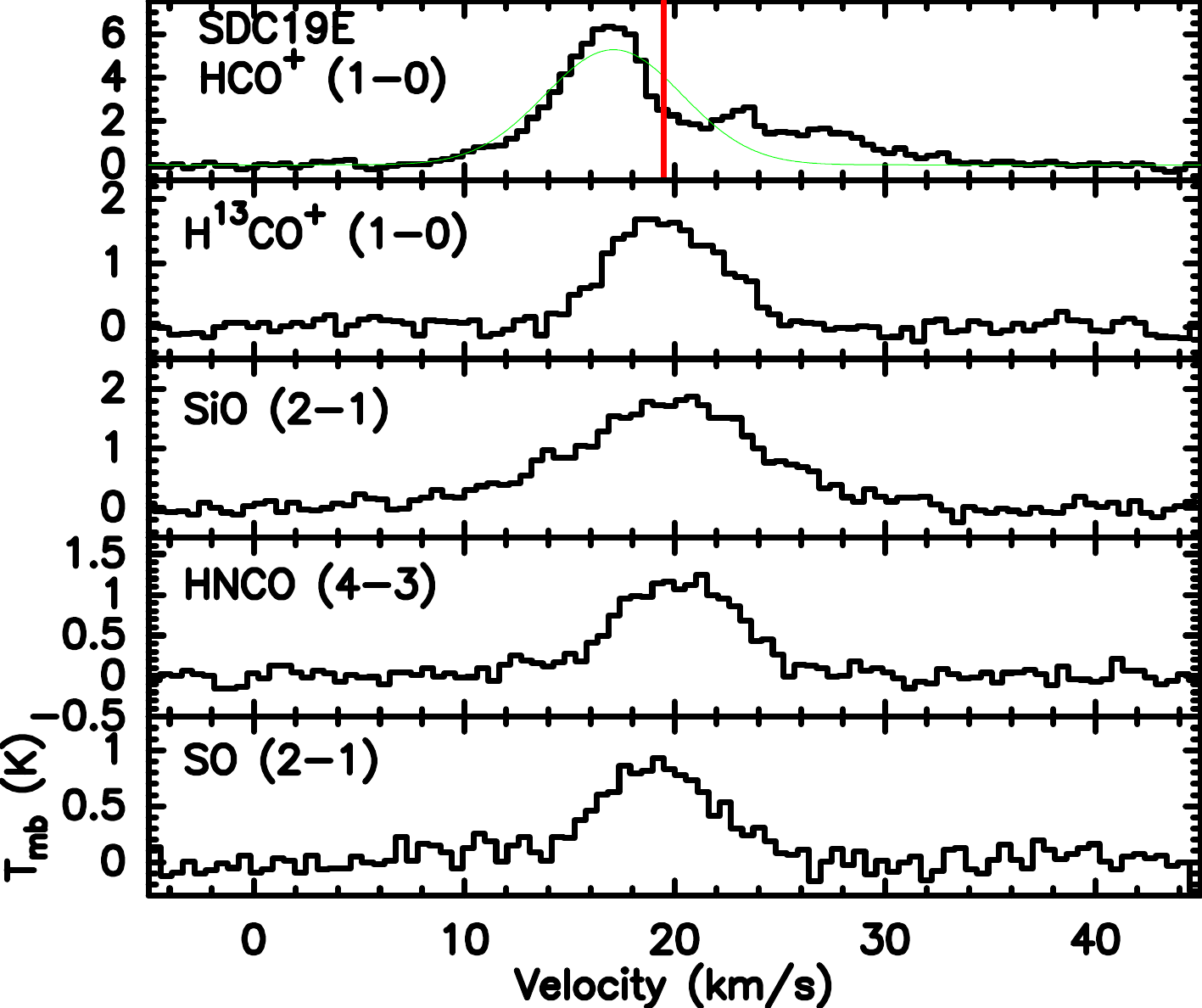}
\includegraphics[width=0.33\textwidth,height=0.39\textwidth,trim={0 0  0 0}, clip]{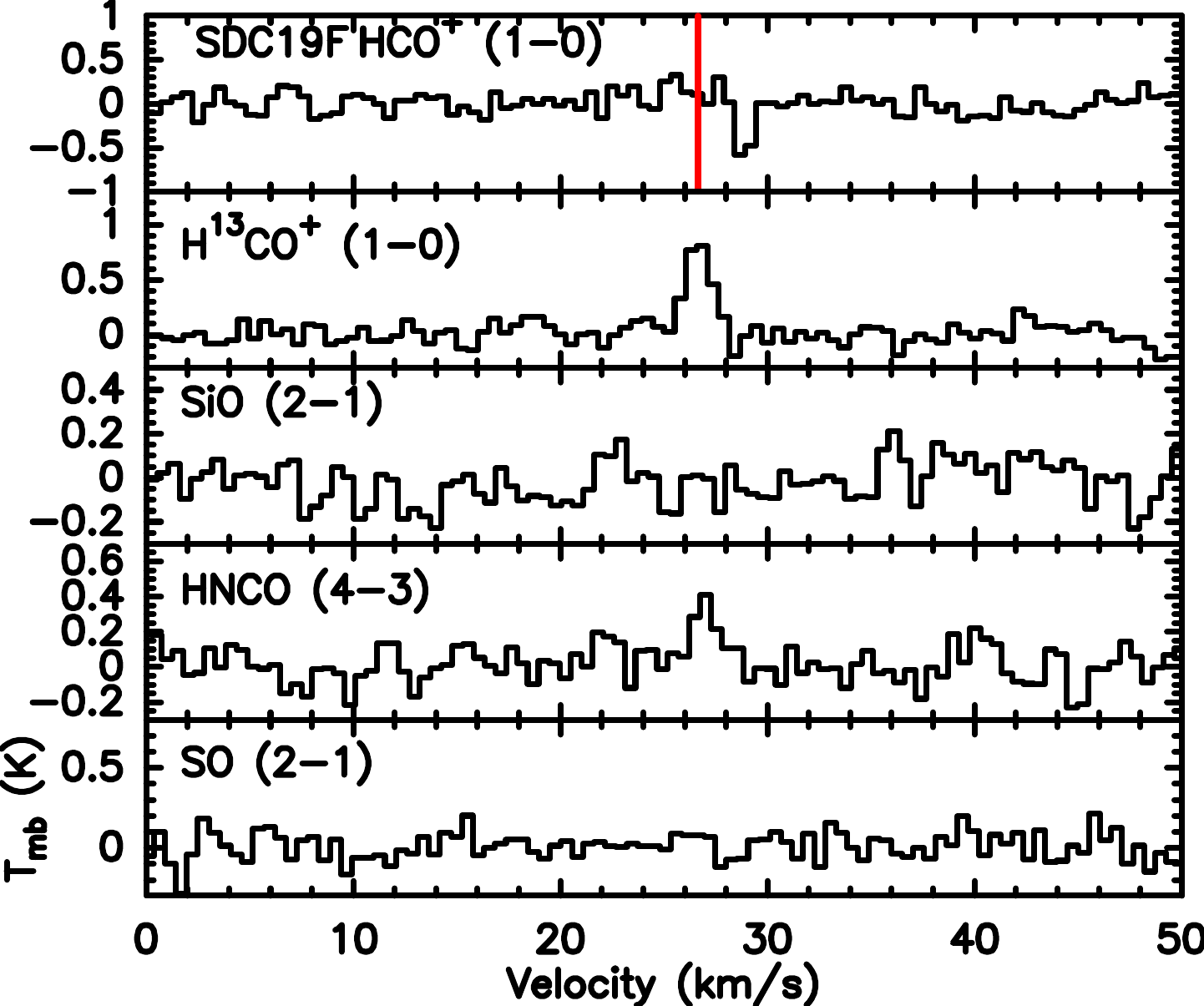}\\[1mm]

\includegraphics[width=0.33\textwidth,height=0.39\textwidth,trim={0 0  0 0}, clip]{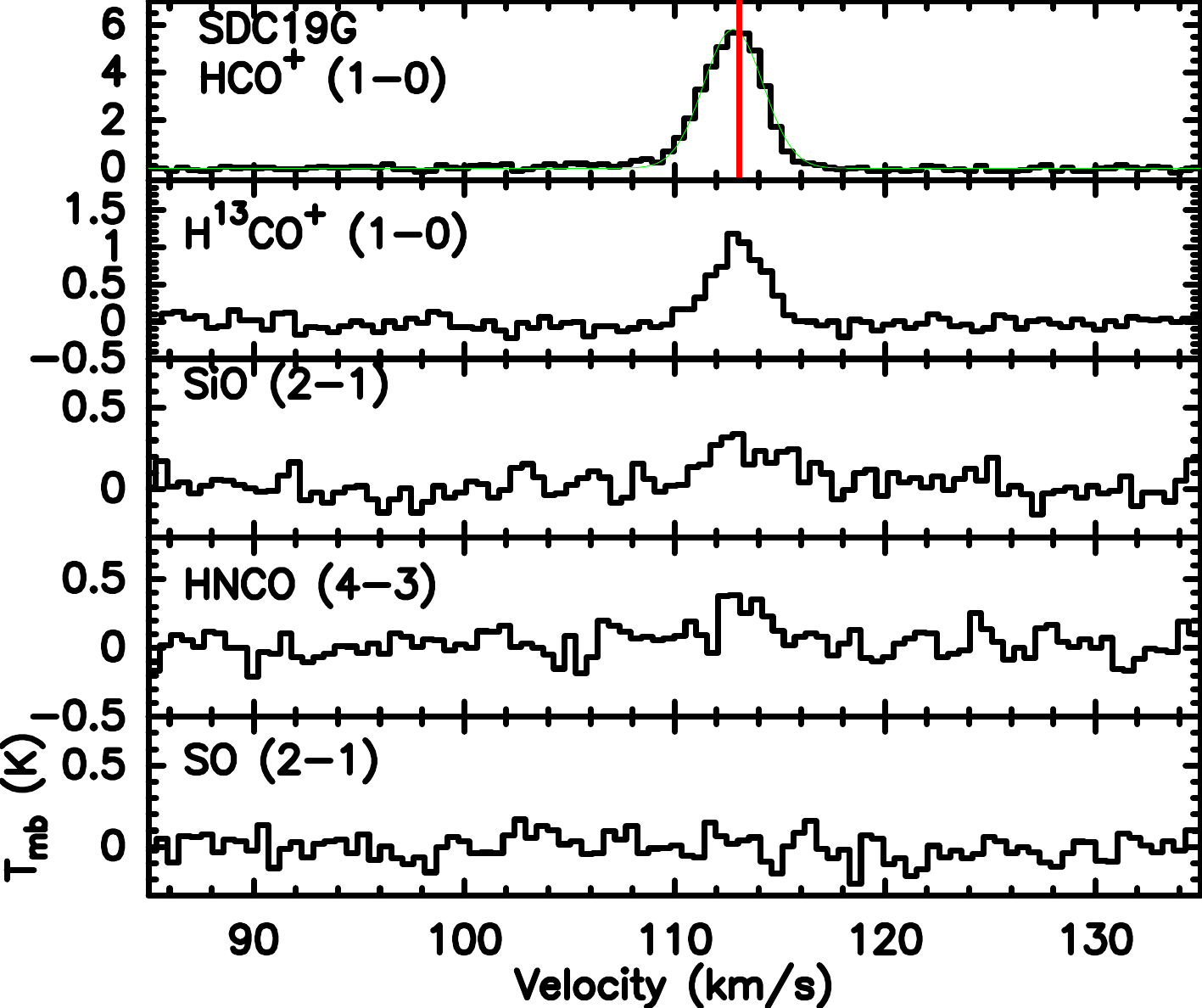}
\includegraphics[width=0.33\textwidth,height=0.39\textwidth,trim={0 0  0 0}, clip]{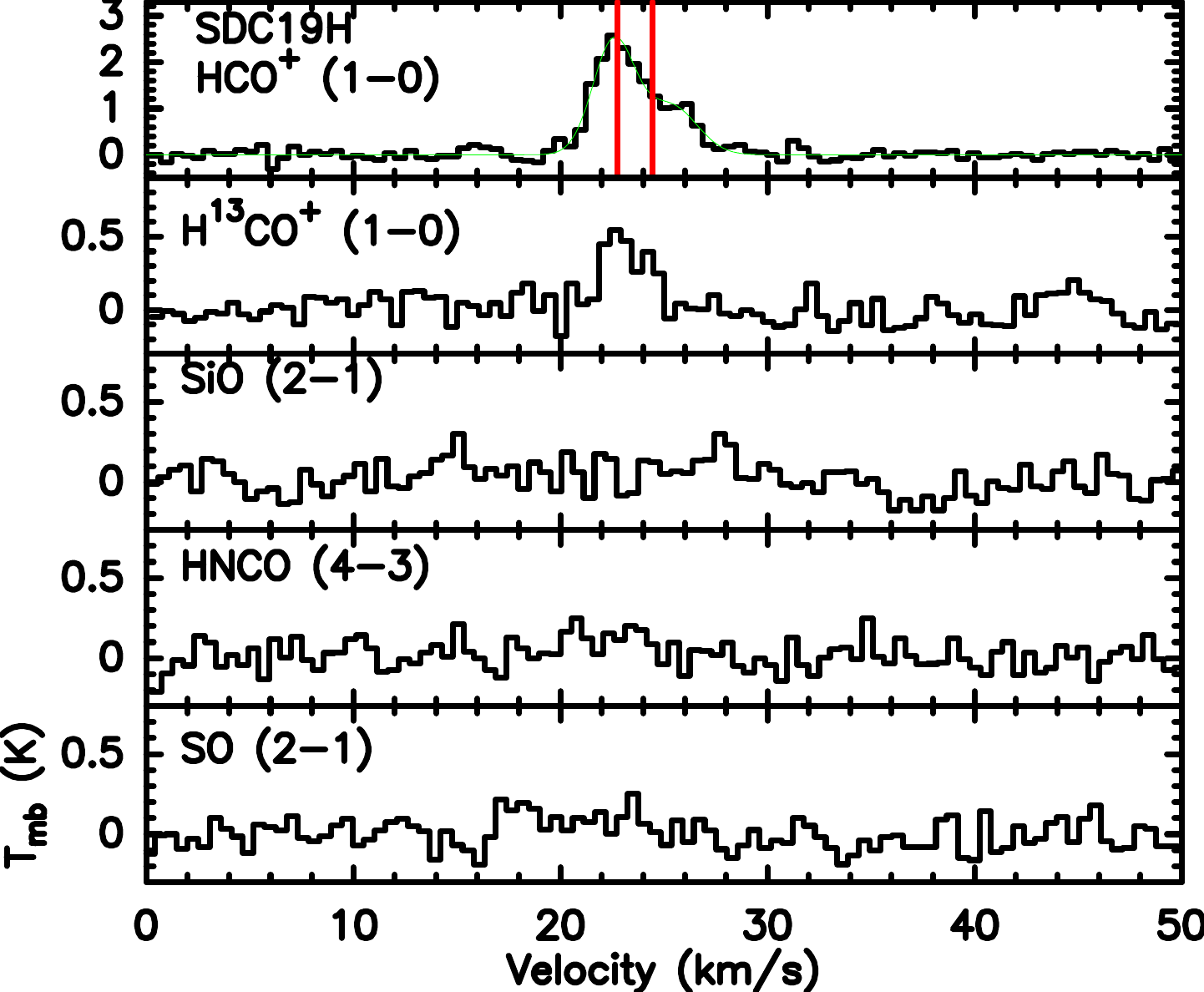}
\includegraphics[width=0.33\textwidth,height=0.39\textwidth,trim={0 0  0 0}, clip]{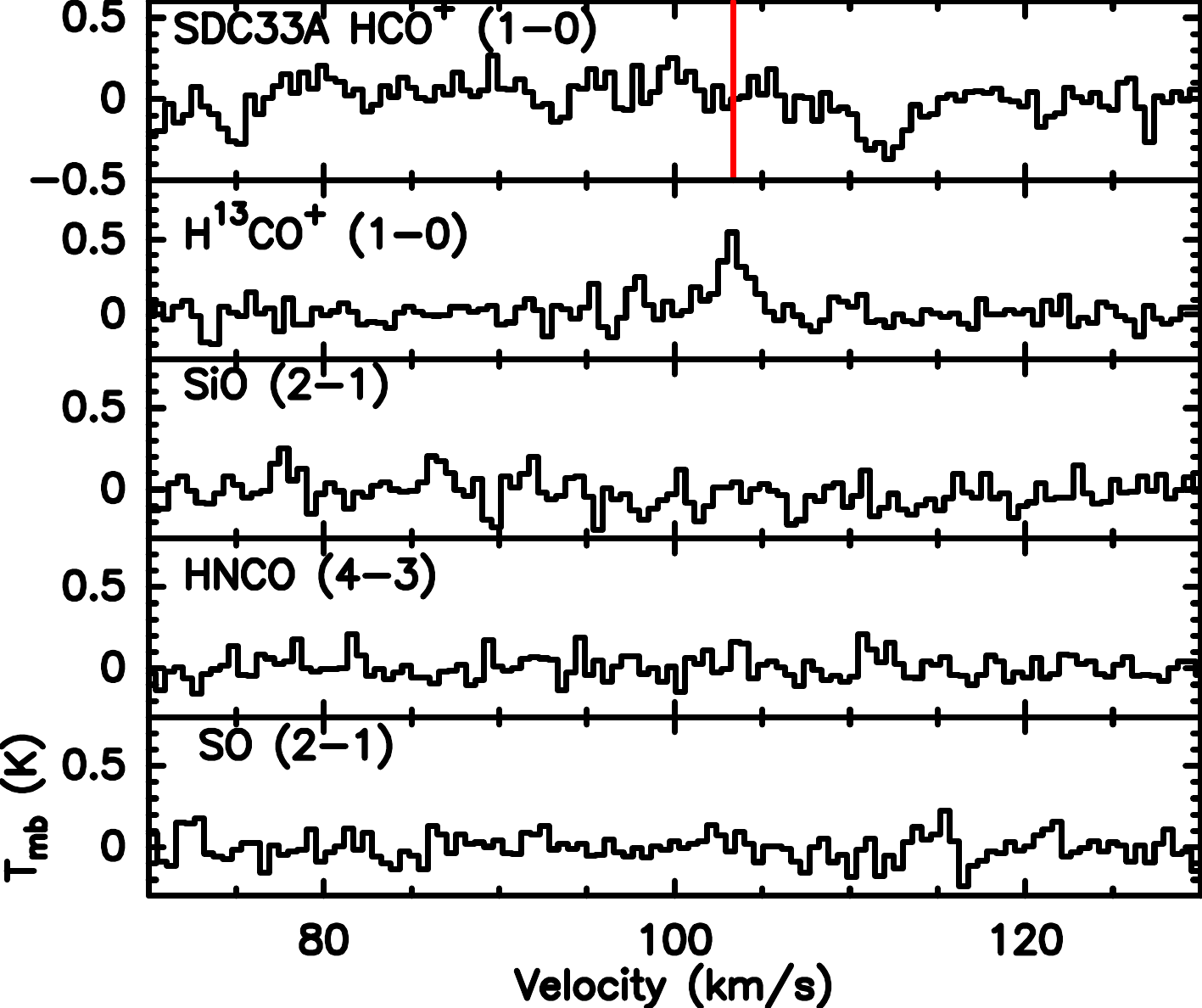}
  \caption{Spectra of the kinematic tracers towards the ALOHA IRDCs clumps. The vertical red lines indicate the centroid velocities derived from Gaussian fits to the optically thin H$^{13}$CO$^+$ profiles. For sources fitted with two H$^{13}$CO$^+$ velocity components, two red lines indicate the corresponding fitted centroid velocities, $v_{1}$ and $v_{2}$. For sources without detected H$^{13}$CO$^+$ emission, no vertical red line is shown. The light-green curve shows the Gaussian model fitted to the HCO$^+$ profile. A single Gaussian is used where one H$^{13}$CO$^+$ component is detected, whereas the sum of two Gaussians is shown where two resolved H$^{13}$CO$^+$ components are present.}
  \label{fig:kinematic_a} 
\end{figure}

 \begin{figure}[H]
   \ContinuedFloat
   \captionsetup{labelsep=period}
  \centering
\includegraphics[width=0.33\textwidth,height=0.43\textwidth,trim={0 0  0 0}, clip]{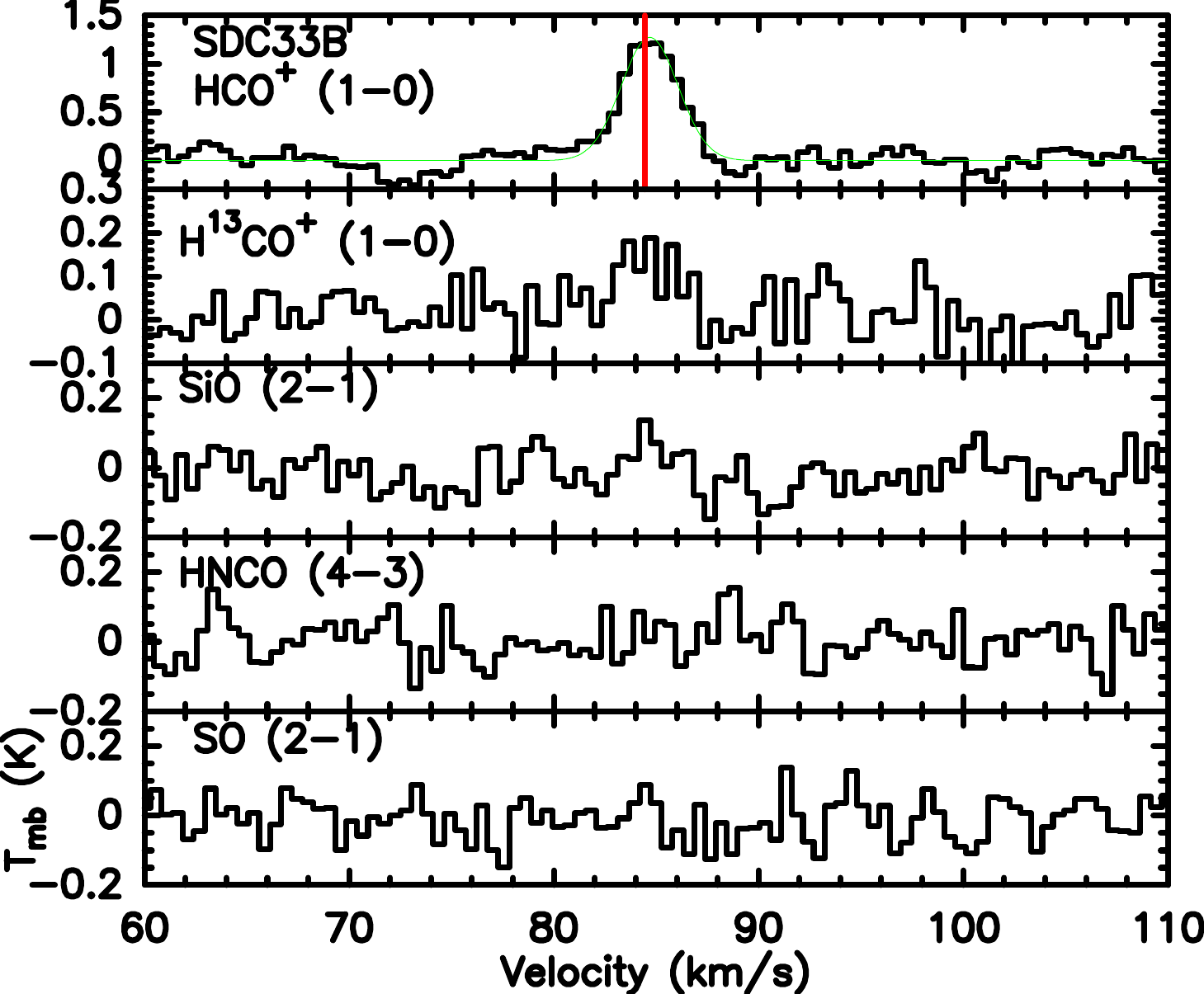}
\includegraphics[width=0.33\textwidth,height=0.43\textwidth,trim={0 0  0 0}, clip]{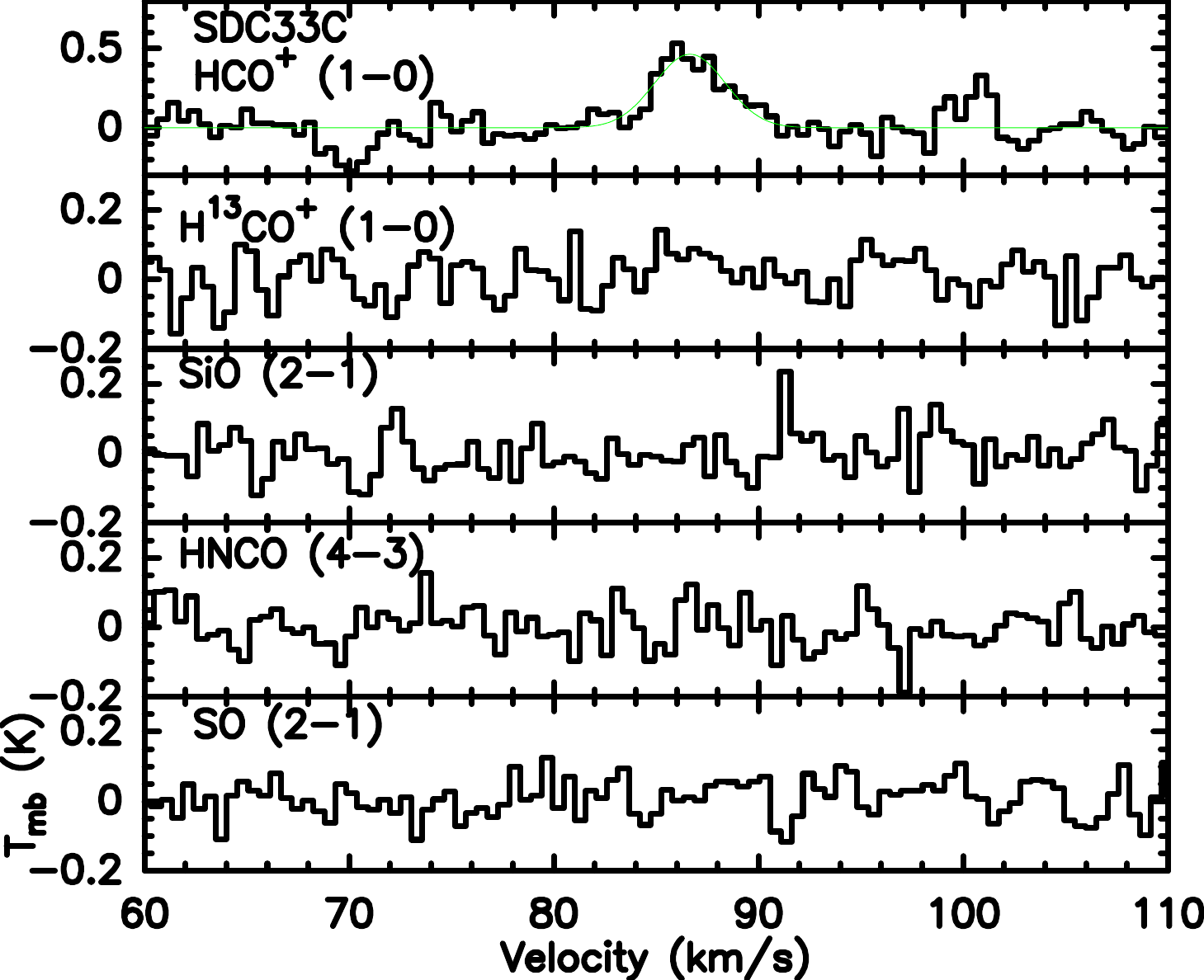}
\includegraphics[width=0.33\textwidth,height=0.43\textwidth,trim={0 0  0 0}, clip]{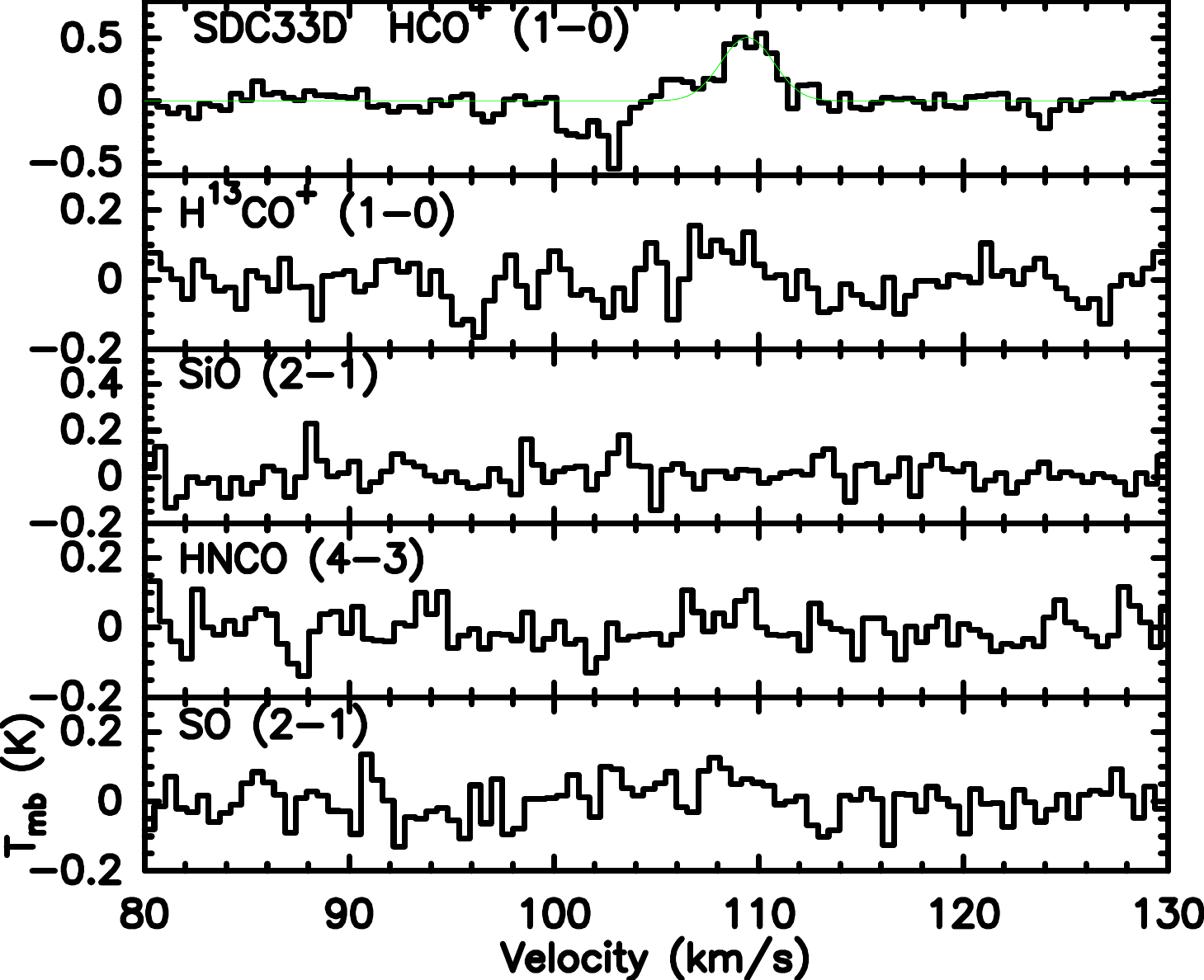}\\[3mm]

\includegraphics[width=0.33\textwidth,height=0.43\textwidth,trim={0 0  0 0}, clip]{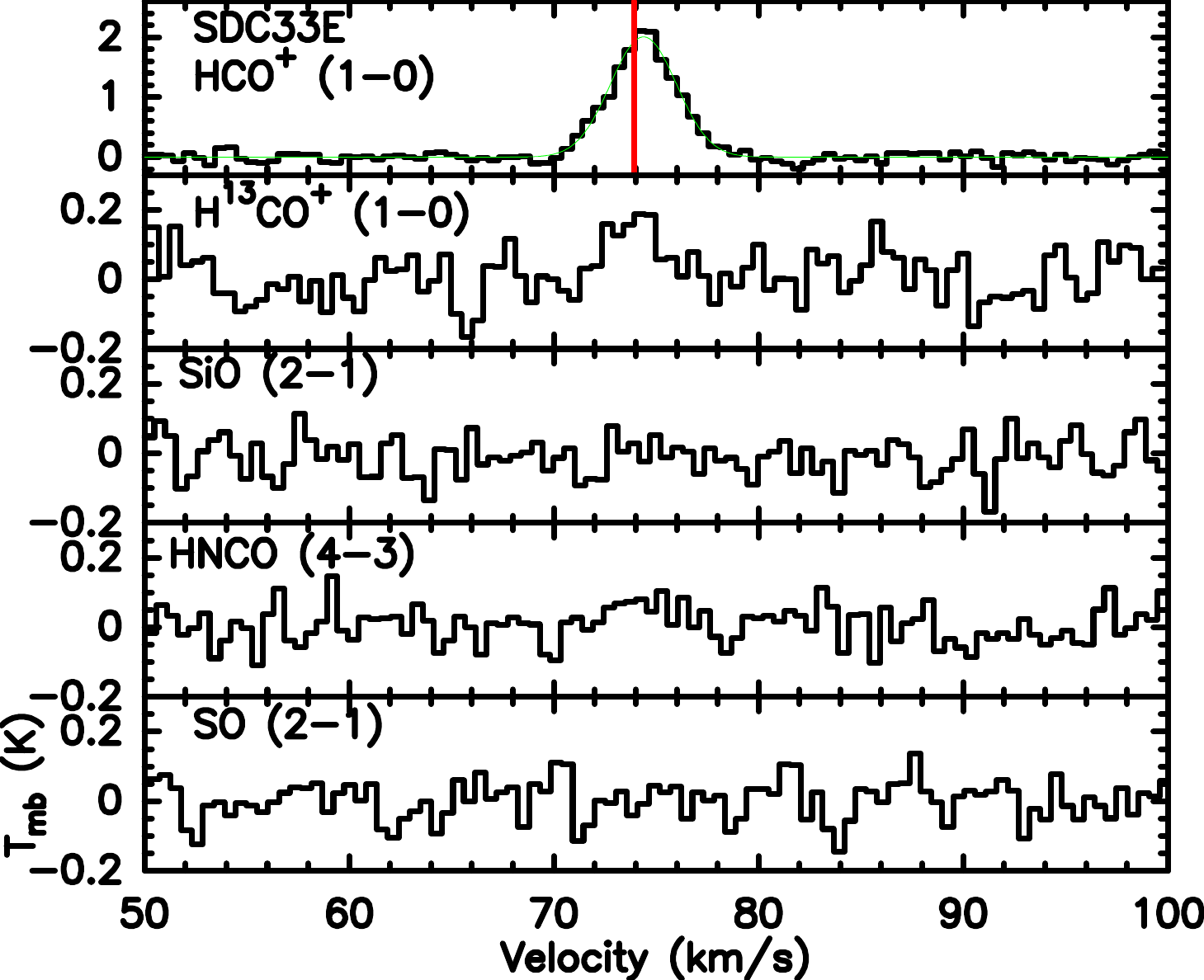}
\includegraphics[width=0.33\textwidth,height=0.43\textwidth,trim={0 0  0 0}, clip]{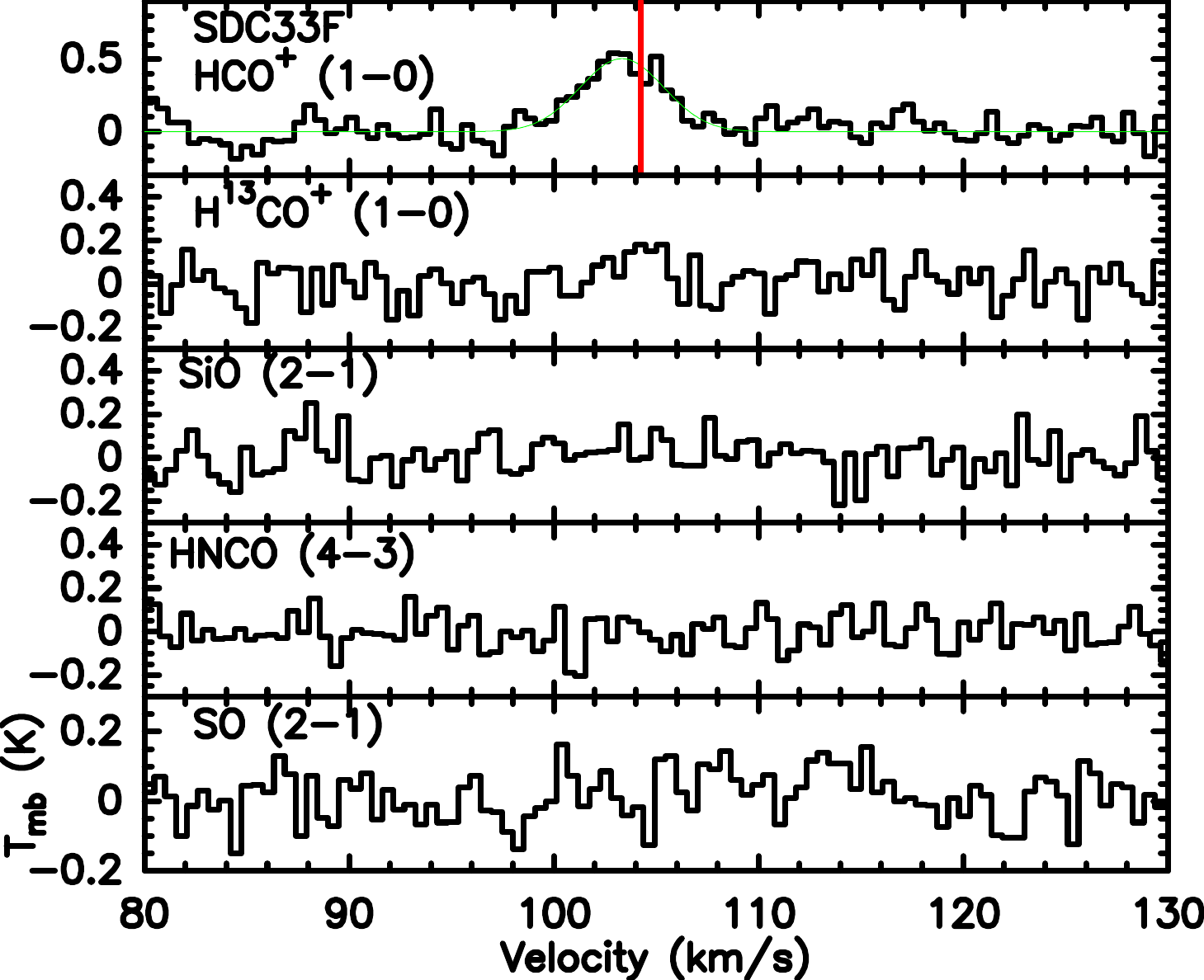}
\includegraphics[width=0.33\textwidth,height=0.43\textwidth,trim={0 0  0 0}, clip]{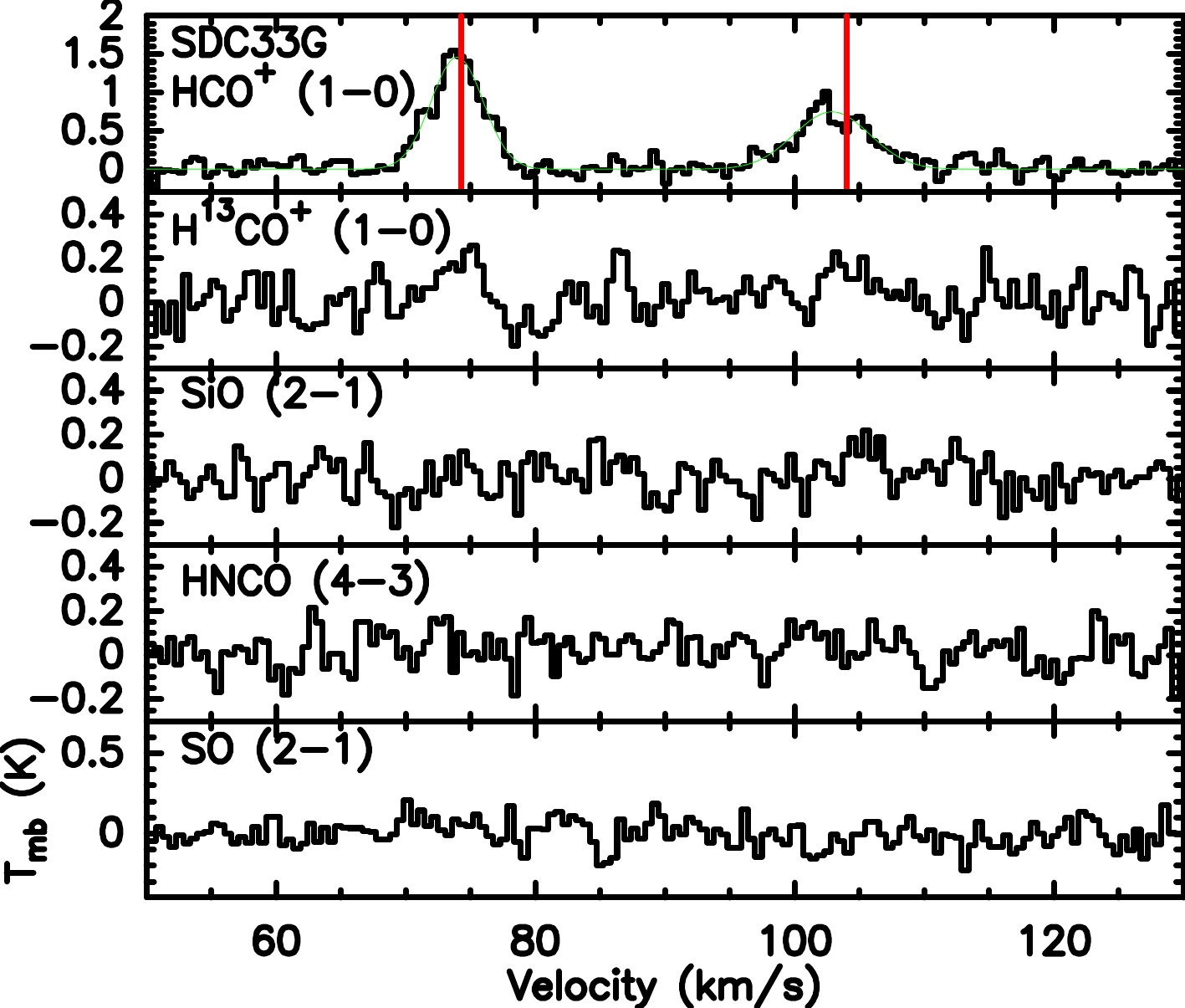}\\[3mm]

\includegraphics[width=0.33\textwidth,height=0.43\textwidth,trim={0 0  0 0}, clip]{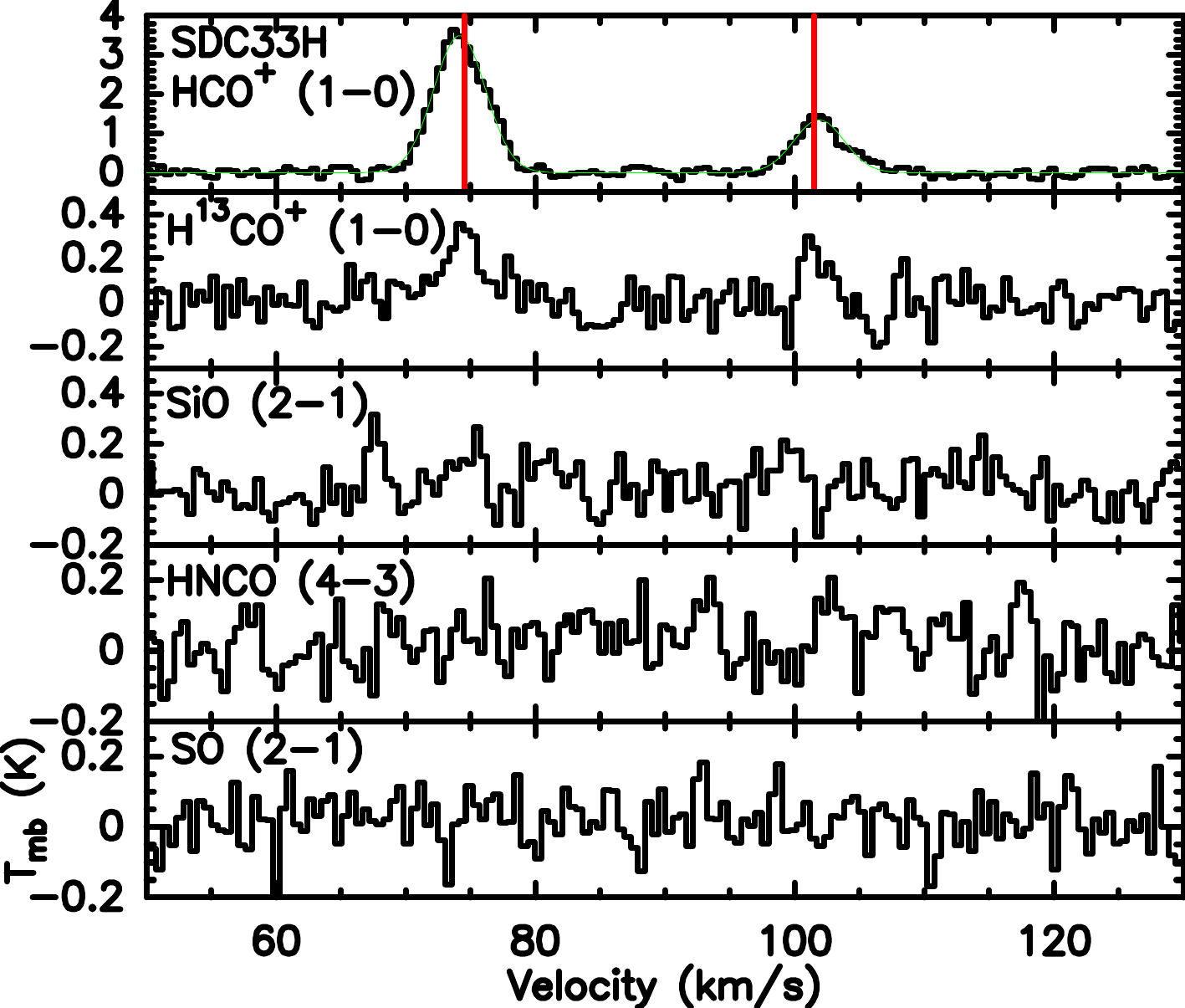}
\includegraphics[width=0.33\textwidth,height=0.43\textwidth,trim={0 0  0 0}, clip]{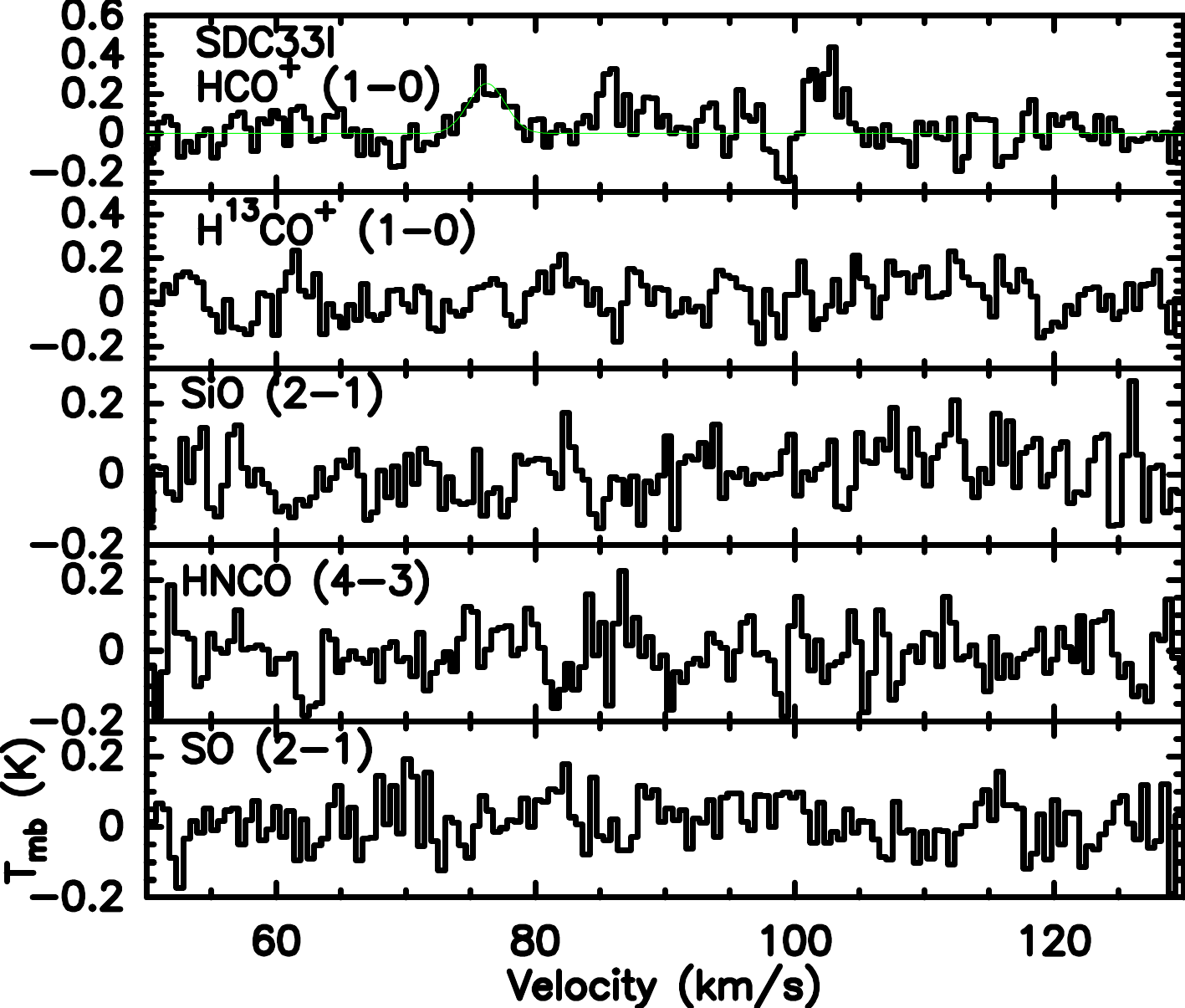}
\includegraphics[width=0.33\textwidth,height=0.43\textwidth,trim={0 0  0 0}, clip]{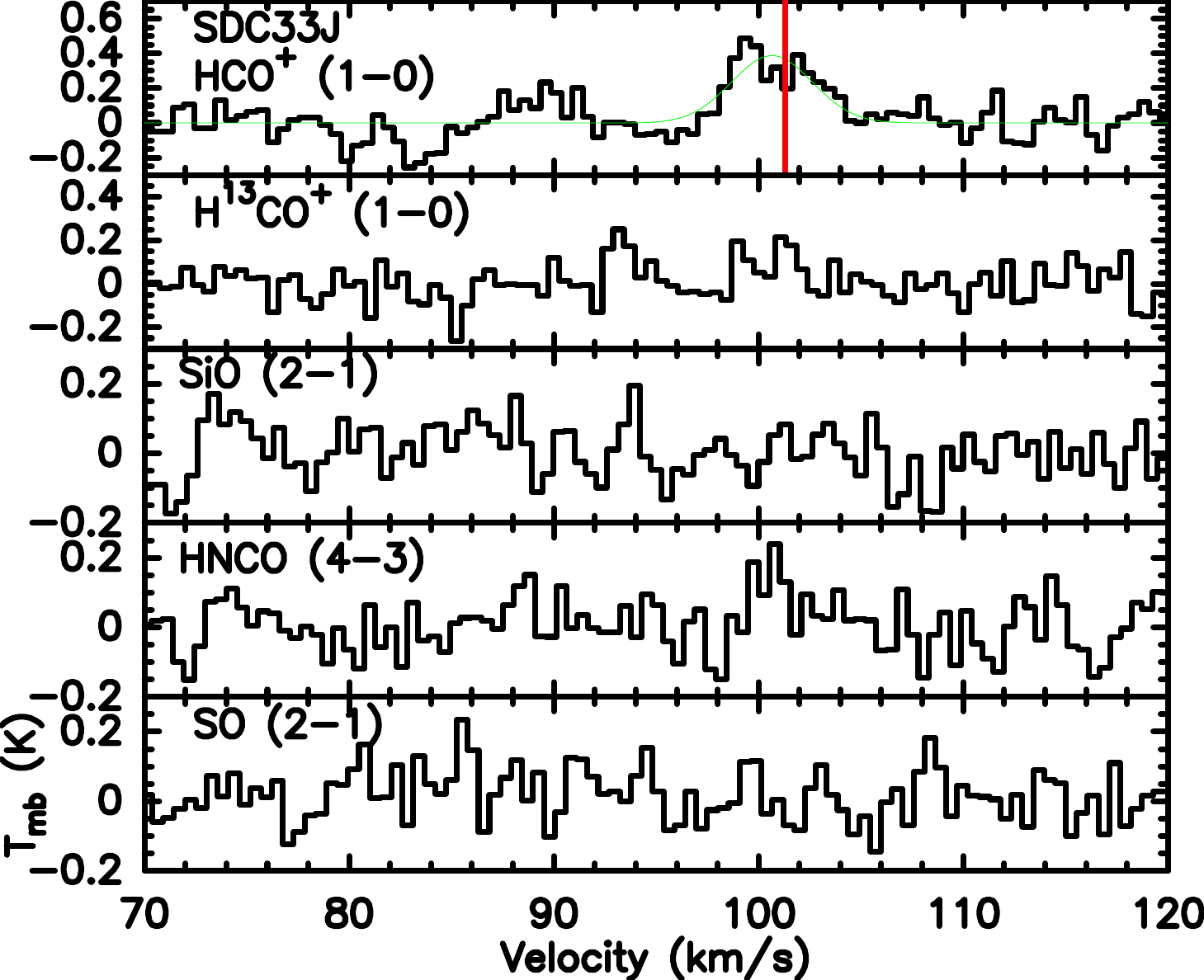}
  \caption[]{Continued.}
  \label{fig:kinematic_a_continued_1} 
\end{figure}

 \begin{figure}[H]
  \ContinuedFloat
  \captionsetup{labelsep=period}
  \centering
\includegraphics[width=0.33\textwidth,height=0.43\textwidth,trim={0 0  0 0}, clip]{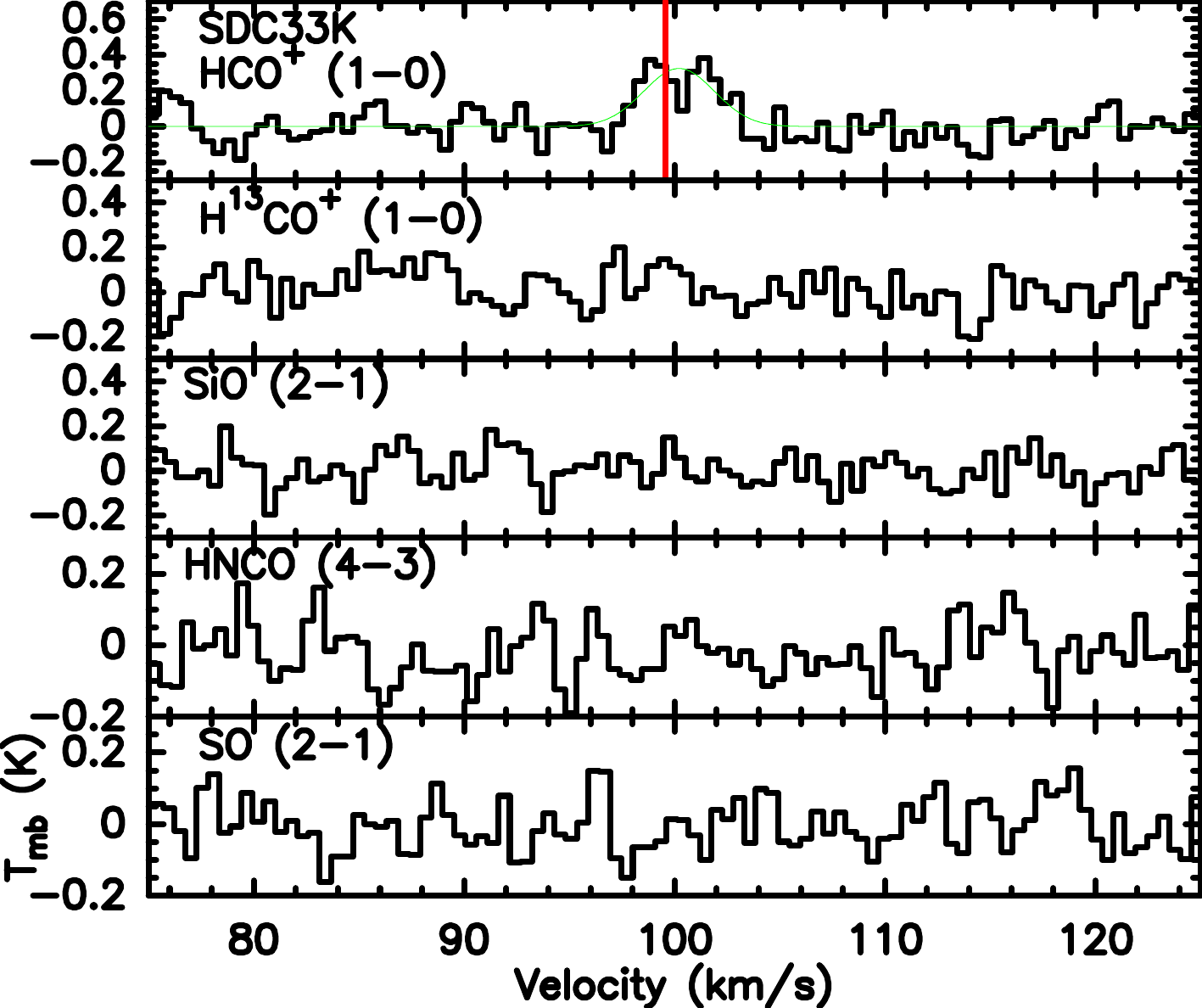}
\includegraphics[width=0.33\textwidth,height=0.43\textwidth,trim={0 0  0 0}, clip]{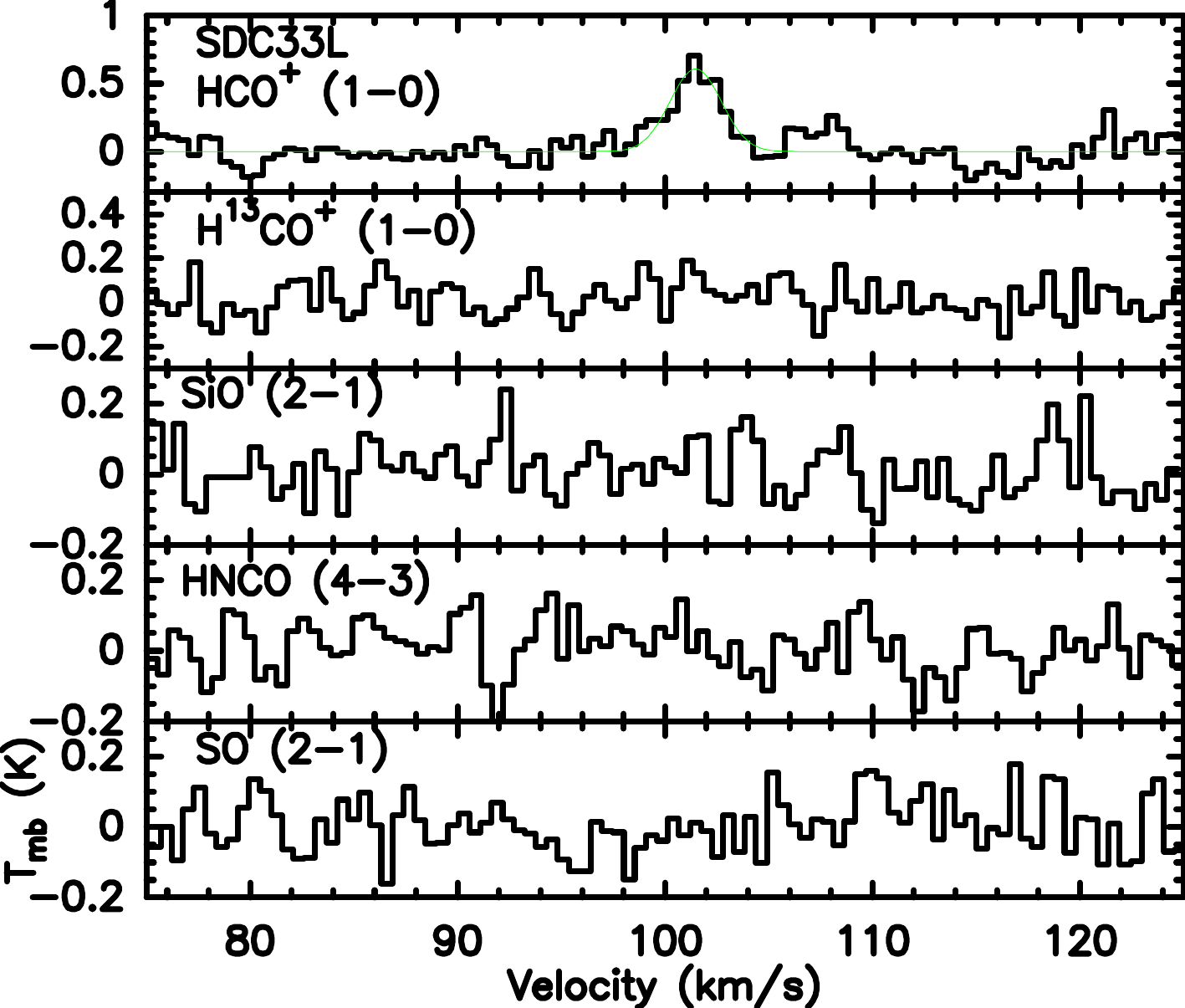}
\includegraphics[width=0.33\textwidth,height=0.43\textwidth,trim={0 0  0 0}, clip]{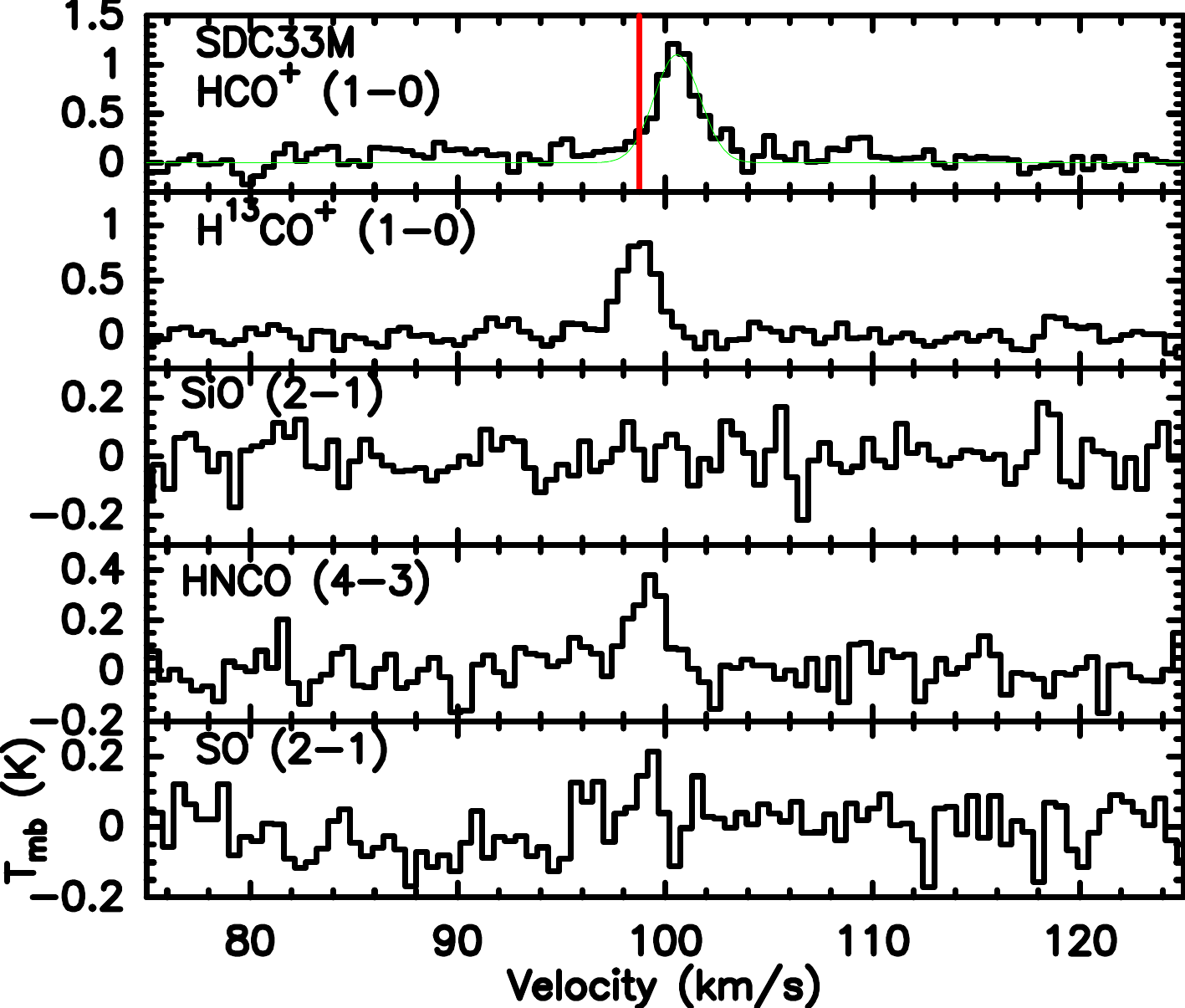}\\[3mm]

\includegraphics[width=0.33\textwidth,height=0.43\textwidth,trim={0 0  0 0}, clip]{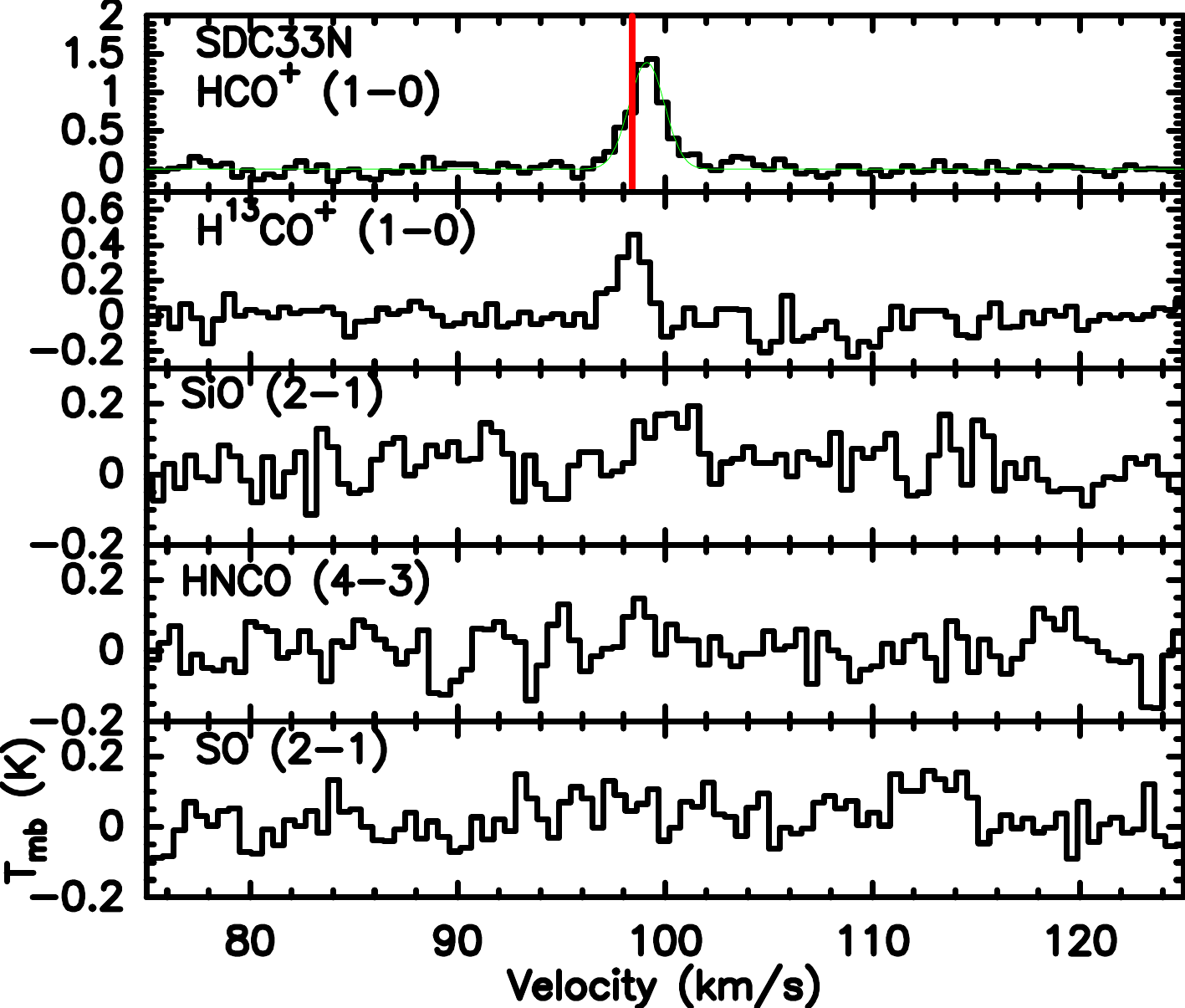}
\includegraphics[width=0.33\textwidth,height=0.43\textwidth,trim={0 0  0 0}, clip]{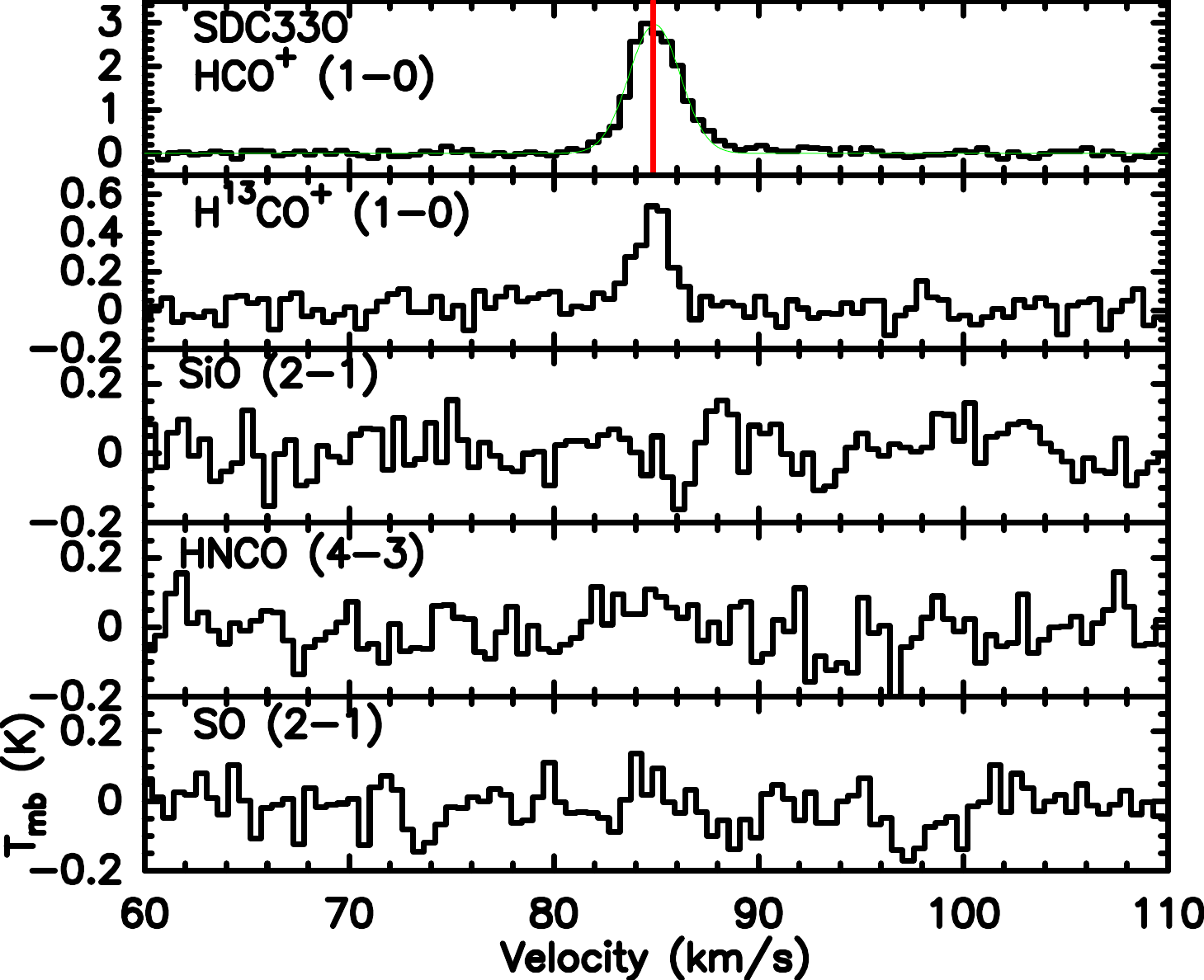}
\includegraphics[width=0.33\textwidth,height=0.43\textwidth,trim={0 0  0 0}, clip]{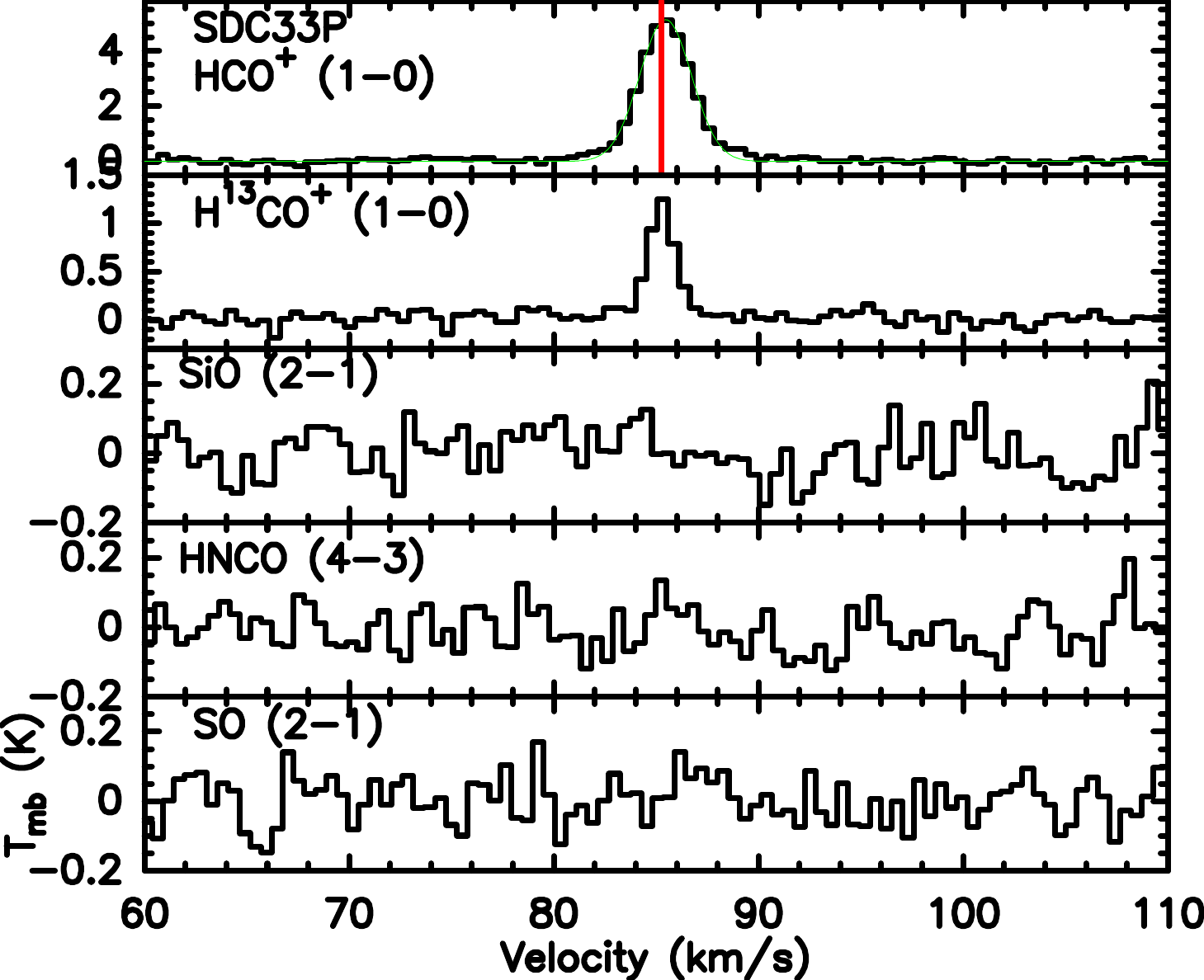}\\[3mm]

\includegraphics[width=0.33\textwidth,height=0.43\textwidth,trim={0 0  0 0}, clip]{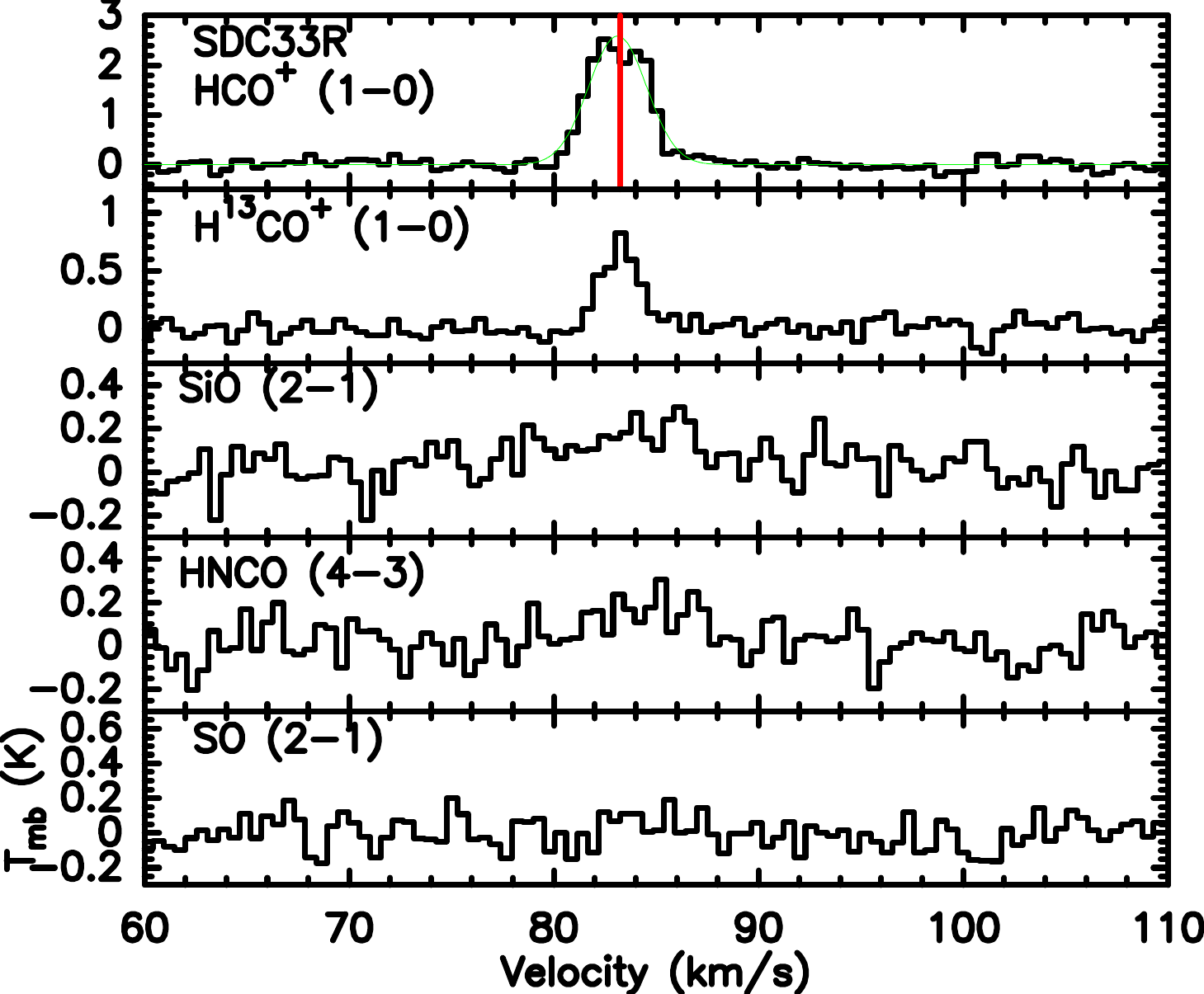}
\includegraphics[width=0.33\textwidth,height=0.43\textwidth,trim={0 0  0 0}, clip]{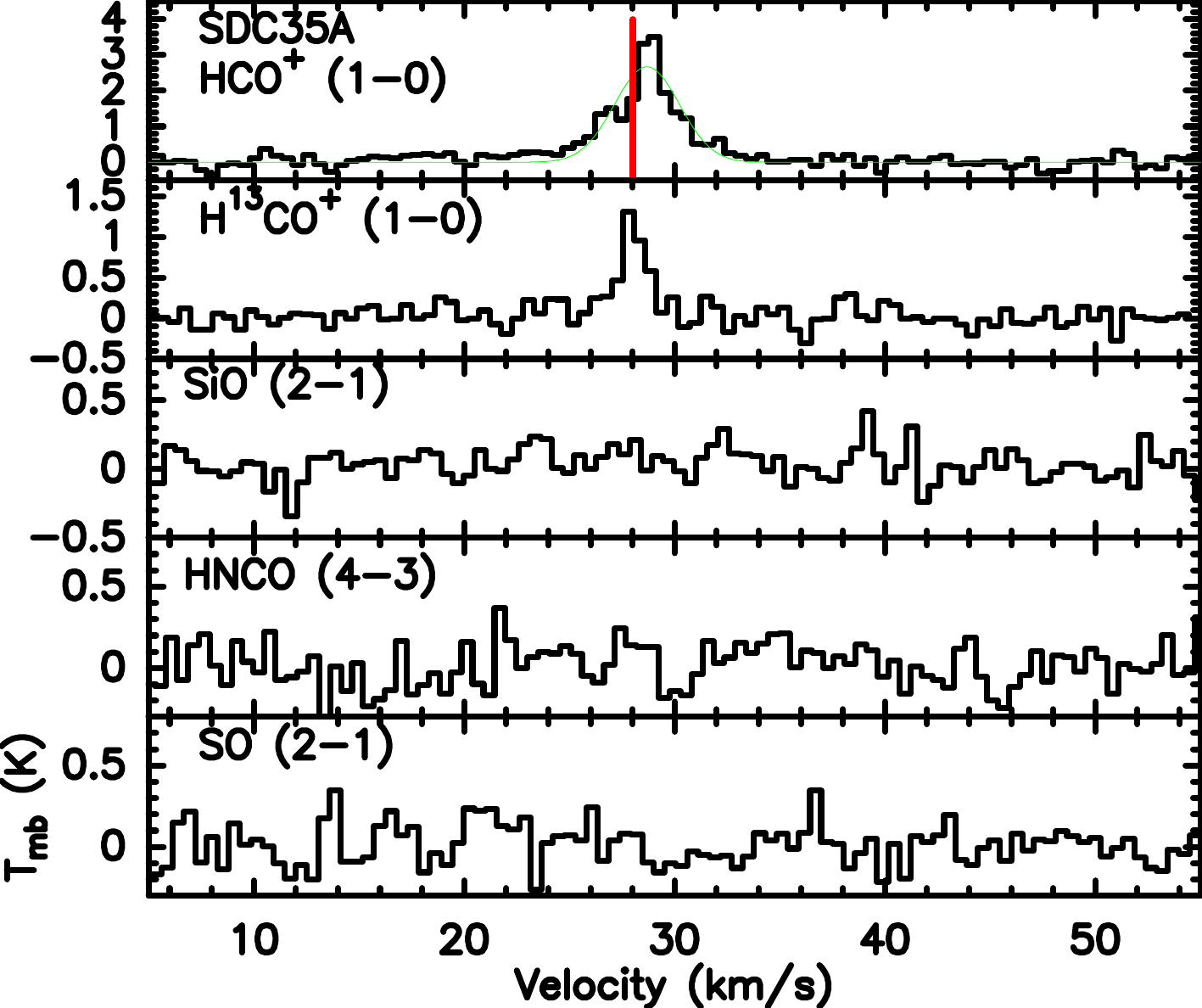}
\includegraphics[width=0.33\textwidth,height=0.43\textwidth,trim={0 0  0 0}, clip]{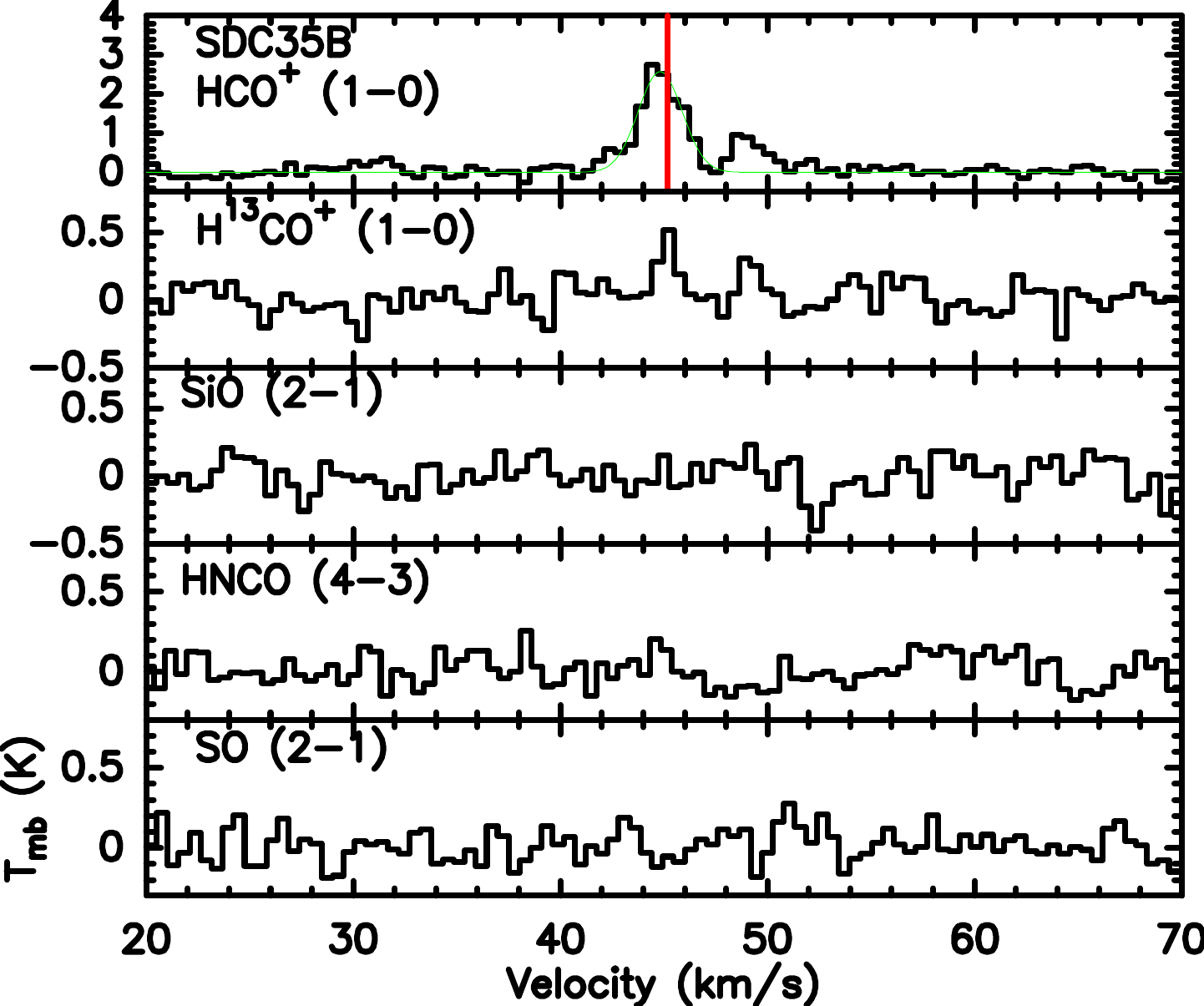}
  \caption[]{Continued.}
  \label{fig:kinematic_a_continued_2} 
\end{figure}

 \begin{figure}[H]
 \ContinuedFloat
 \captionsetup{labelsep=period}
  \centering 
\includegraphics[width=0.33\textwidth,height=0.43\textwidth,trim={0 0  0 0}, clip]{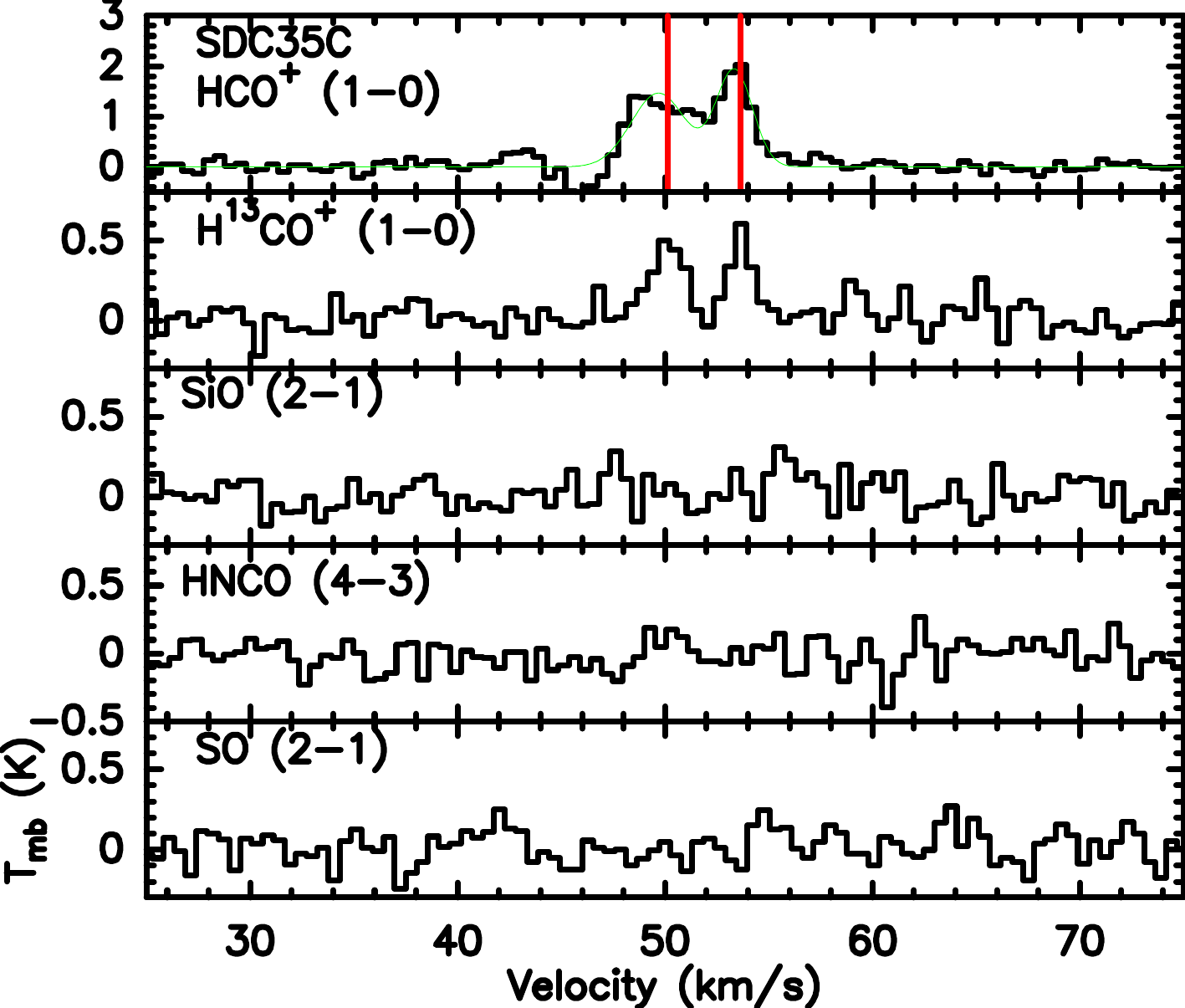}
\includegraphics[width=0.33\textwidth,height=0.43\textwidth,trim={0 0  0 0}, clip]{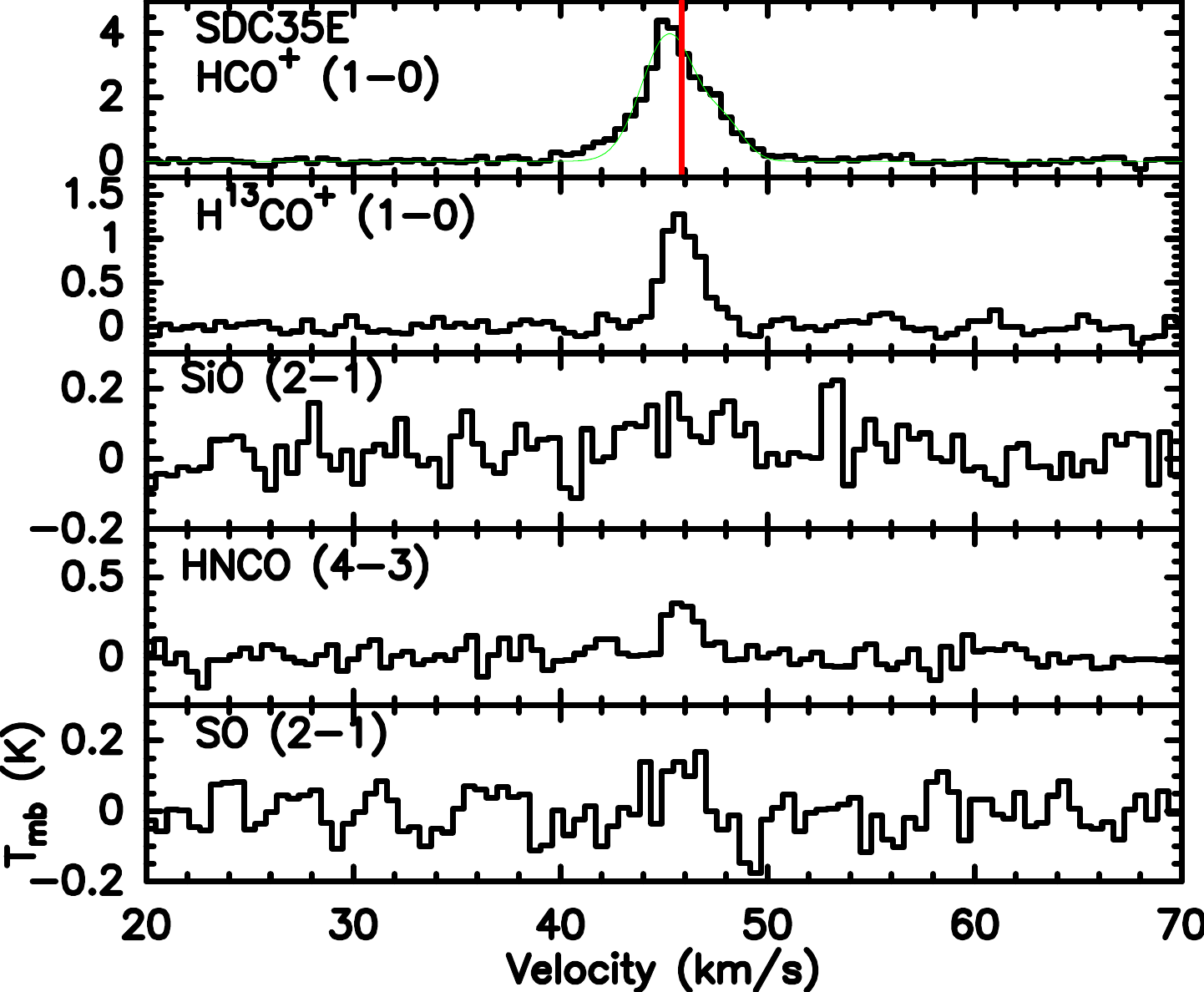}
\includegraphics[width=0.33\textwidth,height=0.43\textwidth,trim={0 0  0 0}, clip]{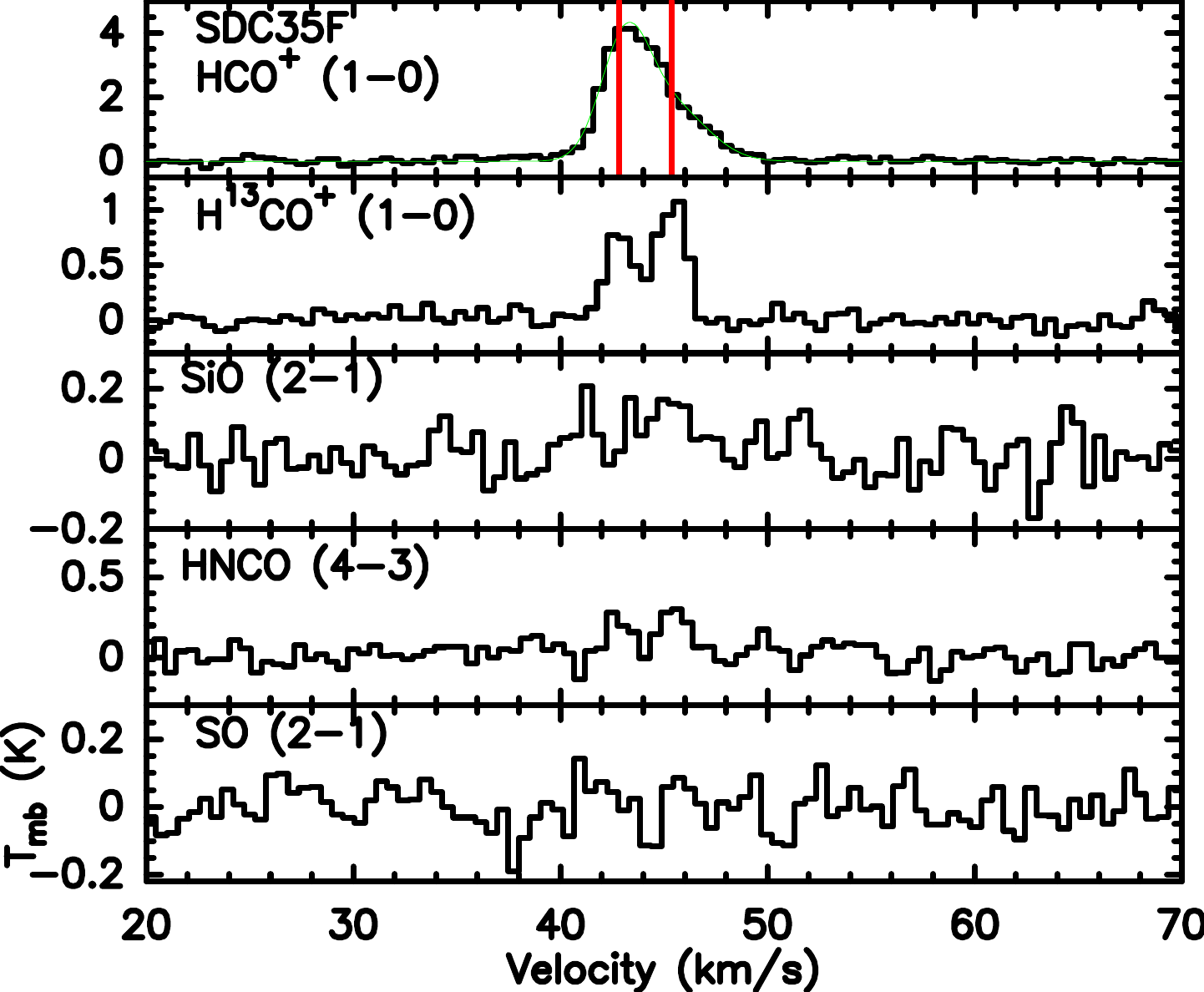}\\[3mm]

\includegraphics[width=0.33\textwidth,height=0.43\textwidth,trim={0 0  0 0}, clip]{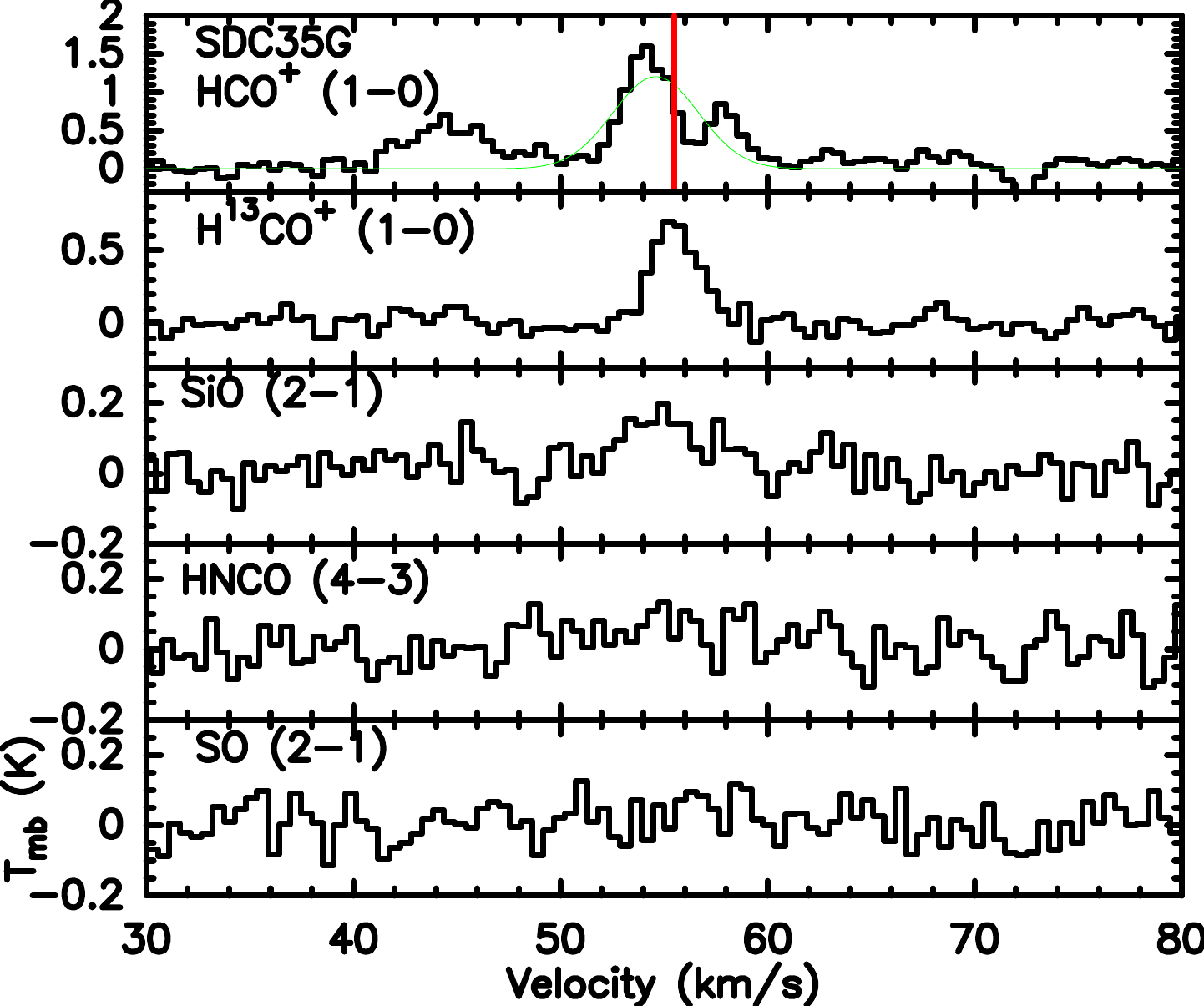}
\includegraphics[width=0.33\textwidth,height=0.43\textwidth,trim={0 0  0 0}, clip]{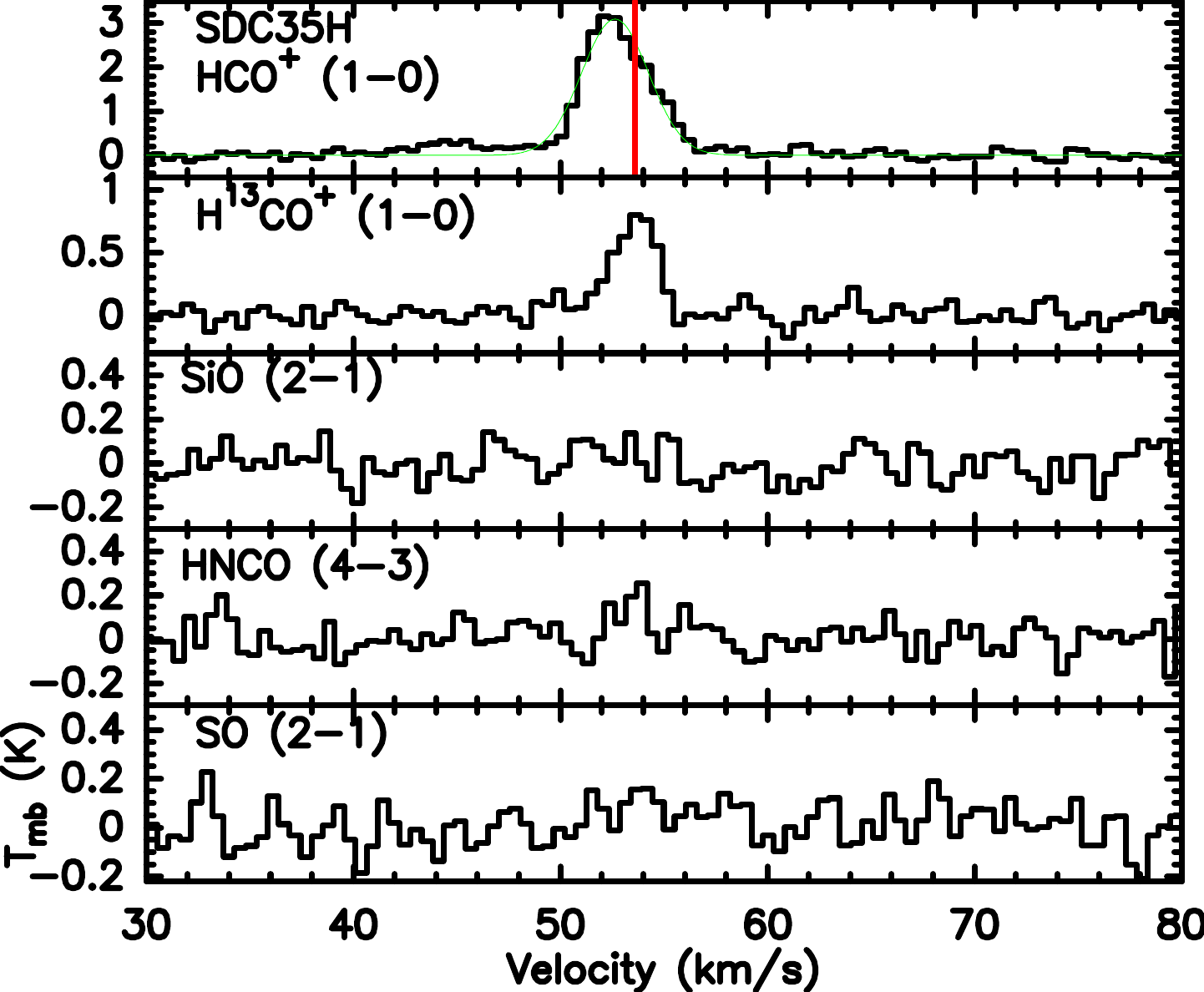}
\includegraphics[width=0.33\textwidth,height=0.43\textwidth,trim={0 0  0 0}, clip]{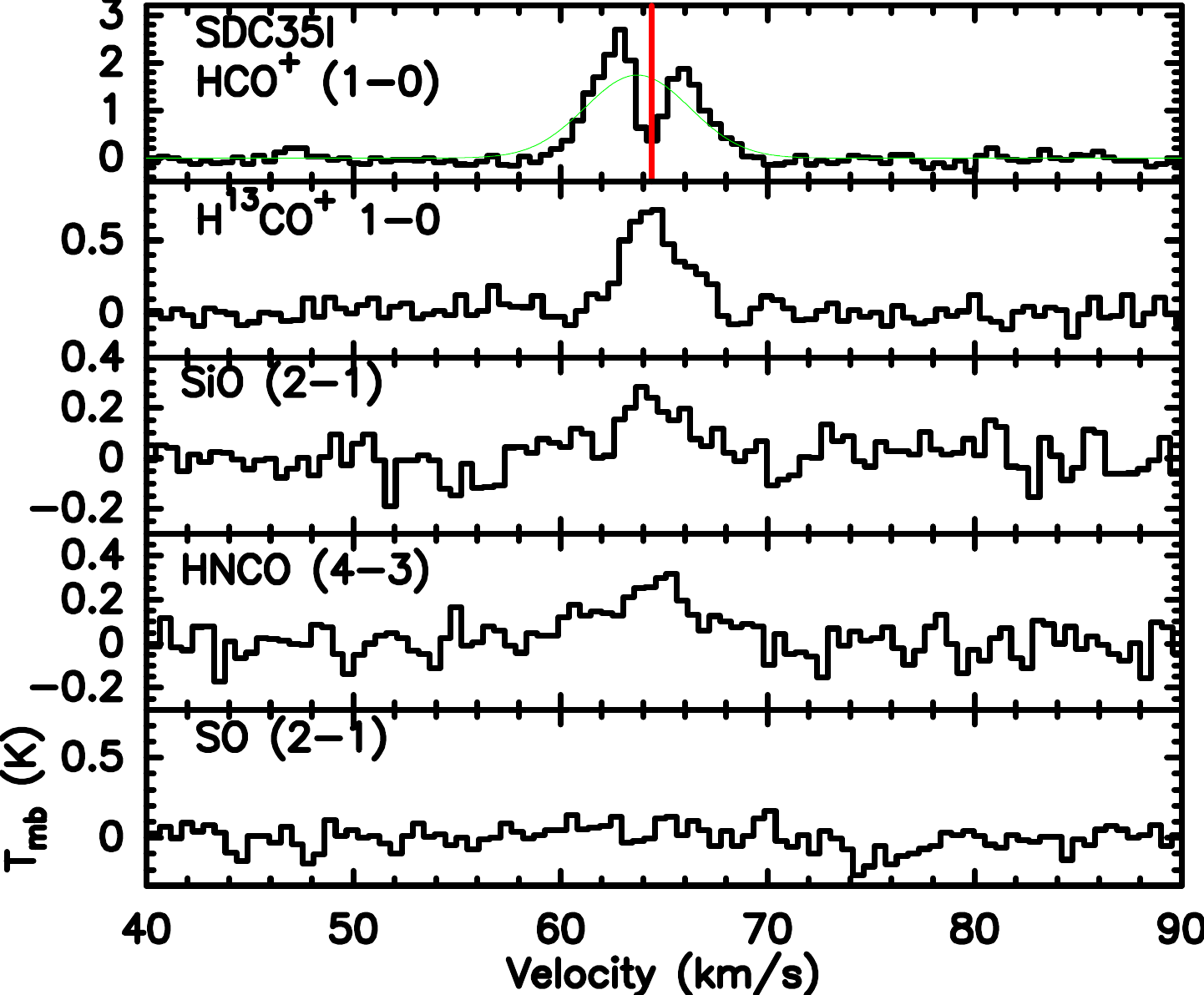}\\[3mm]

\includegraphics[width=0.33\textwidth,height=0.43\textwidth,trim={0 0  0 0}, clip]{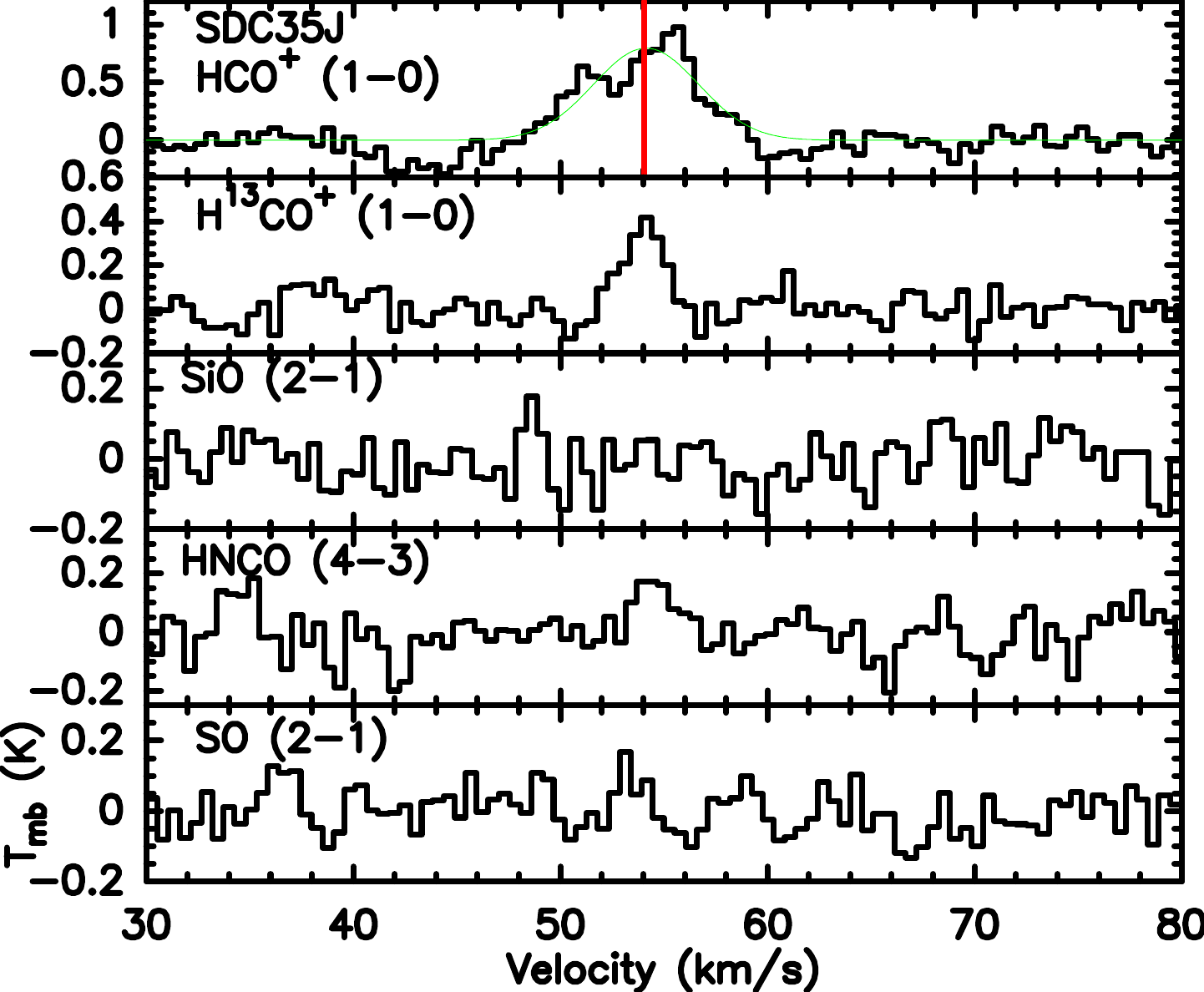}
\includegraphics[width=0.33\textwidth,height=0.43\textwidth,trim={0 0  0 0}, clip]{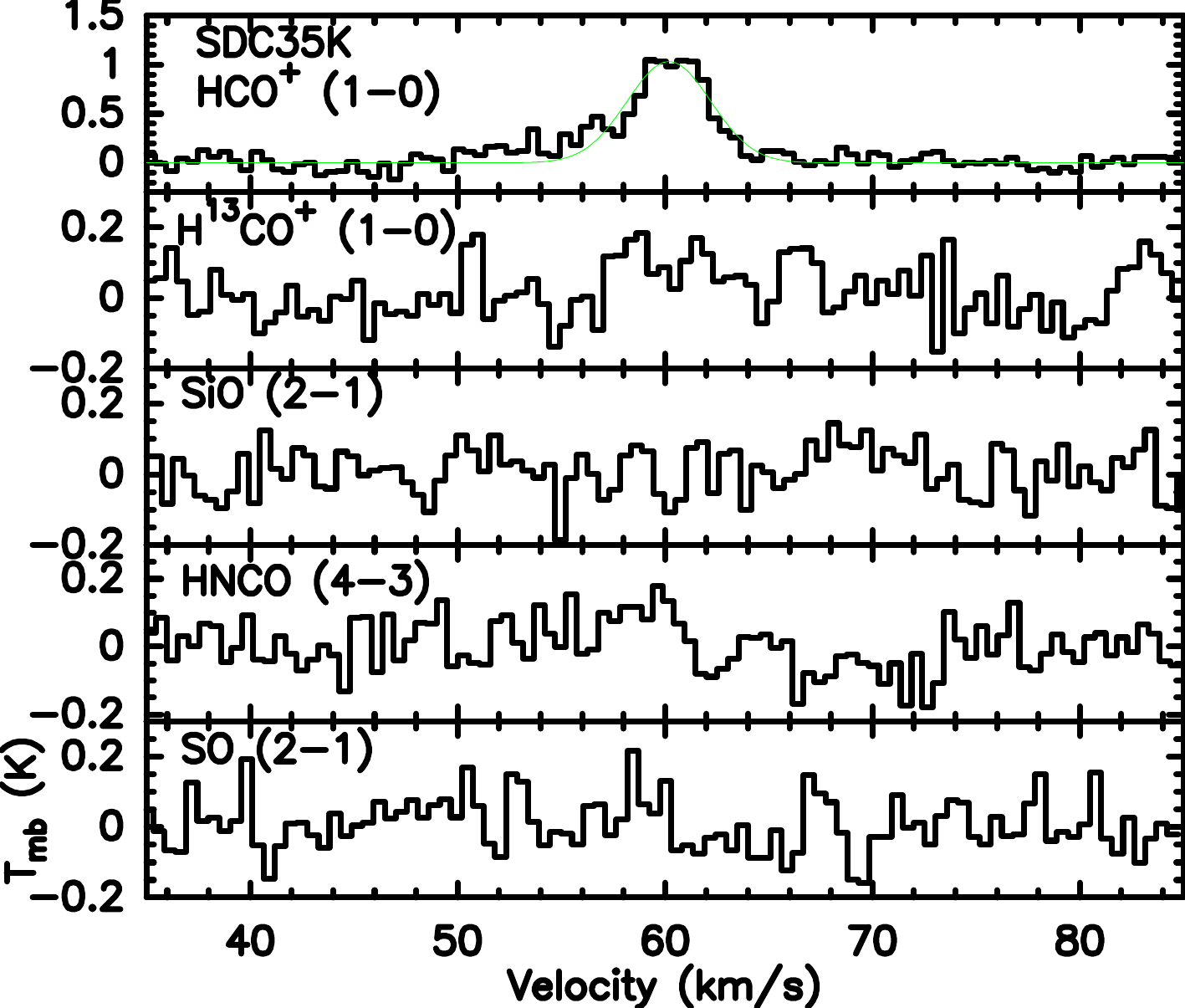}
\includegraphics[width=0.33\textwidth,height=0.43\textwidth,trim={0 0  0 0}, clip]{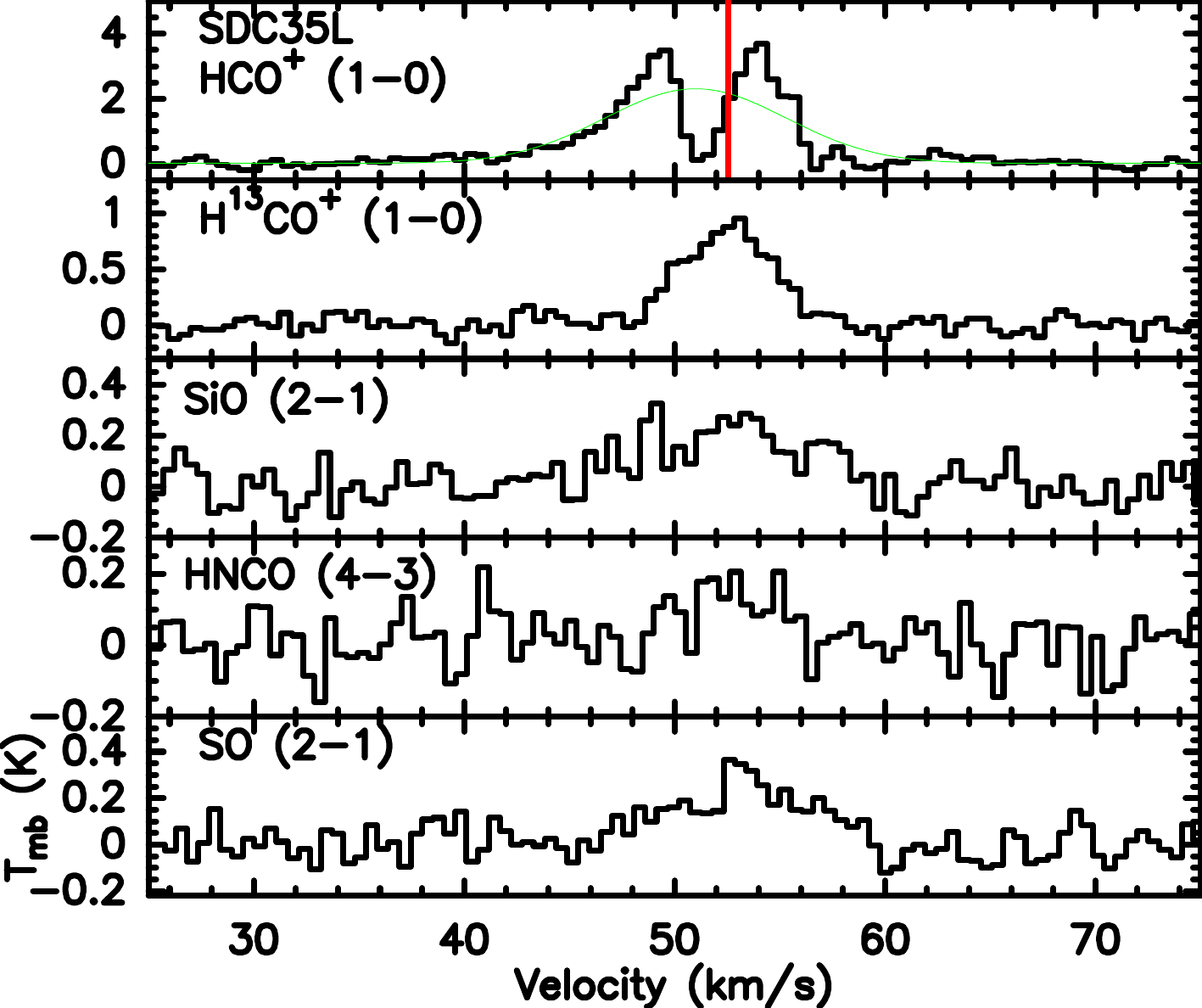}
  \caption[]{Continued.}
  \label{fig:kinematic_a_continued_3} 
\end{figure}

 \begin{figure}[H]
 \ContinuedFloat
 \captionsetup{labelsep=period}
  \centering
\includegraphics[width=0.33\textwidth,height=0.43\textwidth,trim={0 0  0 0}, clip]{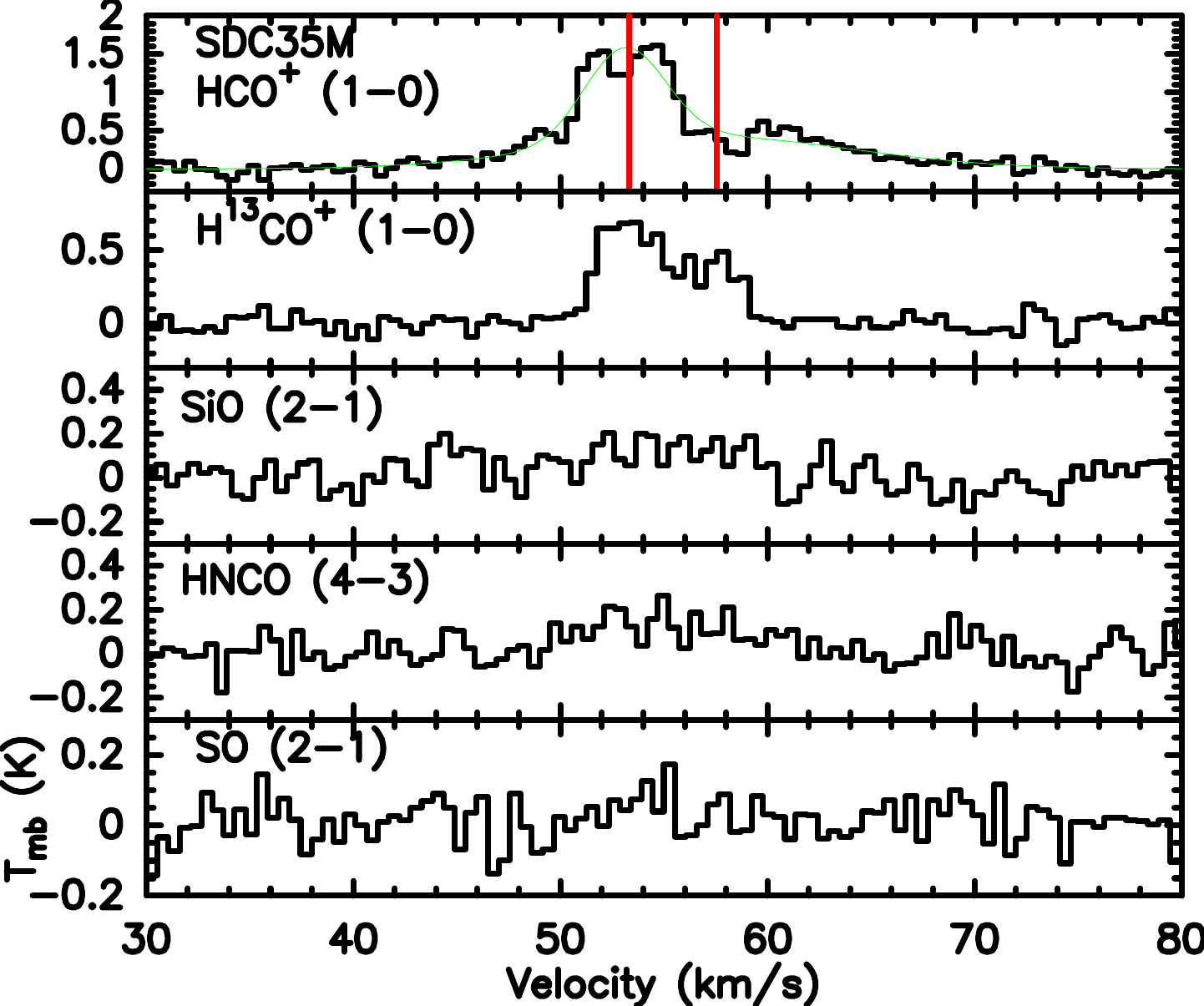}
\includegraphics[width=0.33\textwidth,height=0.43\textwidth,trim={0 0  0 0}, clip]{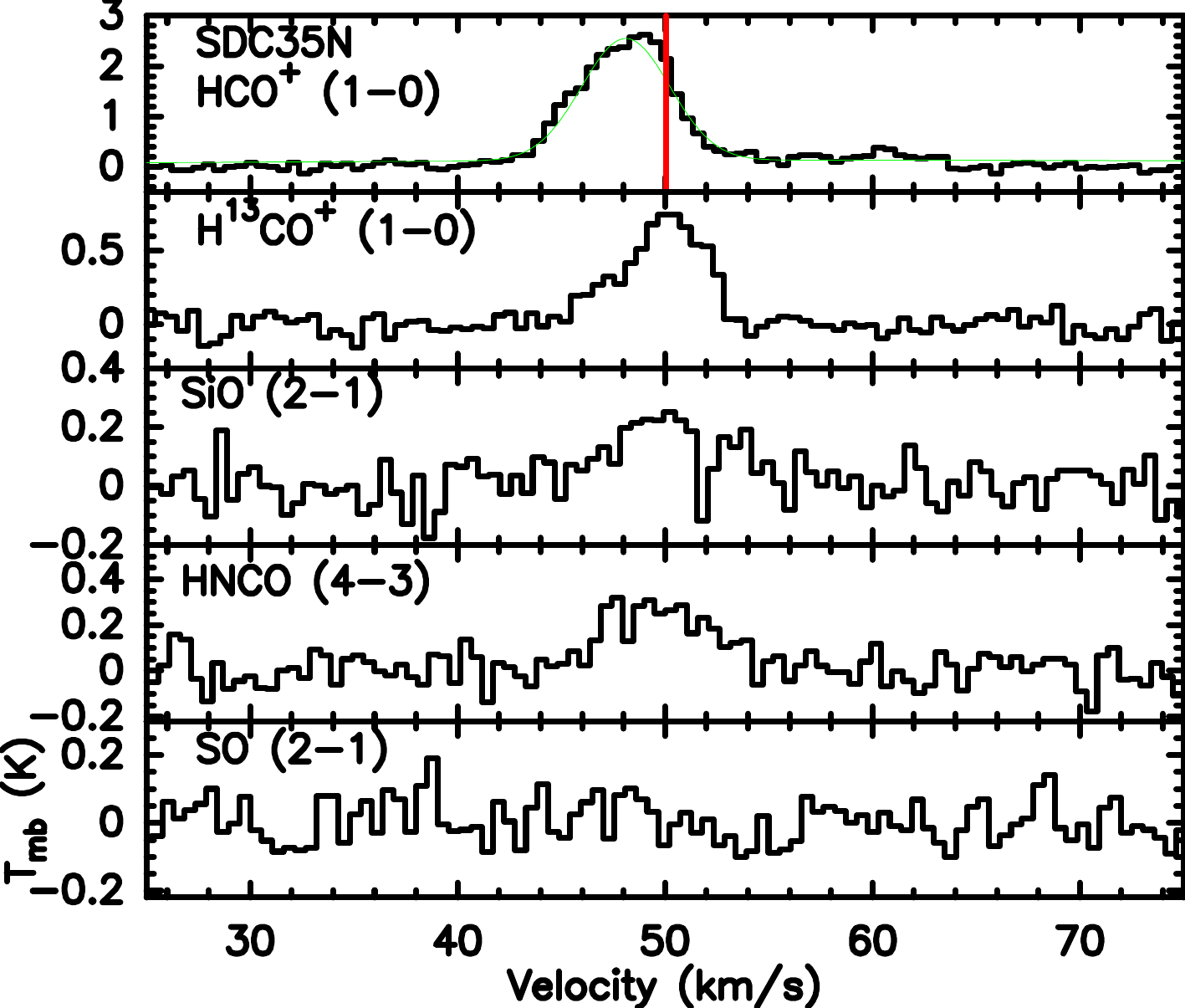}
\includegraphics[width=0.33\textwidth,height=0.43\textwidth,trim={0 0  0 0}, clip]{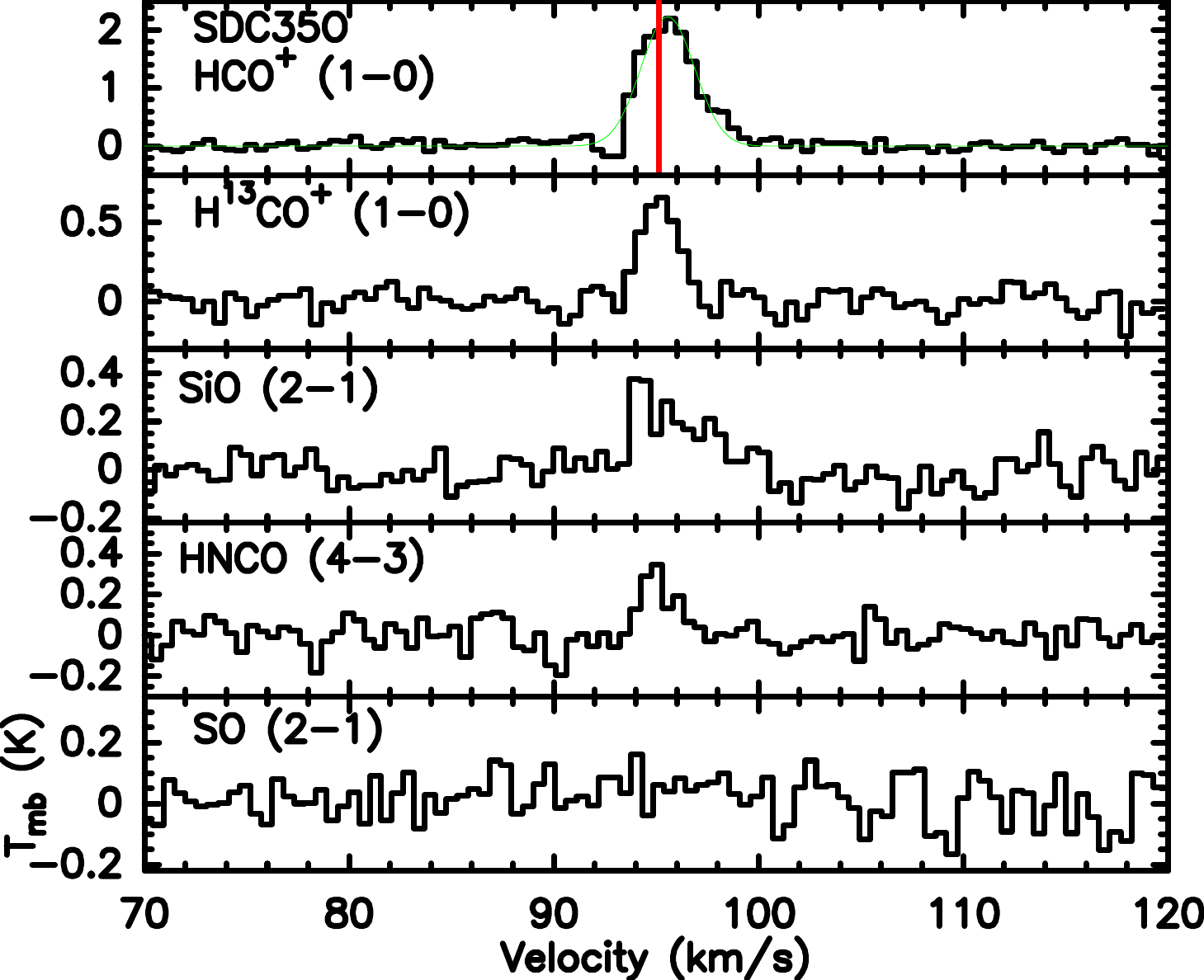}\\[3mm]

\includegraphics[width=0.33\textwidth,height=0.43\textwidth,trim={0 0  0 0}, clip]{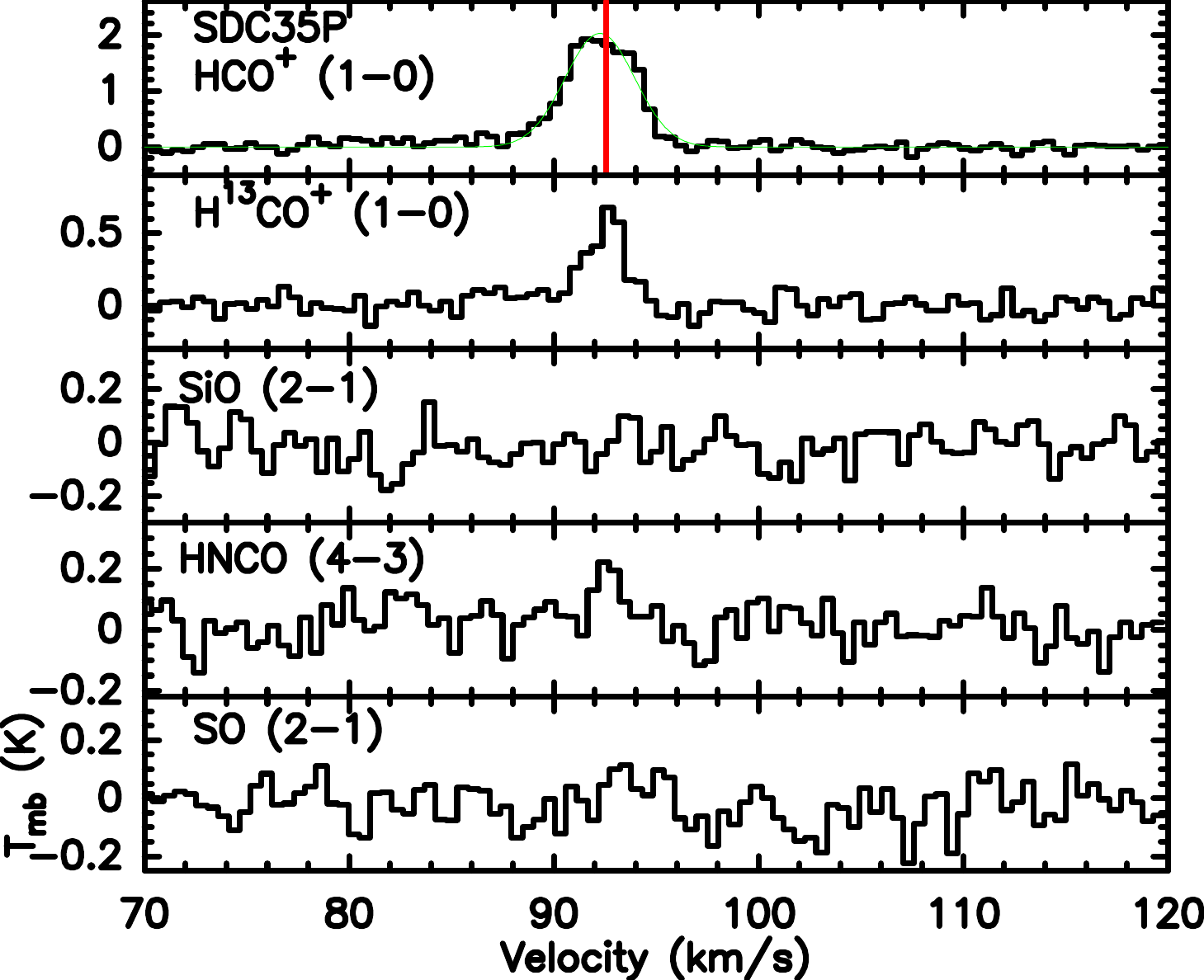}
\includegraphics[width=0.33\textwidth,height=0.43\textwidth,trim={0 0  0 0}, clip]{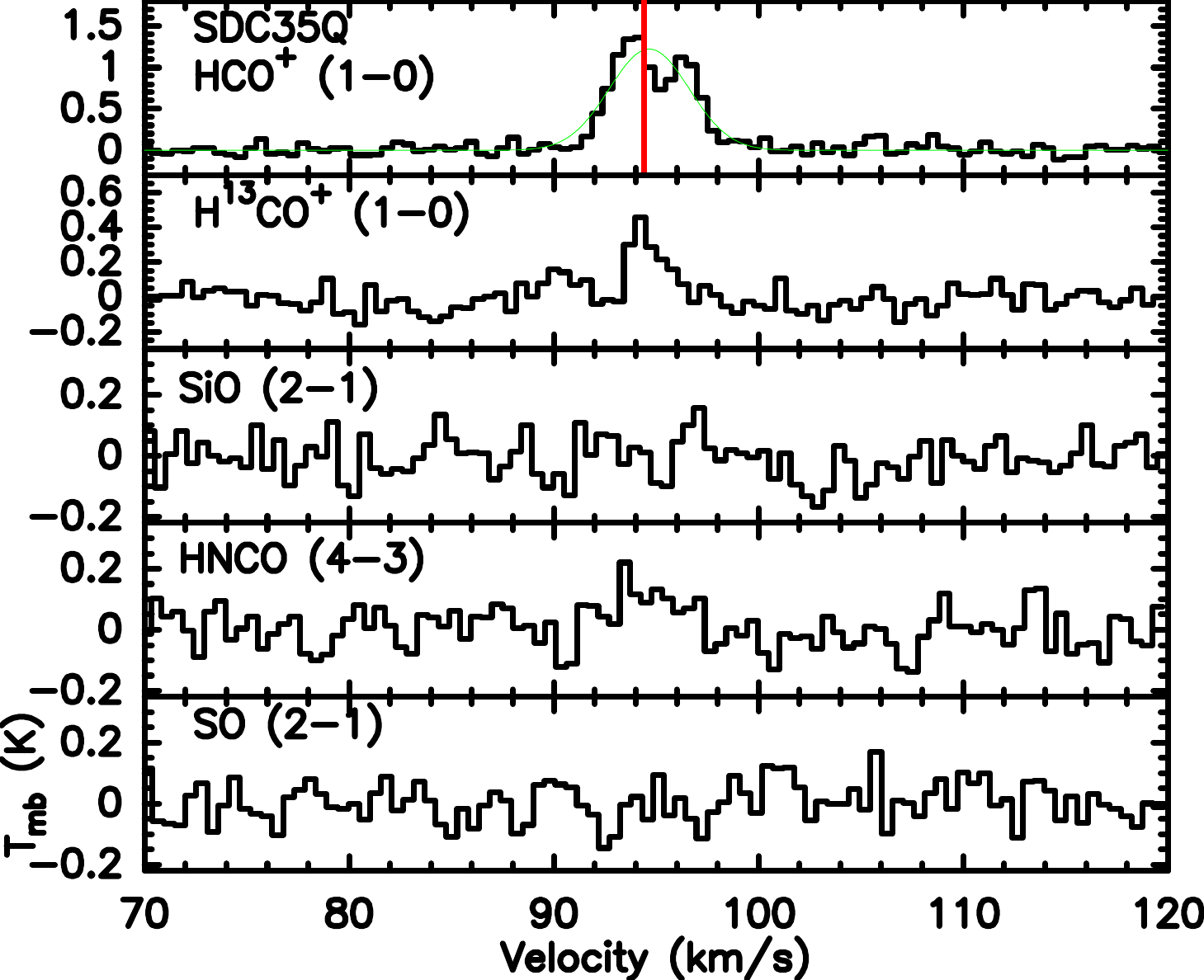}
\includegraphics[width=0.33\textwidth,height=0.43\textwidth,trim={0 0  0 0}, clip]{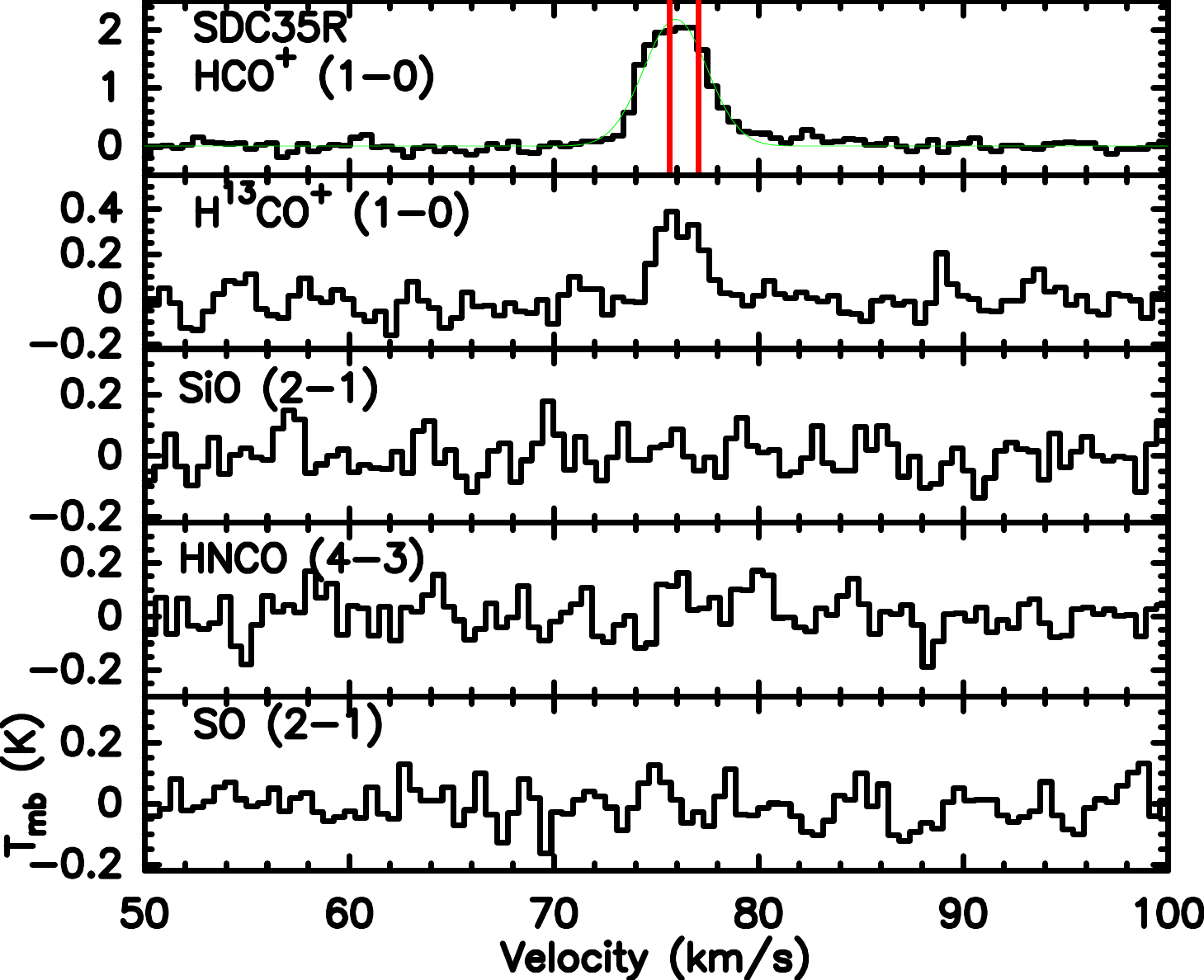}\\[3mm]

\includegraphics[width=0.33\textwidth,height=0.43\textwidth,trim={0 0  0 0}, clip]{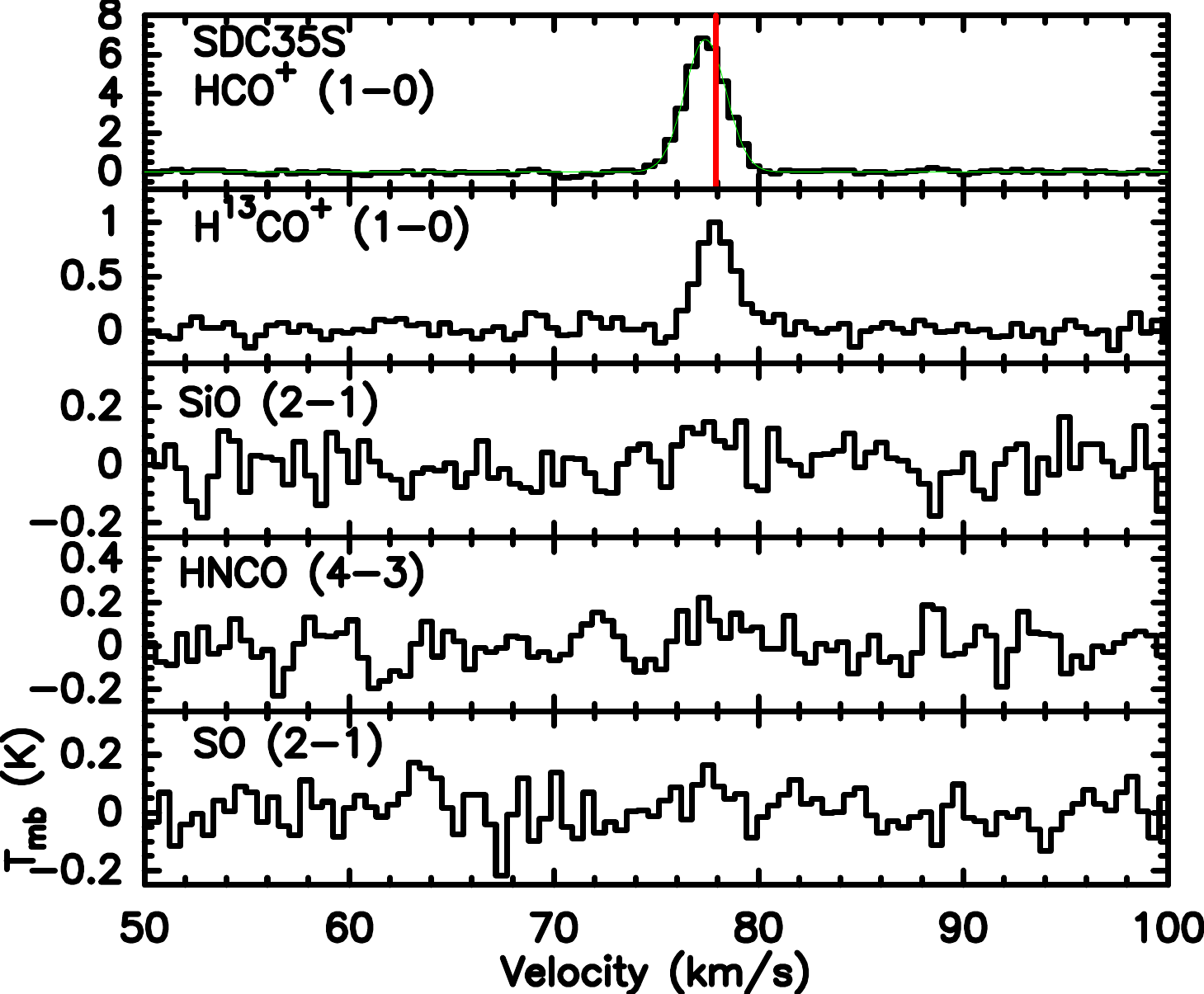}
\includegraphics[width=0.33\textwidth,height=0.43\textwidth,trim={0 0  0 0}, clip]{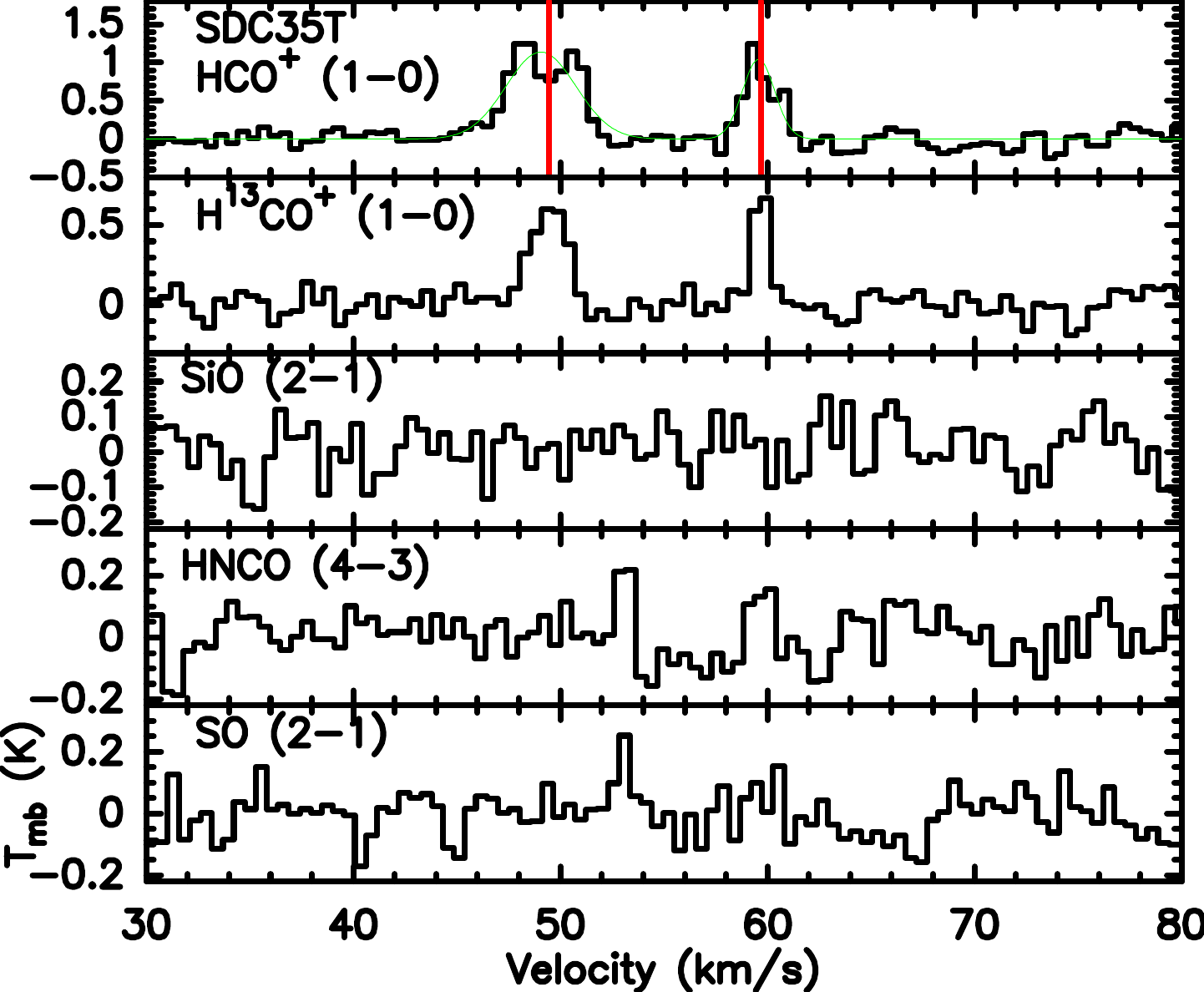}
\includegraphics[width=0.33\textwidth,height=0.43\textwidth,trim={0 0  0 0}, clip]{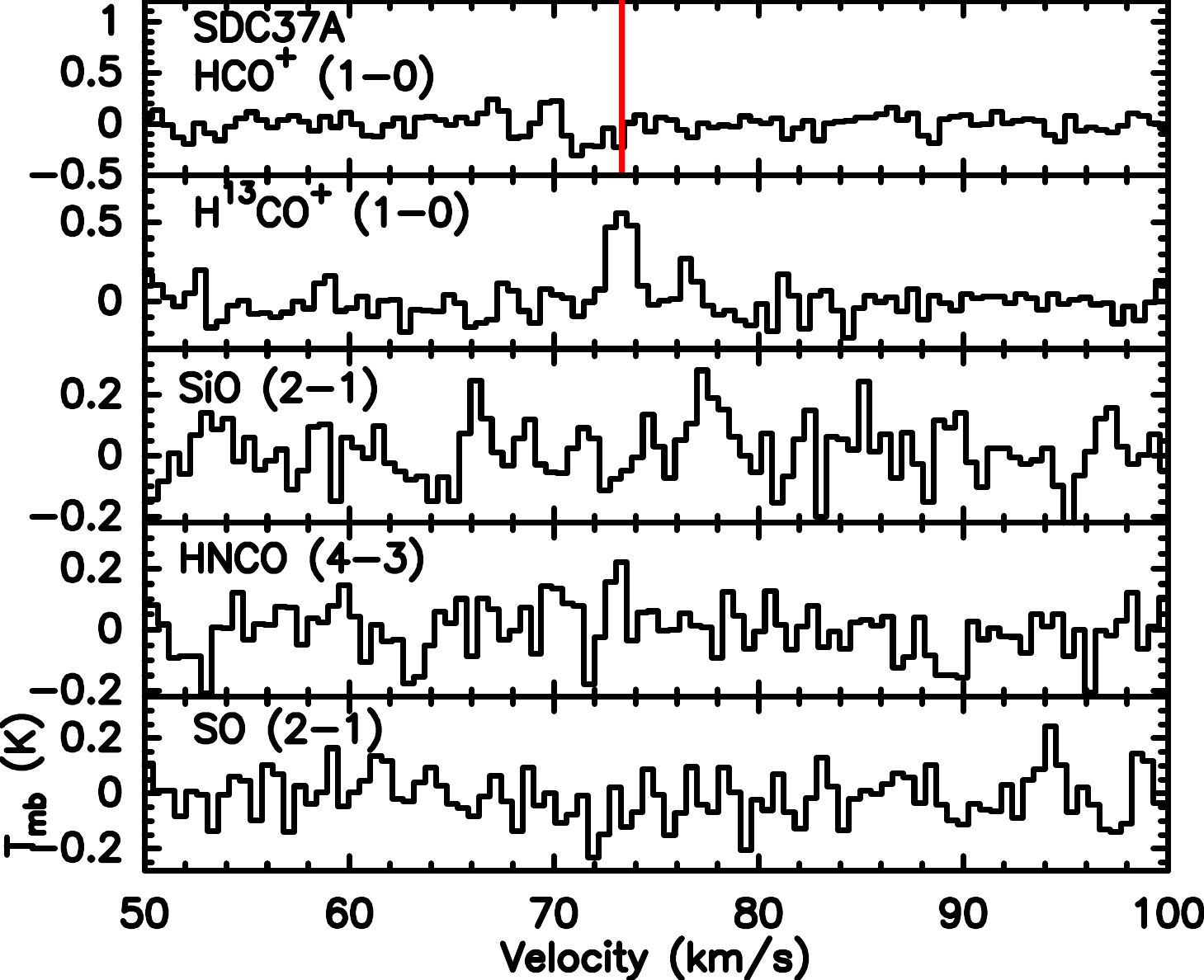}
  \caption[]{Continued.}
  \label{fig:kinematic_a_continued_4} 
\end{figure}

\begin{figure}[H]
\ContinuedFloat
\captionsetup{labelsep=period}
\centering 
\includegraphics[width=0.33\textwidth,height=0.43\textwidth,trim={0 0 0 0},clip]{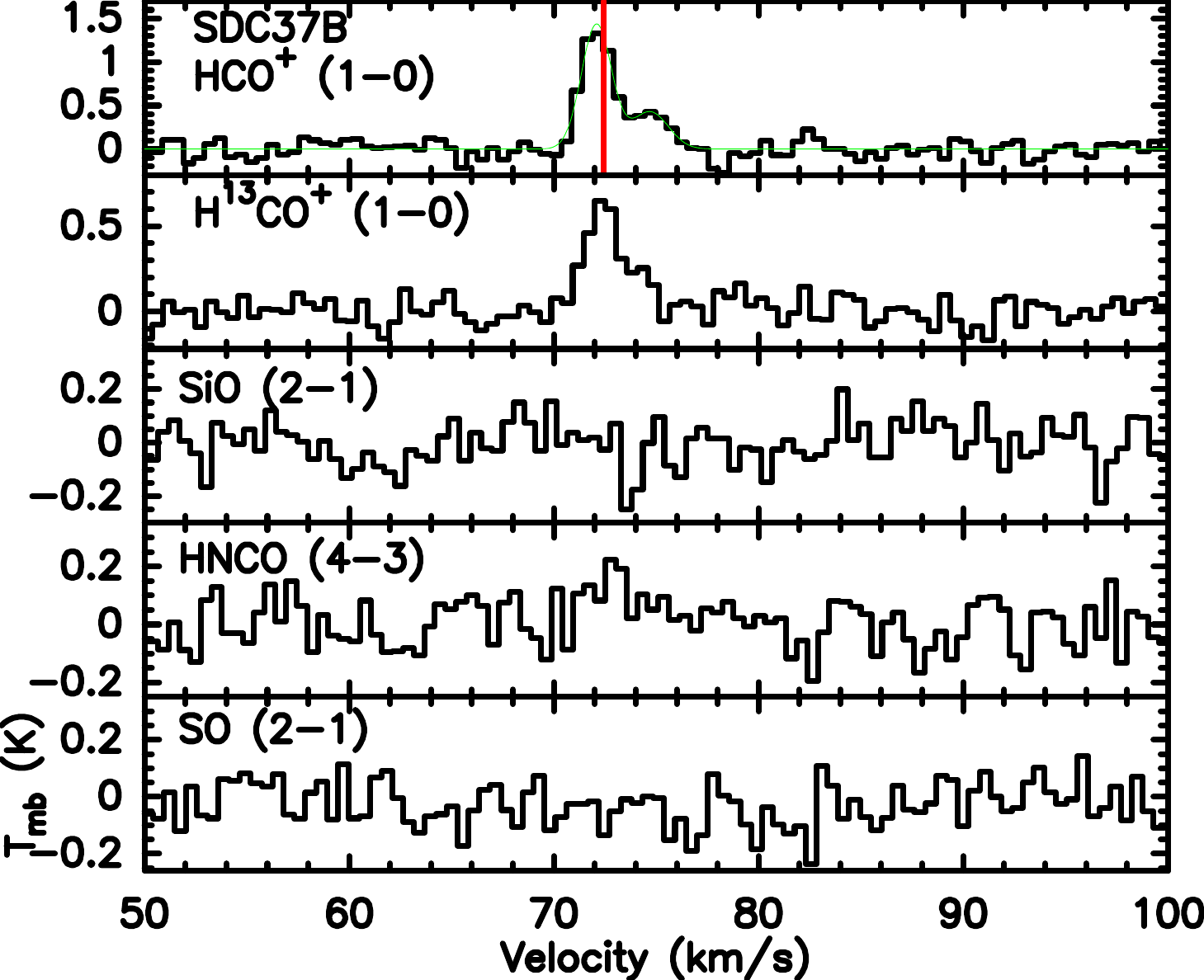}
\includegraphics[width=0.33\textwidth,height=0.43\textwidth,trim={0 0 0 0},clip]{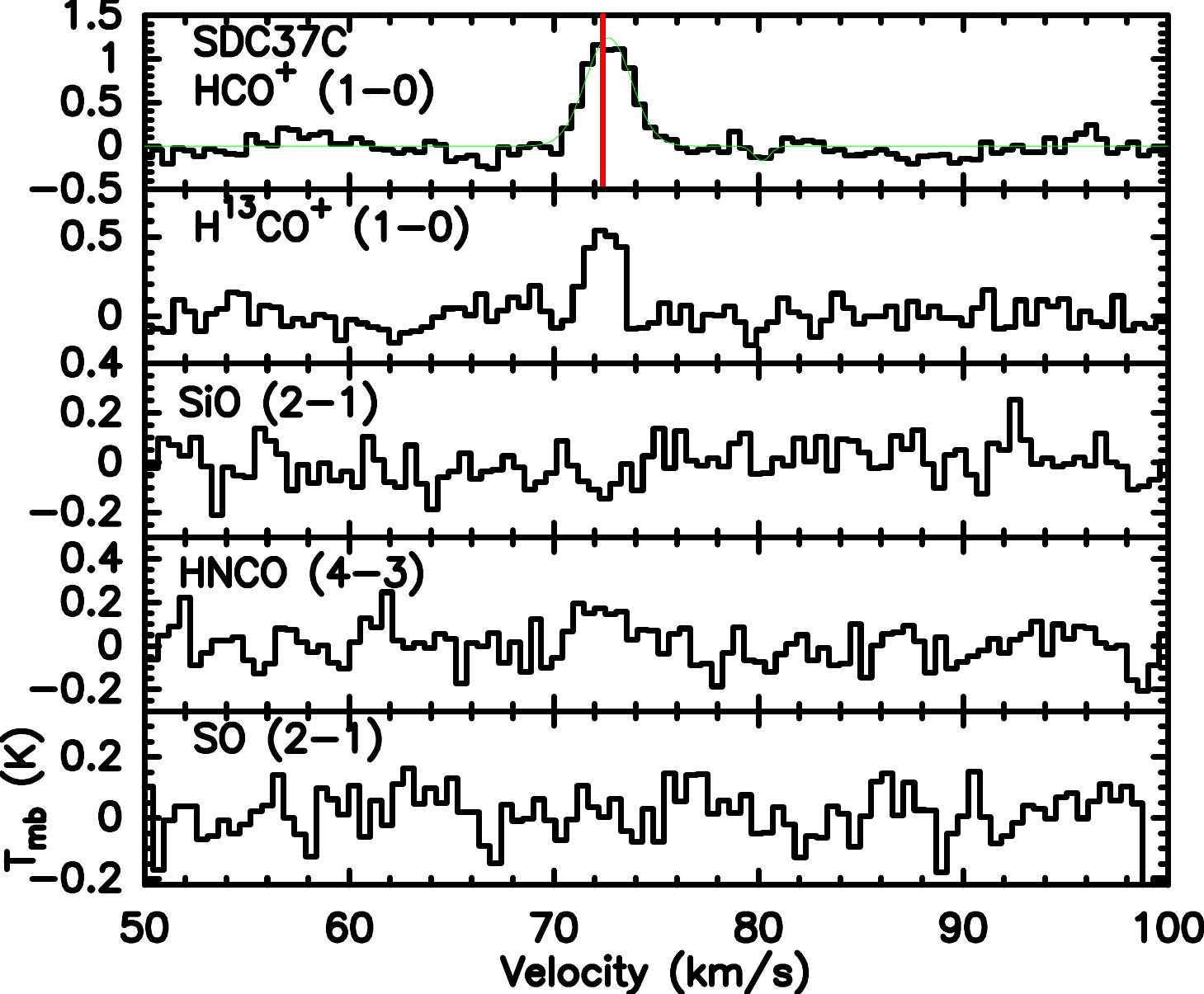}
\includegraphics[width=0.33\textwidth,height=0.43\textwidth,trim={0 0 0 0},clip]{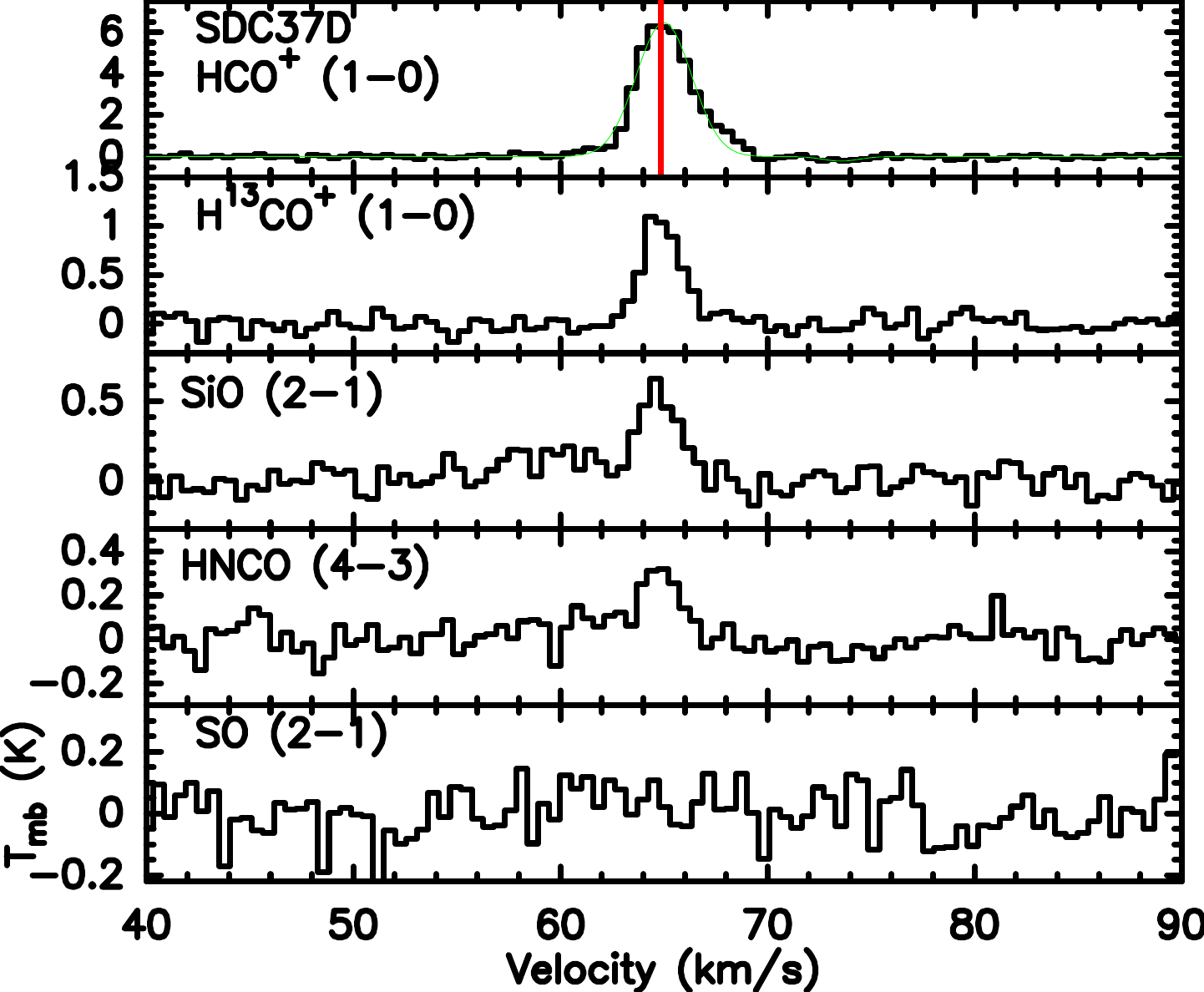}\\[3mm]

\includegraphics[width=0.33\textwidth,height=0.43\textwidth,trim={0 0 0 0},clip]{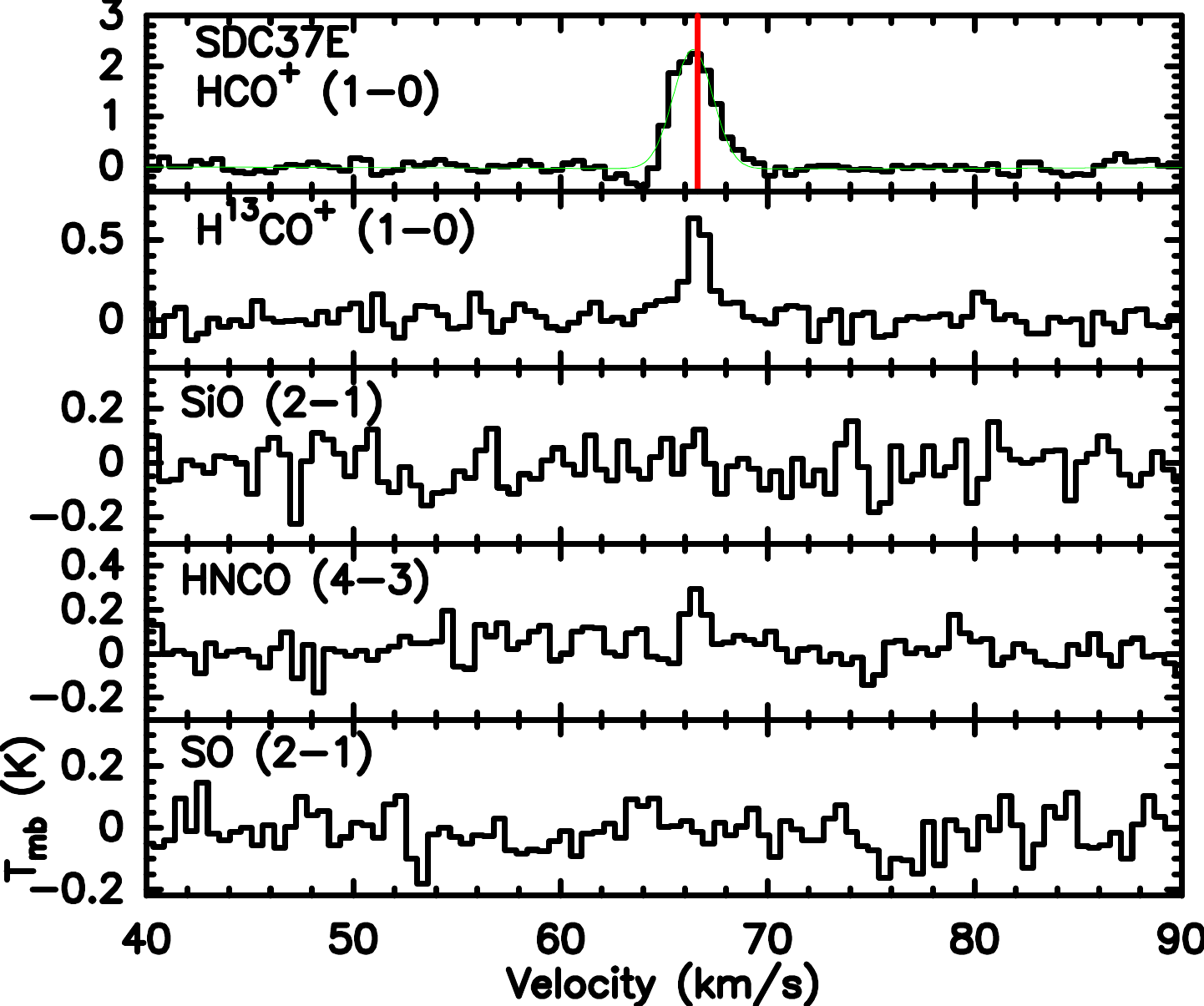}
\includegraphics[width=0.33\textwidth,height=0.43\textwidth,trim={0 0 0 0},clip]{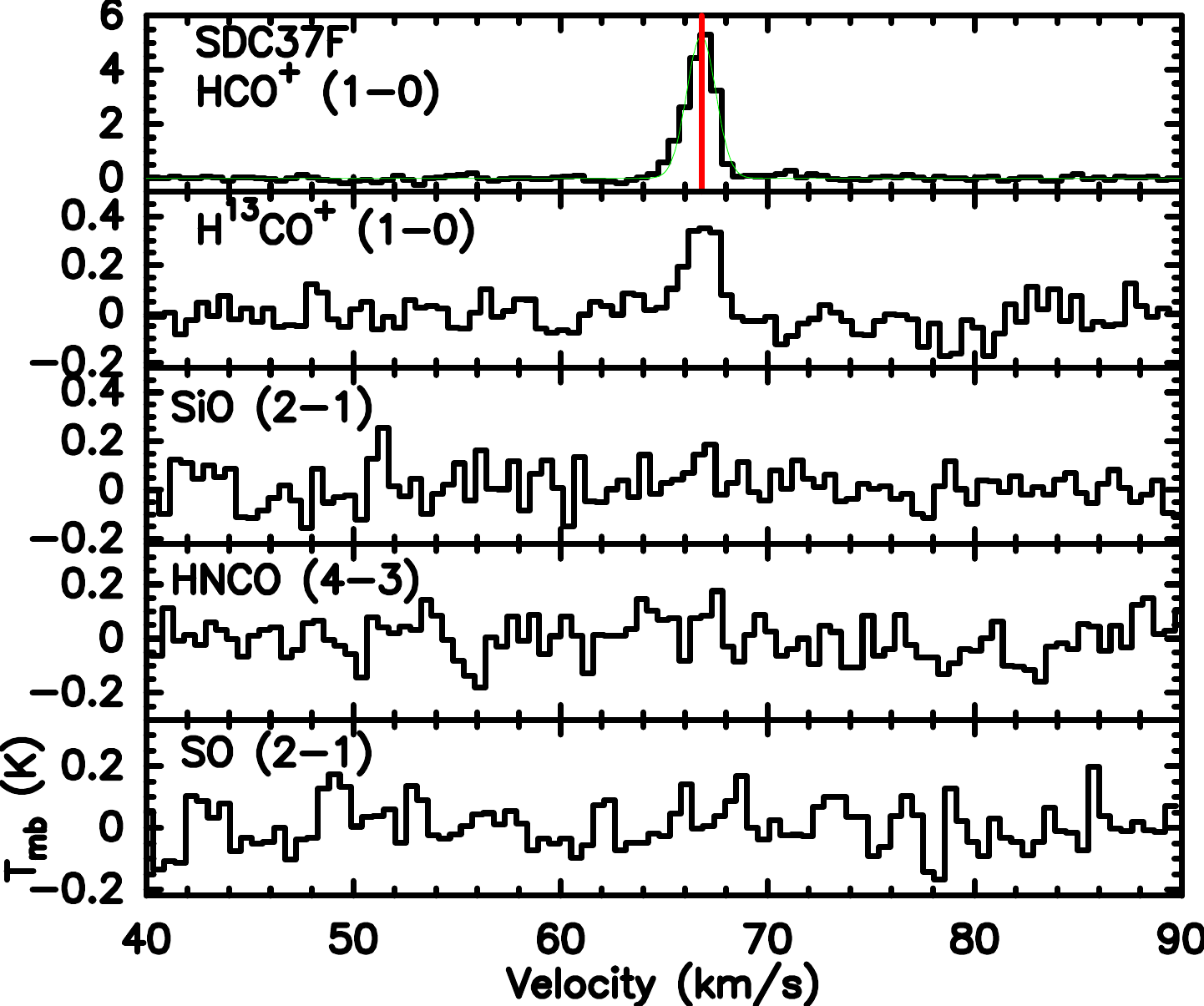}
\includegraphics[width=0.33\textwidth,height=0.43\textwidth,trim={0 0 0 0},clip]{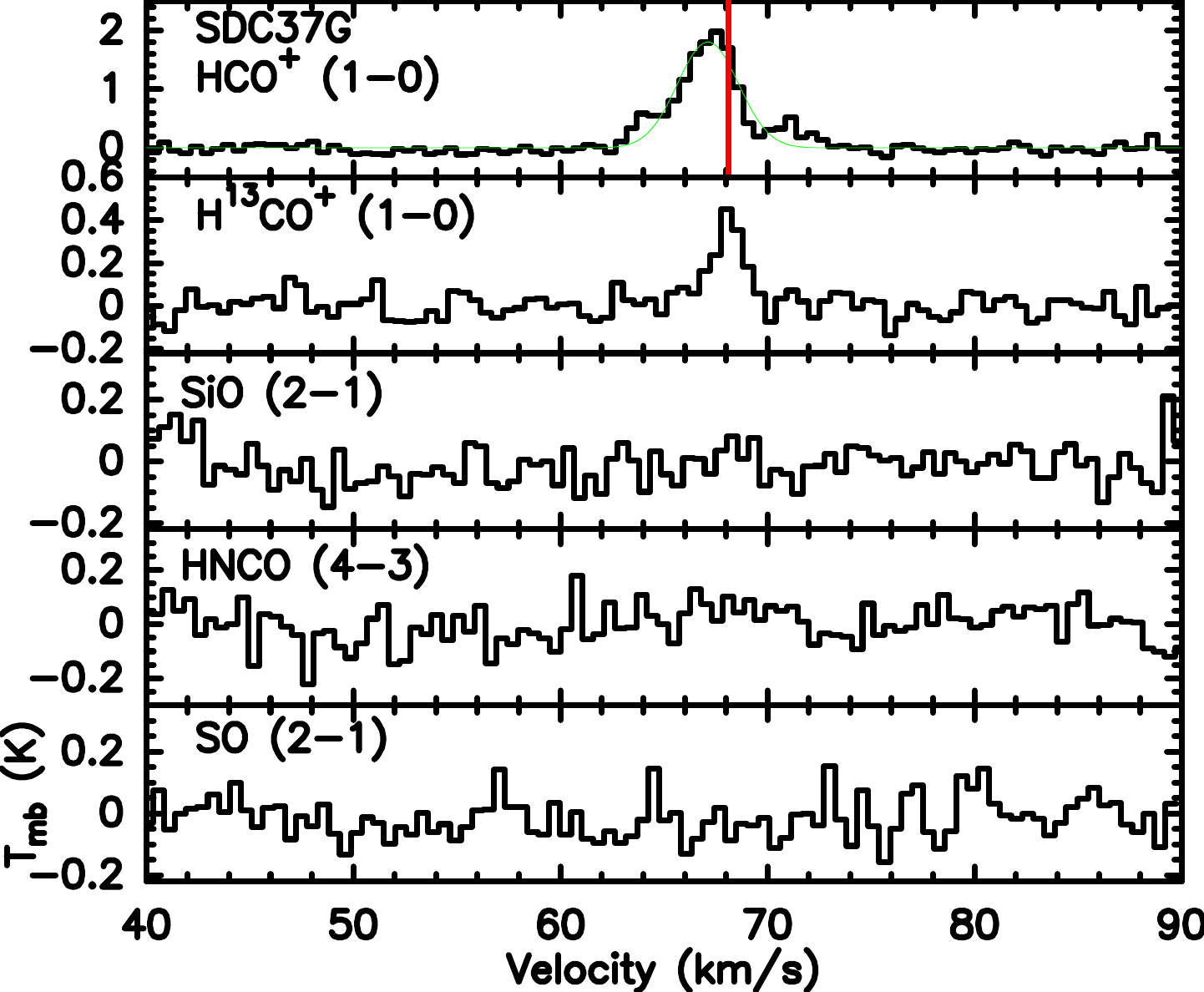}\\[3mm]

\includegraphics[width=0.24\textwidth,height=0.43\textwidth,trim={0 0 0 0},clip]{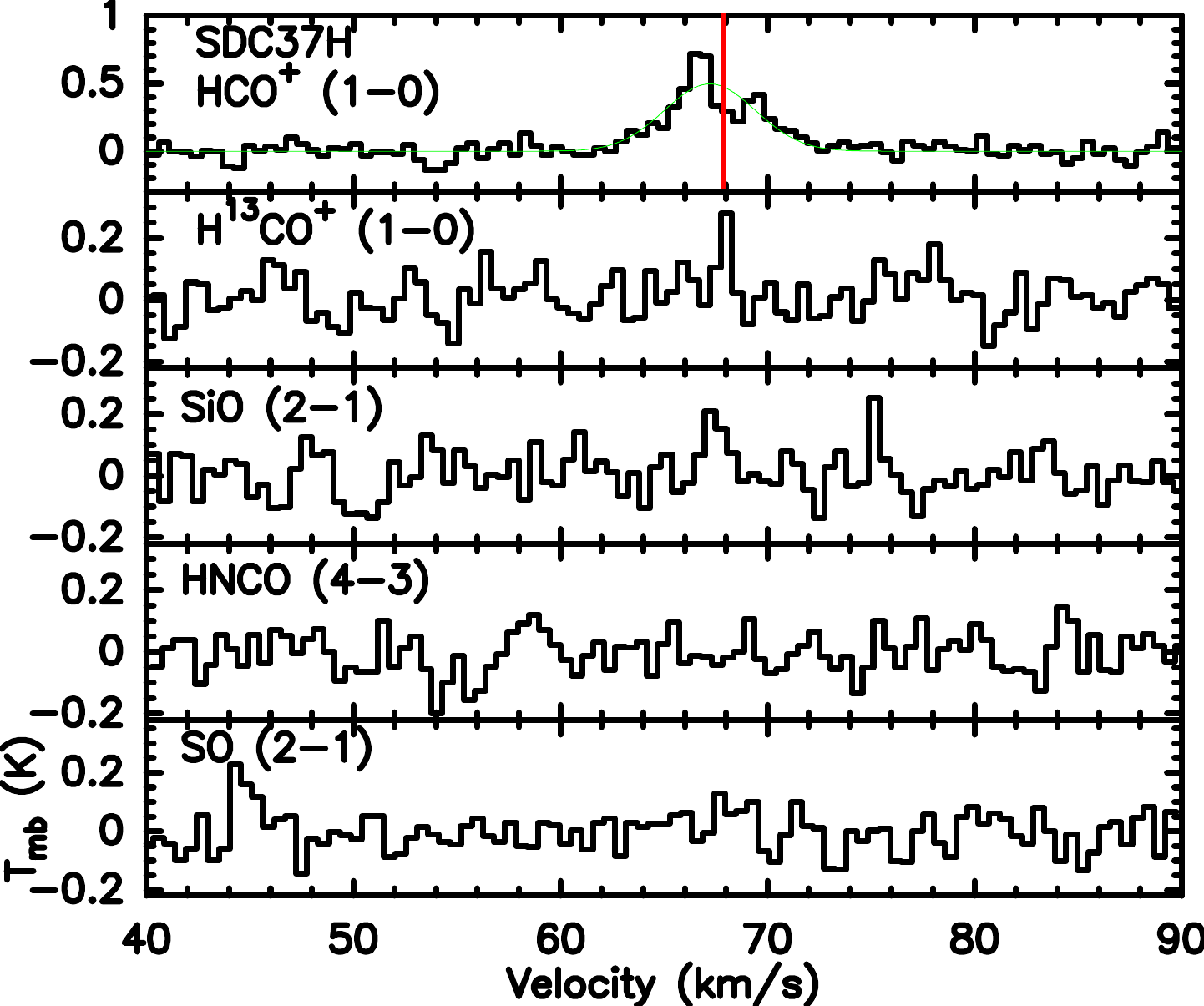}
\includegraphics[width=0.24\textwidth,height=0.43\textwidth,trim={0 0 0 0},clip]{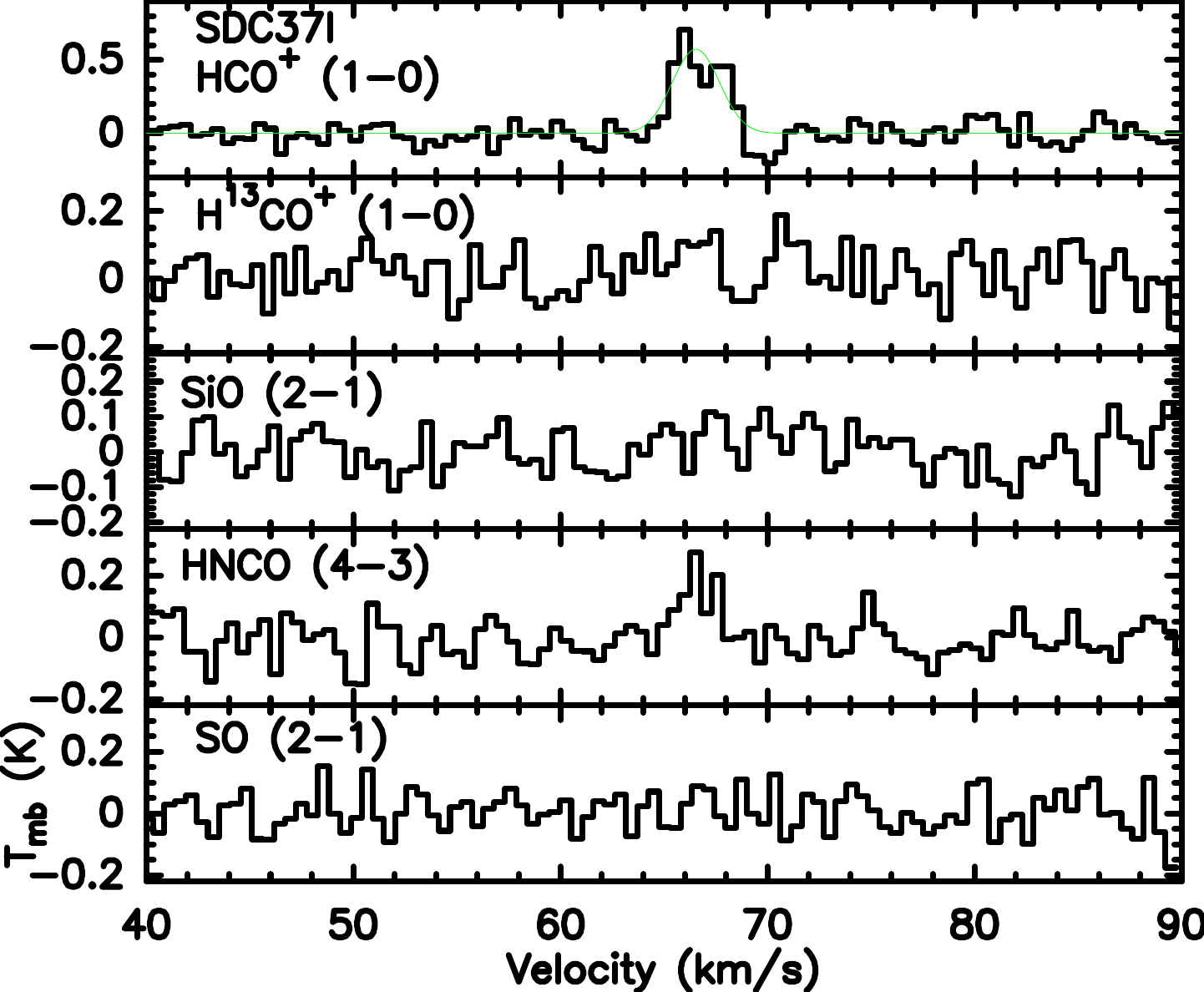}
\includegraphics[width=0.24\textwidth,height=0.43\textwidth,trim={0 0 0 0},clip]{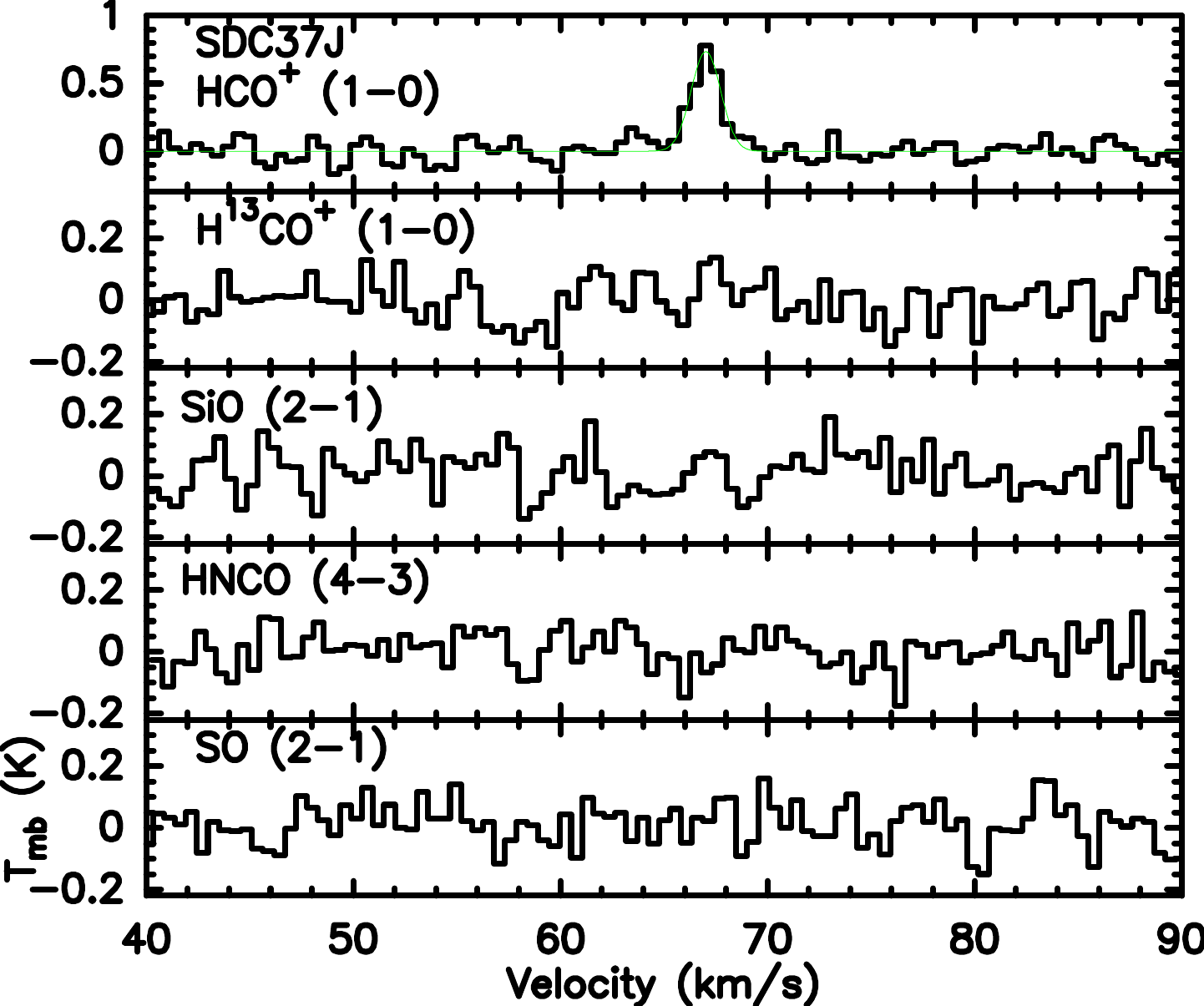}
\includegraphics[width=0.24\textwidth,height=0.43\textwidth,trim={0 0 0 0},clip]{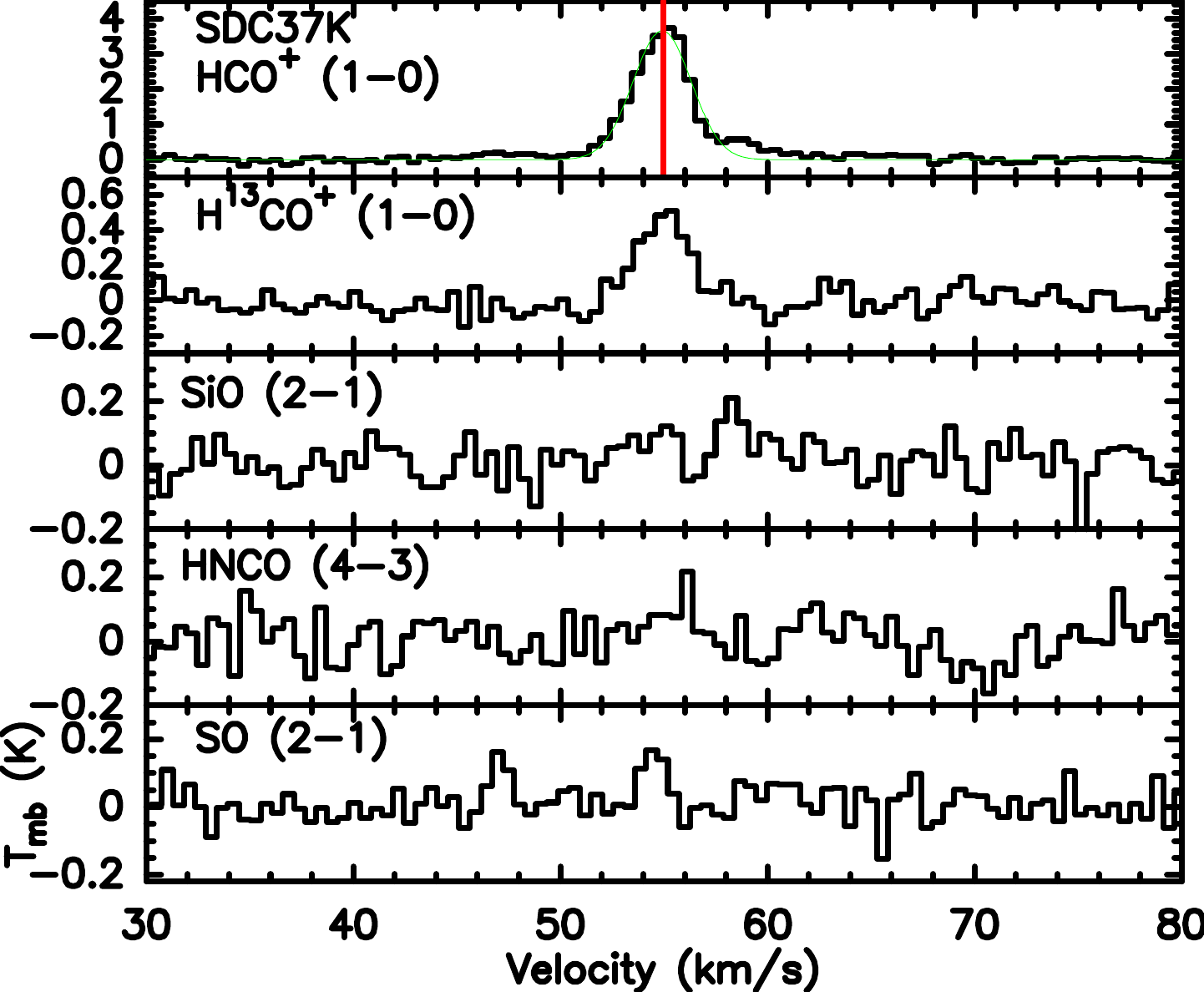}

\caption[]{Continued.}
\label{fig:kinematic_a_continued_5}
\end{figure}

\clearpage
\twocolumn

\end{appendix}

\end{CJK*}

\end{document}